\documentclass[11pt,letterpaper]{article}
\usepackage{authblk} 
\usepackage[titletoc,title]{appendix}
\usepackage{hyperref}
\hypersetup{colorlinks=true,citecolor=blue,linkcolor=blue,filecolor=blue,urlcolor=blue,breaklinks=true}
\usepackage[dvipsnames]{xcolor}
\usepackage{color}
\usepackage[marginal]{footmisc}
\usepackage{url}
\usepackage{cite}

\usepackage{mathtools}
\usepackage{amsmath}
\usepackage{amssymb}
\usepackage[shortlabels]{enumitem}
\usepackage{graphicx,epic,eepic,epsfig,latexsym,verbatim}
\usepackage{dsfont}

\usepackage{framed}
\definecolor{shadecolor}{rgb}{0.9,0.9,0.9}
\usepackage{subcaption}
\usepackage{caption}
\usepackage{float}
\usepackage{tikz}
\usepackage{tcolorbox}
\usepackage{relsize}

\usepackage{pgfplots}
\pgfplotsset{compat=newest}
\usetikzlibrary{plotmarks}
\usetikzlibrary{arrows.meta}
\usepgfplotslibrary{patchplots}

\definecolor{LightGray}{RGB}{220,220,220}
\definecolor{myred}{RGB}{236, 17, 0}
\definecolor{myblue}{RGB}{10, 88, 153}
\definecolor{myorange}{RGB}{236, 137, 0}
\definecolor{mygreen}{RGB}{26, 152, 81}

\usepackage{amsthm}
\newtheorem{definition}{Definition}
\newtheorem{lemma}{Lemma}
\newtheorem{proposition}[lemma]{Proposition}
\newtheorem{corollary}[lemma]{Corollary}

\newtheorem{theorem}{Theorem}

\newtheorem{fact}{Fact}

\def\squareforqed{\hbox{\rlap{$\sqcap$}$\sqcup$}}
\def\qed{\ifmmode\squareforqed\else{\unskip\nobreak\hfil
      \penalty50\hskip1em\null\nobreak\hfil\squareforqed
      \parfillskip=0pt\finalhyphendemerits=0\endgraf}\fi}
\def\endenv{\ifmmode\;\else{\unskip\nobreak\hfil
      \penalty50\hskip1em\null\nobreak\hfil\;
      \parfillskip=0pt\finalhyphendemerits=0\endgraf}\fi}

\newcounter{remark}
\newenvironment{remark}[1][]{\refstepcounter{remark}\par\medskip\noindent%
  \textbf{Remark~\theremark #1} }{\medskip}

\newcounter{example}

\mathchardef\ordinarycolon\mathcode`\:
\mathcode`\:=\string"8000
\def\vcentcolon{\mathrel{\mathop\ordinarycolon}}
\begingroup \catcode`\:=\active
\lowercase{\endgroup
  \let :\vcentcolon
}

\usepackage{colortbl}
\usepackage{pifont}
\definecolor{Gray}{gray}{0.92}
\definecolor{Gray2}{gray}{0.75}
\definecolor{maroon}{cmyk}{0,0.87,0.68,0.32}
\usepackage{booktabs}
\usepackage{makecell}
\usepackage{diagbox}
\usepackage{multirow}

\newcommand{\bra}[1]{\langle#1|}
\newcommand{\ket}[1]{|#1\rangle}

\newcommand{\ketbra}[2]{|#1\rangle\!\langle#2|}
\newcommand{\braket}[2]{\langle#1|#2\rangle}

\newcommand{\tr}{\operatorname{Tr}}
\newcommand{\ox}{\otimes}
\newcommand{\grad}{\operatorname{grad}}
\newcommand{\frob}{\mathrm{F}}
\newcommand{\cA}{{\cal A}}

\newcommand{\cD}{{\cal D}}
\newcommand{\cE}{{\cal E}}

\newcommand{\cH}{{\cal H}}
\newcommand{\cI}{{\cal I}}

\newcommand{\cL}{{\cal L}}
\newcommand{\cM}{{\cal M}}
\newcommand{\cN}{{\cal N}}
\newcommand{\cO}{{\cal O}}
\newcommand{\cP}{{\cal P}}

\newcommand{\cR}{{\cal R}}

\newcommand{\cT}{{\cal T}}
\newcommand{\cV}{{\cal V}}
\newcommand{\cU}{{\cal U}}

\newcommand{\cZ}{{\cal Z}}

\newcommand{\RR}{{{\mathbb R}}}
\newcommand{\CC}{{{\mathbb C}}}
\newcommand{\FF}{{{\mathbb F}}}
\newcommand{\NN}{{{\mathbb N}}}

\newcommand{\1}{{\mathds{1}}}
\newcommand{\supp}{{\operatorname{supp}}}

\def\ve{\varepsilon}

\makeatletter
\newcommand*\rel@kern[1]{\kern#1\dimexpr\macc@kerna}
\newcommand*\widebar[1]{%
  \begingroup
  \def\mathaccent##1##2{%
    \rel@kern{0.8}%
    \overline{\rel@kern{-0.8}\macc@nucleus\rel@kern{0.2}}%
    \rel@kern{-0.2}%
  }%
  \macc@depth\@ne
  \let\math@bgroup\@empty \let\math@egroup\macc@set@skewchar
  \mathsurround\z@ \frozen@everymath{\mathgroup\macc@group\relax}%
  \macc@set@skewchar\relax
  \let\mathaccentV\macc@nested@a
  \macc@nested@a\relax111{#1}%
  \endgroup
}
\makeatother

\DeclareMathOperator{\tangent}{T}

\usepackage{algorithm}
\usepackage{algorithmic}

\usepackage{mathrsfs}

\newcommand{\density}{\mathscr{D}}

\newcommand{\suchthat}{\text{\rm s.t.}}

\newcommand{\CPTP}{\text{\rm CPTP}}

\newcommand{\CP}{\text{\rm CP}}

\newcommand{\dg}{\operatorname{dg}}

\usepackage{stmaryrd}

\usepackage{pgfplots}

\usepackage{microtype}

\usepackage{fullpage}
\allowdisplaybreaks

\begin{document}

\title{\textbf{Geometric optimization for quantum communication}}

\author[1]{Chengkai Zhu }
\author[2]{Hongyu Mao}
\author[2]{Kun Fang \thanks{kunfang@cuhk.edu.cn}}
\author[1]{Xin Wang \thanks{felixxinwang@hkust-gz.edu.cn}}
\affil[1]{\small Thrust of Artificial Intelligence, Information Hub,\protect\\  The Hong Kong University of Science and Technology (Guangzhou), Guangdong 511453, China}
\affil[2]{\small School of Data Science, The Chinese University of Hong Kong, Shenzhen, Guangdong, 518172, China}
\date{\today}
\maketitle

\begin{abstract}
Determining the ultimate limits of quantum communication, such as the quantum capacity of a channel and the distillable entanglement of a shared state, remains a central challenge in quantum information theory, primarily due to the phenomenon of superadditivity. This work develops Riemannian optimization methods to establish significantly tighter, computable two-sided bounds on these fundamental quantities. 
For upper bounds, our method systematically searches for state and channel extensions that minimize known information-theoretic bounds. We achieve this by parameterizing the space of all possible extensions as a Stiefel manifold, enabling a universal search that overcomes the limitations of ad-hoc constructions. Combined with an improved upper bound on the one-way distillable entanglement based on a refined continuity bound on quantum conditional entropy, our approach yields new state-of-the-art upper bounds on the quantum capacity of the qubit depolarizing channel for large values of the depolarizing parameter, strictly improving the previously best-known bounds. 
For lower bounds, we introduce Riemannian optimization methods to compute multi-shot coherent information. We establish lower bounds on the one-way distillable entanglement by parameterizing quantum instruments on the unitary manifold, and on the quantum capacity by parameterizing code states with a product of unitary manifolds. Numerical results for noisy entangled states and different channels demonstrate that our methods successfully unlock superadditive gains, improving previous results. Together, these findings establish Riemannian optimization as a principled and powerful tool for navigating the complex landscape of quantum communication limits. Furthermore, we prove that amortization does not enhance the channel coherent information, thereby closing a potential avenue for improving capacity lower bounds in general. This result can be of independent interest.
\end{abstract}

\newpage
\setcounter{tocdepth}{2}
{
\hypersetup{linkcolor=black}
\tableofcontents
}

\section{Introduction}

Since its inception, a central goal of quantum information theory has been to determine the ultimate physical limits on processing and transmitting information encoded in quantum systems. This pursuit extends Shannon's classical theory to the quantum realm~\cite{wilde2011classical,wilde2013quantum}, where the principles of superposition and entanglement introduce fundamentally new challenges and opportunities. Entanglement, in particular, has been identified as the key resource in quantum information processing protocols~\cite{Horodecki_2009}, e.g., quantum teleportation~\cite{Bennett1993a}, superdense coding~\cite{Bennett1992}, and quantum key distribution~\cite{Ekert1991,Renner2006}; and it underpins the potential advantages of quantum communication over classical methods~\cite{Bennett2002}. Consequently, two of the most vital problems in the field are quantifying the usable entanglement in a bipartite state $\rho_{AB}$ and determining the ultimate capacity of a quantum channel $\cN_{A\to B}$ to transmit it.

The operational measures for these tasks are the \textit{distillable entanglement} $D(\rho_{AB})$~\cite{Bennett1996,Bennett1996b}, and the \textit{quantum capacity} $Q(\cN_{A\to B})$~\cite{Graeme2010}, respectively. The former quantifies the ultimate rate at which maximally entangled states can be extracted from independent and identically distributed (i.i.d.) copies of $\rho_{AB}$ through local operations and classical communication (LOCC). When only one-way communication from Alice to Bob is allowed, we have the so-called \textit{one-way distillable entanglement} $D_{\to}(\cdot)$. The latter quantifies the ultimate rate for reliable quantum communication through a channel $\cN_{A\to B}$. These quantities are inextricably linked, as the quantum capacity could also be understood as the rate at which entanglement can be generated between a sender and receiver by using the channel.

Despite their fundamental importance, computing $D_{\to}(\cdot)$ and $Q(\cdot)$ for general states and channels remains a notoriously difficult problem. Both quantities are proved to be given via a regularization over many copies of the resource, i.e.,
\begin{equation}\label{Eq:intro_reg}
    D_{\to}(\rho_{AB}) = \lim_{n\to \infty} \frac{1}{n} D_{\to}^{(1)}(\rho_{AB}^{\ox n}),\quad \text{and} \quad Q(\cN_{A\to B}) = \lim_{n\to \infty} \frac{1}{n}Q^{(1)}(\cN_{A\to B}^{\ox n}),
\end{equation}
where $D_{\to}^{(1)}(\cdot)$ and $Q^{(1)}(\cdot)$ are \textit{single-shot} quantities based on the \textit{coherent information}. The regularization can be removed, making $D_{\to}(\cdot)$ and $Q(\cdot)$ solvable, for classes of \textit{degradable states}~\cite{Leditzky2018usefulsates} and \textit{degradable channels}~\cite{Devetak2005} whose $D_{\to}^{(1)}(\cdot)$ and $Q^{(1)}(\cdot)$ are shown to be additive. However, it is necessary in general because of the phenomenon of \textit{superadditivity}, where the coherent information of a joint state or channel can be strictly greater than the sum of its parts~\cite{Felix2018,Youngrong2018,Youngrong2019,Noh_2020,Vikesh2021,Felix2023,Wu2025}. This non-additivity makes the direct computation of the limits intractable. The computation of $D_{\to}(\cdot)$ and $Q(\cdot)$, even for qubit isotropic states and depolarizing channels, still remains a long-standing open problem in quantum information theory. Thereby, research has bifurcated into two threads: 
\begin{itemize}
    \item deriving upper bounds that prove fundamental limits on performance;
    \item establishing lower bounds by constructing explicit coding schemes.
\end{itemize}

For upper bounds, a prominent one on the distillable entanglement is the Rains bound~\cite{Rains2001}, which has inspired related information-theoretic upper bounds on the quantum capacity, including the Rains information~\cite{tomamichel2017strong}, max-Rains information~\cite{Wang2019}, and geometric-Rains information~\cite{fangGeometricRenyiDivergence2021}. Another major line of research leverages the concept of \textit{degradability}. Since the quantum capacity of a degradable channel (and correspondingly, one-way distillable entanglement of a degradable state) is known, which is a single-letter coherent information, continuity bounds based on \textit{approximate degradability} have been established for both one-way distillable entanglement and quantum capacity~\cite{Sutter2014approxepChannel,Leditzky2018usefulsates,jabbour2024tightening}. A related approach, which has also yielded competitive upper bounds, is to decompose a state or channel into a convex combination of a degradable and an antidegradable part~\cite{Leditzky2018usefulsates,Zhu_2024}.

Moreover, another intuitive and powerful strategy for deriving upper bounds is based on the construction of state or channel \textit{extensions}. This approach relies on the principle that for any extension $\hat{\rho}_{ABF}$ of a state $\rho_{AB}$, or $\widehat{\cN}_{A\to BF}$ of a channel $\cN_{A\to B}$, the monotonicity
\begin{equation}
    D_{\to}(\rho_{AB}) \leq D_{\to}(\hat{\rho}_{ABF}),\quad \text{and} \quad Q(\cN_{A\to B}) \leq Q(\widehat{\cN}_{A\to BF})
\end{equation}
must hold. This follows directly from the data processing inequality, as the receiver can freely discard the auxiliary system $F$. The challenge thus shifts to finding an \textit{optimal extension} that minimizes $D_{\to}(\cdot)$ or $Q(\cdot)$ of the extended object. A leading technique for this purpose is the so-called \textit{flag extension} method~\cite{Leditzky2018usefulsates,wang2021pursuing,fanizzaQuantumFlagsNew2020,kianvashBoundingQuantumCapacity2022}, which has proven highly effective for structured channels, e.g., Pauli channels~\cite{kianvashBoundingQuantumCapacity2022}. Notably, the state-of-the-art (SOTA) upper bounds on the quantum capacity of depolarizing channels are given by the flag extension method in~\cite{kianvashBoundingQuantumCapacity2022}. However, the flag extension method faces a fundamental optimization bottleneck. The construction is tied to a specific convex decomposition of the state or channel (e.g., a Kraus representation), and the subsequent optimization is performed over a chosen ansatz for the `flag' states. The efficacy of the method is therefore highly dependent on these initial choices, for which no systematic \textit{a priori} guidelines exist. It has been unclear how to unlock the full potential of the extension-based approach in a general, systematic manner.

For lower bounds, on the other hand, the most well-known ones on $D_{\to}^{(1)}(\cdot)$ and $Q^{(1)}(\cdot)$ are given by the \textit{coherent information} of a state~\cite{{devetakDistillationSecretKey2005}} or a channel~\cite{bennettMixedstateEntanglementQuantum1996}, known as the \textit{hashing bounds}. Particularly, a positive lower bound on the quantum capacity is a constructive proof that quantum communication is possible, i.e., there exists a \textit{quantum error correction code} for the noise modeled by that channel. Due to the superadditivity, one needs to calculate the $n$-shot coherent information to pursue the ultimate performance for a blocklength of $n$, where the optimization is over all possible input \textit{code states}. Historically, techniques in quantum error correction have been used to design highly structured codes~\cite{Shor1996,DiVincenzo1998a,Smith2007,Fern2008}. For certain channels, e.g., Pauli channels, exploiting the symmetries of these codes can make the coherent information calculation tractable~\cite{Johannes2021}. More recently, some permutation-invariant code states have been proposed to achieve a large blocklength evaluation of the coherent information and an improved capacity threshold~\cite{bhalerao2025}. However, these methods are often tailored to specific symmetries and code families. Other approaches have also been explored using neural network states as an ansatz~\cite{Bausch_2020} to find code states.

In this work, we directly tackle the challenges in computing $D_\to(\cdot)$ and $Q(\cdot)$ by introducing a comprehensive framework based on \textit{Riemannian optimization}. We develop distinct but philosophically aligned geometric approaches to obtain computable upper and lower bounds on these quantities. An overview of our results is illustrated in Figure~\ref{fig:overview} below.

\begin{figure}[H]
    \centering
    \includegraphics[width=1\linewidth]{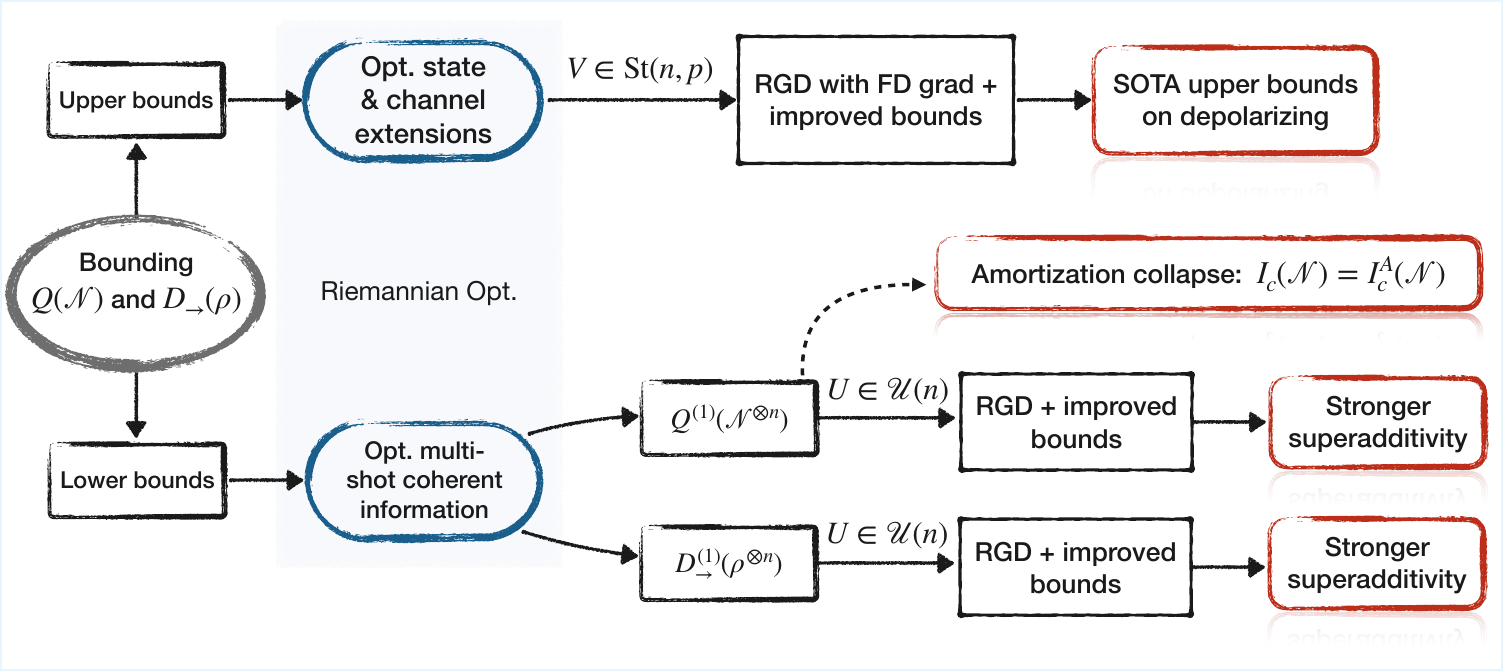}
    \caption{An overview of our geometric optimization methods for computing two-sided bounds on the quantum capacity and the one-way distillable entanglement.}
    \label{fig:overview}
\end{figure}

\subsection{Main results}
For upper bounds, we reframe the search for optimal state and channel extensions as a tractable Riemannian optimization problem. By characterizing the space of all valid extensions as a complex Stiefel manifold, our method systematically discovers the tightest possible bounds, moving beyond the limitations of previous ad-hoc flag constructions. We further enhance this framework by incorporating a recently improved continuity bound for conditional entropy~\cite{Koenraad2024,Mario2025} to derive a tighter bound on one-way distillable entanglement in terms of a state's approximate degradability (cf.~Theorem~\ref{thm:oneway_upperbound}), a result analogous to the upper bound on quantum capacity in~\cite[Proposition 10]{Mario2025}. We analyze the continuity property of the target objective function in our optimization, and use a finite-difference scheme to implement the \textit{Riemannian gradient descent} (RGD) algorithm to optimize our bounds (cf.~Section~\ref{sec:num_grad}).

The synergy between our optimization framework and the continuity bounds yields a powerful methodology for computing the tightest known upper bounds on $D_{\to}(\cdot)$ and $Q(\cdot)$. For the qubit isotropic state, our method produces an upper bound on one-way distillable entanglement that is strictly tighter than the best previously known result reported in \cite{kianvashBoundingQuantumCapacity2022}, thereby establishing an SOTA upper bound for the quantum capacity of the qubit depolarizing channel (cf.~Section~\ref{sec:improve_iso}). We demonstrate the versatility of our approach on other noisy states and channels, consistently achieving tighter estimates than existing methods.

For lower bounds on the one-way distillable entanglement, we formulate the computation of $D^{(1)}_{\to}(\cdot)$ as an optimization problem over the complex unitary manifold (cf.~Section~\ref{sec:lower_oneway_DE}). Then, we develop an RGD algorithm to estimate $D^{(1)}_{\to}(\cdot)$ for a multi-copy of a target state to obtain tighter lower bounds on $D_{\to}(\cdot)$ than the hashing bound. Our numerical results demonstrate the efficacy of our algorithm, which clearly shows the superadditivity of $D_{\to}^{(1)}(\cdot)$ for various bipartite states.

For lower bounds on the quantum capacity, we propose a scalable parameterization of the input code states using an interleaved local unitary ansatz. This transforms the optimization problem from a search over an exponentially large sphere to a tractable optimization over a product of low-dimensional unitary manifolds (cf.~Section~\ref{sec:lower_qcap}). We develop an RGD algorithm to optimize the channel coherent information (cf.~Corollary~\ref{cor:local_U_Riemangrad}), which allows for the efficient exploration of entangled input states. Numerical results show that our methods can reveal stronger superadditivity of the channel coherent information compared to previous code state constructions. Finally, we prove that amortization of the underlying channel coherent information does not enhance the original channel coherent information (cf.~Theorem~\ref{thm:Ic_amort}), closing a potential pathway for improving single-letter lower bounds.

All implementations of Riemannian algorithms in this work are based on a MATLAB toolbox \texttt{Manopt}~\cite{MATLAB,boumal2014manopt}, together with the software \texttt{CVX}~\cite{cvx} and \texttt{QETLAB}~\cite{qetlab}. The numerical experiments are performed on a 2.4 GHz AMD EPYC 7532 32-core Processor with 256 GB RAM under MATLAB R2024a. All MATLAB codes used to obtain numerical results, as well as all data, are publicly available at~\cite{coderepo}.

\subsection{Structure of the paper}
The remainder of this paper is organized as follows. 
In Section~\ref{sec:pre}, we introduce some notation we used. We first develop the framework for upper bounds in Section~\ref{sec:upper}, beginning with our unified framework on state and channel extensions in Section~\ref{sec:extension}. Section~\ref{sec:upper_oneway_DE} presents upper bounds on the one-way distillable entanglement by combining the improved continuity bound with the state extension method. Section~\ref{sec:upper_qcapacity} presents upper bounds on the quantum capacity using the channel extension method. Section~\ref{sec:improve_iso} establishes our improved upper bound on the quantum capacity of qubit depolarizing channels, and Section~\ref{sec:upper_eg} includes further examples on states and channels. 

Then, we tackle the lower bounds in Section~\ref{sec:lower}. Section~\ref{sec:lower_oneway_DE} introduces our Riemannian optimization method for computing lower bounds on the one-way distillable entanglement of bipartite states. Section~\ref{sec:lower_qcap} introduces our Riemannian optimization method for computing lower bounds on the quantum capacity. Section~\ref{sec:superadd_state} and Section~\ref{sec:superadd_channel} present numerical results on demonstrating concrete superadditivity for states and channels, respectively, via our proposed methods. Section~\ref{sec:amort} presents the result on amortized channel coherent information. Finally, Section~\ref{sec:conclu} provides concluding remarks and outlooks.

\section{Preliminaries}\label{sec:pre}
A quantum system, denoted by a capital letter such as $A$, is modeled by a finite-dimensional Hilbert space $\cH_A$. The dimension of this space is denoted by $|A|$. The sets of linear operators and positive semidefinite operators on $\cH_A$ are denoted by $\cL(A)$ and $\cP(A)$, respectively. The set of \emph{quantum states} on system $A$ is denoted as $\density(A)\coloneqq \{\rho \in \cP(A):\tr\rho = 1\}$. A sub-normalized state is a positive semidefinite operator with trace no greater than one. The identity operator on system $A$ is denoted by $\1_A$. The trace norm of $X$ is given by $\|X\|_1 \coloneqq \tr \sqrt{X^\dagger X}$, and its Frobenius norm is $\|X\|_\frob \coloneqq \sqrt{\tr X^\dag X}$. The operator norm $\|X\|_\infty$ is defined as the maximum eigenvalue of $\sqrt{X^\dagger X}$. The support of a Hermitian operator $X$ is denoted by $\supp(X)$. 

The von Neumann entropy of a quantum state is defined as $H(\rho) \coloneqq -\tr(\rho \log \rho)$ where all logarithms in this work are taken in base two. The coherent information of a bipartite quantum state $\rho_{AB}$ is defined by $I(A\rangle B)_{\rho} \coloneqq H(\rho_B) - H(\rho_{AB})$ and the conditional entropy by $H(A|B)_\rho \coloneqq -I(A\rangle B)_\rho$. For $p\in [0,1]$, the binary entropy is defined by $h(p) \coloneqq -p \log p - (1-p) \log(1-p)$ and the bosonic entropy is defined by $g(p)\coloneqq \left(1+p\right)h\left(p/(1+p)\right)$.  

A \emph{quantum channel} $\cN_{A\to B}$ is a completely positive and trace-preserving (CPTP) linear map from $\cL(A)$ to $\cL(B)$. A \emph{subchannel} $\cM_{A\to B}$ is a completely positive and trace non-increasing linear map from $\cL(A)$ to $\cL(B)$. We denote by $\CP(A : B)$ the set of completely positive (CP) maps from $A$ to $B$, and by $\CPTP (A : B)$ the set of all CPTP maps. The \emph{diamond norm} of a linear map is defined by $\|\cE\|_\diamond := \sup_{\rho\in \density(RA)}\|\cE_{A\to B}(\rho_{RA})\|_1$. We will drop the identity map $\cI$ and identity operator $\1$ if they are clear from the context. The \emph{Choi matrix} of a linear map $\cN_{A'\to B}$ is defined as $J_{\cN} = (\cI_{A}\ox\cN_{A'\to B})(\ketbra{\Gamma}{\Gamma}_{A'A})$, where $\ket{\Gamma}_{A'A} = \sum_{i} \ket{i}_{A'}\ket{i}_A$ is the unnormalized maximally entangled state. For every quantum channel $\cN_{A\to B}$, there exists an isometry $V:\cH_A\to \cH_B\ox \cH_E$ such that $\cN(\rho) = \tr_E (V\rho V^\dag)$ where $\cH_E$ is an environment space and $V$ is called a \emph{Stinespring isometry}. The \textit{complementary channel} of $\cN$ is denoted by $\cN^c \in \CPTP(A:E)$ that acts as $\cN^c(\rho) = \tr_B (V\rho V^\dagger)$. The action of a quantum channel also has a Kraus representation, i.e., $\cN(\cdot) = \sum_i K_i^{} (\cdot) K_i^\dagger$ with $\sum_i K_i^\dagger K_i^{} = \1$.

\section{Upper bounds on quantum communication}\label{sec:upper}
In this section, we develop Riemannian optimization methods for estimating upper bounds on the one-way distillable entanglement and the quantum capacity. First, we introduce our methods for constructing states and channel extensions.

\subsection{Extensions of quantum states and channels}\label{sec:extension}
A common strategy for constructing extensions of states or channels is referred to as the \textit{flag extension} method, which relies on a convex decomposition of the state or channel~\cite{Leditzky2018usefulsates,wang2021pursuing,fanizzaQuantumFlagsNew2020,kianvashBoundingQuantumCapacity2022}. For a given bipartite quantum state $\rho_{AB}$, one first decomposes it into $\rho_{AB} = \sum_{i} p_i \omega_{AB}^{(i)}$ where $\{\omega_{AB}^{(i)}\}_i$ are normalized states. An extension is then constructed by appending an auxiliary \textit{flag} state $\tau_F^{(i)}$ to each $\omega_{AB}^{(i)}$, i.e.,
\begin{equation}
    \hat{\rho}_{ABF} \coloneqq \sum_{i} p_i \omega_{AB}^{(i)} \ox \tau_F^{(i)}.
\end{equation}
By construction, tracing out the flag system $F$ recovers the original state. A similar procedure applies to a quantum channel $\cN_{A\to B}$. Using a Kraus representation $\cN_{A\to B}(\cdot) = \sum_i p_i K_i(\cdot)K_i^\dag$, one can define an extended channel where the action of each Kraus operator is correlated with a corresponding flag state, i.e.,
\begin{equation}
    \widehat{\cN}_{A\to BF}(\cdot) \coloneqq \sum_i p_i K_i(\cdot)K_i^\dag \ox \tau_F^{(i)}.
\end{equation}
It is clear that $\tr_F \widehat{\cN}_{A\to BF}(\rho_{A}) = \cN_{A\to B}(\rho_{A})$ for any input state $\rho_A$. The central challenge of this method is that the subsequent optimization over the flag state ensemble $\{\tau_F^{(i)}\}_i$ is fundamentally tied to the initial choice of decomposition. To establish a more general framework that avoids this dependency, we begin by formally defining state and channel extensions.

\begin{definition}[State extension]
For any given bipartite quantum state $\rho_{AB}$, a quantum state $\hat{\rho}_{ABF}\in\mathscr{D}(ABF)$ is called an extended state of $\rho_{AB}$ if $\tr_F \hat{\rho}_{ABF} = \rho_{AB}$.
\end{definition}

\begin{definition}[Channel extension]
For any given quantum channel $\cN\in\CPTP(A:B)$, a quantum channel $\widehat{\cN}\in\CPTP(A:BF)$ is called an extended channel of $\cN$ if 
\begin{equation}
    \cN(\rho) = \tr_F \widehat{\cN}(\rho),~\forall \rho\in\mathscr{D}(A).
\end{equation}
\end{definition}

After direct observation, we can gather the following facts about any state or channel extension.

\begin{shaded}
\begin{fact}\label{fact:state_ext}
For any bipartite quantum state $\rho_{AB}$, $\hat{\rho}_{ABF}$ is an extended state of $\rho_{AB}$ if and only if there is an environment system $R$ and an isometry $V:\cH_E\to \cH_R\ox \cH_F$ such that $\hat{\rho}_{ABF} = \tr_{R} (V \ketbra{\phi}{\phi} V^\dag)$ where $\ket{\phi}_{ABE}$ is a purification of $\rho_{AB}$.
\end{fact}
\end{shaded}

\begin{proof}
For the if-part, assume $\hat{\rho}_{ABF} = \tr_{R} (V \ketbra{\phi}{\phi} V^\dag)$ where $V:\cH_E\to \cH_R\ox \cH_F$. We can check that
\begin{equation}
\begin{aligned}
\tr_F \hat{\rho}_{ABF} &= \tr_{RF} \big[(\1_{AB}\ox V) \ketbra{\phi}{\phi} (\1_{AB}\ox V^\dag)\big]\\
&= \tr_{E} \big[(\1_{AB}\ox V^\dag V) \ketbra{\phi}{\phi} \big] = \tr_E\ketbra{\phi}{\phi} = \rho_{AB}.
\end{aligned}
\end{equation}
For the only if-part, assume $\hat{\rho}_{ABF}$ is an extension of $\rho_{AB}$, i.e., $\tr_F \hat{\rho}_{ABF} = \rho_{AB}$. Let $\ket{\phi}_{ABE}$ and $\ket{\psi}_{ABFR}$ be purifications of $\rho_{AB}$ and $\hat{\rho}_{ABF}$, respectively. The fact that $\tr_{FR}\ketbra{\psi}{\psi}_{ABFR} = \rho_{AB}$ means that $\ket{\psi}_{ABFR}$ is also a purification of $\rho_{AB}$. Then there must exist an isometry $V:\cH_{E}\to \cH_F\ox\cH_R$ such that $\ket{\psi}_{ABFR} = (\1_{AB}\ox V)\ket{\phi}_{ABE}$. Hence, it is obvious to express $\hat{\rho}_{ABF}$ as $\hat{\rho}_{ABF} = \tr_{R} (V \ketbra{\phi}{\phi} V^\dag)$ which completes the proof.
\end{proof}

\begin{shaded}
\begin{fact}\label{fact:channel_ext}
For any given quantum channel $\cN \in \CPTP(A : B)$, a quantum channel $\widehat{\cN}\in\CPTP(A:BF)$ is an extended channel of $\cN$ if and only if there is an environment system $R$ and an isometry $V: \cH_E \to \cH_R\ox \cH_F$ such that $W = (\1_B \ox V)U$ is a Stinespring isometry of $\widehat{\cN}$ where $U:\cH_A\to \cH_B\ox \cH_E$ is a Stinespring isometry of $\cN$. 
\end{fact}
\end{shaded}

\begin{proof}
The if-part is again a direct calculation. Assume $\widehat{\cN}_{A\to BF}$ has an Stinespring isometry $W = (\1_B \ox V)U$. We have that
\begin{equation}
\begin{aligned}
\tr_F\widehat{\cN}(\rho) &= \tr_{FR}\big[(\1_B \ox V)U\rho U^\dag(\1_B \ox V^\dag)\big]\\
&= \tr_E\big[(\1_B \ox V^\dag V)U\rho U^\dag\big] = \tr_E[U\rho U^\dag] = \cN(\rho).
\end{aligned}
\end{equation}
For the only if-part, assume $\widehat{\cN}\in\CPTP(A:BF)$ is an extended channel of $\cN$ which has an isometry $U_{\cN}:\cH_A\to \cH_B\ox\cH_E$, and $U_{\widehat{\cN}}:\cH_A\to \cH_B\ox \cH_R\ox \cH_F$ is an isometry of $\widehat \cN$. We have that $\cN(\rho) = \tr_{FR}[U_{\widehat{\cN}}\rho U_{\widehat{\cN}}^\dag]$ which shows $U_{\widehat{\cN}}$ is an isometry of $\cN$ as well. Then there exists an isometry $V:\cH_E\to \cH_R\ox \cH_F$ such that $U_{\widehat{\cN}}=(\1_B\ox V)U_{\cN}$, which completes the proof.
\end{proof}

\subsubsection{Riemannian optimization for extensions}\label{sec:RiemanOpt}
Based on the two facts above, the key insight of our work is that the space of all valid extensions for a given state or channel can be endowed with a \textit{Riemannian manifold} structure. Consequently, the problem of optimizing a chosen figure of merit over this space can be cast as a Riemannian optimization problem, allowing for the direct application of its powerful numerical tools; see Ref.~\cite{absil2008optimization} for an overview.

Taking the state extension as an example, for a given bipartite quantum state $\rho_{AB}$, suppose the figure of merit we are interested in is to minimize a function $f(\rho_{ABF})$ where $\rho_{ABF}$ is any extension of $\rho_{AB}$. Then using Fact~\ref{fact:state_ext}, for any fixed $|F|,|R|\in\NN_+$, taking $\rho_{ABF} = \tr_{R} (V \ketbra{\phi}{\phi} V^\dag)$ into the objective, the problem 
\begin{equation}
    \min_{V \in \mathrm{St}(|FR|,|E|)} f(V)
\end{equation}
reduces to a constrained optimization problem over the isometries. The feasible region of these isometries is the \textit{complex Stiefel manifold} 
\begin{equation}
\mathrm{St}(|FR|,|E|) \coloneqq \big\{V\in\CC^{|FR|\times |E|}: V^\dag V = \1\big\},
\end{equation}
a well-studied Riemannian manifold~\cite{atiyah1960complex}. This naturally leads us to the framework of \textit{Riemannian optimization}, which leverages the geometric structure of the manifold to generate iterates that remain within the feasible set~\cite{absil2008optimization}. Optimization problems on Riemannian manifolds have appeared in various areas of quantum information sciences (see, e.g., Refs.~\cite{Luchnikov_2021,Casanova2024,Kotil_2024,Hsu2024,Zhu_2025,Le2025riemannianquantum,zhu2025R,Li2025} as a very incomplete list).

A central element of many Riemannian optimization methods, e.g., Riemannian gradient descent (RGD)~\cite{Bonnabel_2013}, is the \textit{Riemannian gradient}. This gradient provides the direction of steepest ascent in the \textit{tangent space} at a given point on the manifold, guiding the optimization process. However, a key challenge in many problems we would encounter is that the objective function $f(V)$ is inherently non-smooth, meaning its gradient may not be defined at all points. This necessitates the use of optimization techniques suited for non-smooth problems, such as subgradient or gradient-free methods. We will further see details in Section~\ref{sec:num_grad}.

\subsection{Upper bounds on the one-way distillable entanglement}\label{sec:upper_oneway_DE}

In this section, we first establish an improved continuity bound on the one-way distillable entanglement based on the recently improved continuity bound on the quantum conditional entropy as shown in Lemma~\ref{lem:better_cont_bound}. Then, we apply the Riemannian optimization methods for searching extensions of a given state to give tighter upper bounds on the one-way distillable entanglement, combining with the continuity bound.

For the task of entanglement distillation from $n$ i.i.d. copies of a bipartite state $\rho_{AB}$, the objective is to transform the initial state $\rho_{AB}^{\ox n}$ into $m$ copies of a Bell state $\Phi$ via LOCC, such that the fidelity with the target state $\Phi^{\ox m}$ approaches one in the asymptotic limit ($n \to \infty$). When the classical communication is restricted from Alice to Bob, the protocol class is known as \textit{one-way LOCC}. The corresponding optimal rate of distillation is the \textit{one-way distillable entanglement}, denoted $D_{\to}(\rho_{A|B})$, where the notation $A|B$ emphasizes that the restricted communication direction (only from $A$ to $B$ is possible). A seminal work by Devetak and Winter~\cite{devetakDistillationSecretKey2005} provides a characterization of this quantity via a regularized formula
\begin{align}
    D_{\to}(\rho_{A|B}) = \lim_{n\to \infty} \frac{1}{n} D_{\to}^{(1)}(\rho_{A|B}^{\ox n}),
\end{align}
where $D_{\to}^{(1)}(\rho_{A|B})$ can be expressed as
\begin{align}\label{Eq:qinstr_oneway}
    D_{\to}^{(1)}(\rho_{A|B}) = \max_{\cT} I(A'\rangle BM)_{\cT(\rho_{AB})}.
\end{align}
Here $\cT$ is a quantum instrument, i.e.,
\begin{align}
    \cT: A \to A'M, \quad \cT(\rho_A) := \sum_{j=1}^m T_j(\rho_{A}) \ox \ketbra{j}{j},
\end{align}
where for each $j,~T_j\in\CP(A:A')$ and $\sum_j T_j \in \CPTP(A:A')$, and $M$ is a classical system to store measurement results with $|M|=m$. Equivalently, Eq.~\eqref{Eq:qinstr_oneway} can be expressed as
\begin{align}
    D_{\to}^{(1)}(\rho_{A|B}) = \max_{\cT} \sum_{j=1}^m \lambda_j I(A'\rangle B)_{\rho_j},
\end{align}
where $\lambda_j$ is equal to the probability of outcome $j$, i.e., $\lambda_j = \tr[T_j(\rho_{AB})]$, $\rho_j = 1/\lambda_jT_j(\rho_{AB})$ is the state after measurement with outcome $j$.

A bipartite state $\rho_{AB}$ is termed \textit{degradable} if there exists a quantum channel $\cM_{B\to E}$ such that $\cM_{B\to E}(\rho_{AB}) = \tr_B \ketbra{\phi}{\phi}_{ABE}$, where $\ket{\phi}_{ABE}$ is a purification of $\rho_{AB}$~\cite{Leditzky2018usefulsates}. The significance of this property is that the coherent information is additive for any degradable state, which simplifies the regularized expression for one-way distillable entanglement, i.e., $D_{\to}(\rho_{AB}) = I(A\rangle B)_{\rho}$~\cite{Leditzky2018usefulsates}. In the same work, the approximate degradability was introduced as follows.

\begin{definition}[Degradability parameter of bipartite states~\cite{Leditzky2018usefulsates}]
Let $\rho \in \density(AB)$ be a bipartite quantum state with purification $\ket{\rho}_{ABE}$, where $E$ denotes the environment system. We say that $\rho$ is $\ve$-degradable if there exists a quantum channel $\cM \in \CPTP(B:E')$ such that $E\cong E'$ and
\begin{align}
\frac{1}{2}\big\|\rho_{AE} - \cM_{B\to E'}(\rho_{AB})\big\|_1 \leq \ve.
\end{align}
The channel $\cM$ is referred to as an $\ve$-degrading channel of $\rho_{AB}$. The degradability parameter of $\rho_{AB}$ is defined as
\begin{align}
\dg(\rho) \coloneqq \min_{\cM \in \CPTP(B:E')} \frac{1}{2}\big\|\rho_{AE} - \cM_{B\to E'}(\rho_{AB})\big\|_{1}.
\end{align}
\end{definition} 

Notably, the degradability parameter of a given state can be efficiently computed via semidefinite programming (SDP)~\cite[Lemma 2.10]{Leditzky2018usefulsates}. Based on the approximate degradability, an efficiently computable upper bound on the one-way distillable entanglement was established in~\cite[Theorem 2.12]{Leditzky2018usefulsates}. We remark that this upper bound is essentially also based on the continuity bound of the conditional entropy, known as the Alicki-Fannes inequality~\cite{Alicki_2004}, and was improved by Winter later~\cite{winterTightUniformContinuity2016}. Recently, this continuity bound has been further improved with an additional marginal restriction on the considered states in~\cite[Theorem 5]{Koenraad2024} and~\cite[Theorem 5]{Mario2025}.

\begin{lemma}[Continuity of the conditional entropy \cite{winterTightUniformContinuity2016}]\label{lem:winter_cont_bound}
For any two quantum states $\rho, \sigma \in \density(AB)$, if $\frac{1}{2} \|\rho - \sigma\|_{1} \leq \ve \leq 1$, it holds that
\begin{align}\label{eq: winter, continuity bounds of the conditional entropy}
|H(A|B)_\rho-H(A|B)_\sigma|\leq \ve\log(|A|^2)+g(\ve).
\end{align}
\end{lemma}

\begin{lemma}[Continuity of the conditional entropy~\cite{Koenraad2024,Mario2025}]\label{lem:better_cont_bound}
For two quantum states $\rho, \sigma \in \density(AB)$ with equal marginals $\rho_B = \sigma_B$ and $ \frac{1}{2}\|\rho - \sigma\|_1 = \ve$ for some $\ve\in[0,1]$, it holds that
\begin{align}
\left| H(A|B)_{\rho} - H(A|B)_{\sigma}\right| \leq \ve \log(\left| A \right|^2 - 1) + h(\ve).
\end{align}
\end{lemma}

\begin{figure}[h]
    \centering
    \begin{tikzpicture}
    \node[anchor=south west,inner sep=0] (image) at (0,0) {\includegraphics[width=0.85\linewidth]{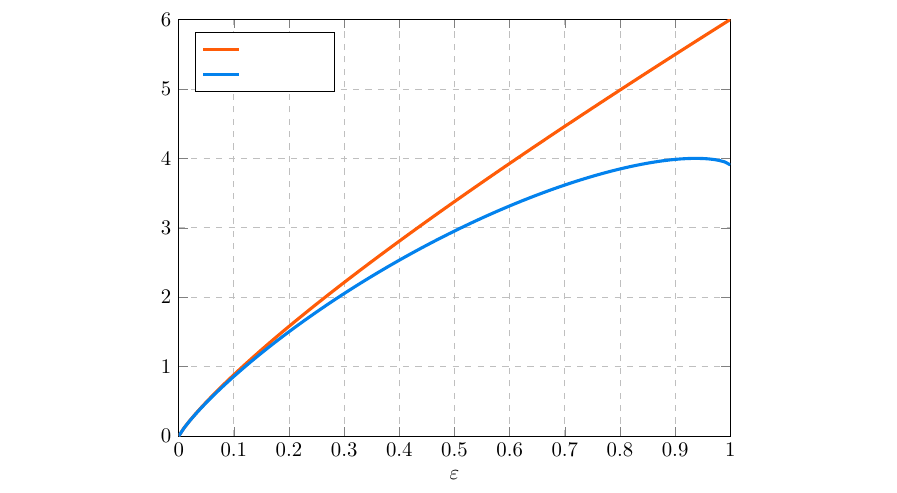}};
    \node[anchor=south west,inner sep=0] at (3.82,6.85){\footnotesize Lemma~\ref{lem:winter_cont_bound}};
    \node[anchor=south west,inner sep=0] at (3.82,6.45){\footnotesize Lemma~\ref{lem:better_cont_bound}};
    \end{tikzpicture}
    \caption{Continuity bounds comparison, $\left| A \right| = 4$.}
    \label{fig:tail_comparison}
\end{figure}

\noindent In Figure~\ref{fig:tail_comparison}, we can observe the difference between the two continuity bounds through a numerical experiment ($\left| A \right| = 4$).

Building upon the improved continuity bound on conditional entropy, we improve the upper bound on the one-way distillable entanglement in Theorem~\ref{thm:oneway_upperbound}. A parallel result for the upper bound on quantum capacity has been shown in~\cite[Proposition 10]{Mario2025}. Before stating it, we introduce a modified version of the coherent information, which mimics the similar definition for channels used in~\cite[Proposition 3.2]{Sutter2014approxepChannel}.

\begin{figure}[H]
  \centering 
  \includegraphics[width=0.45\textwidth]{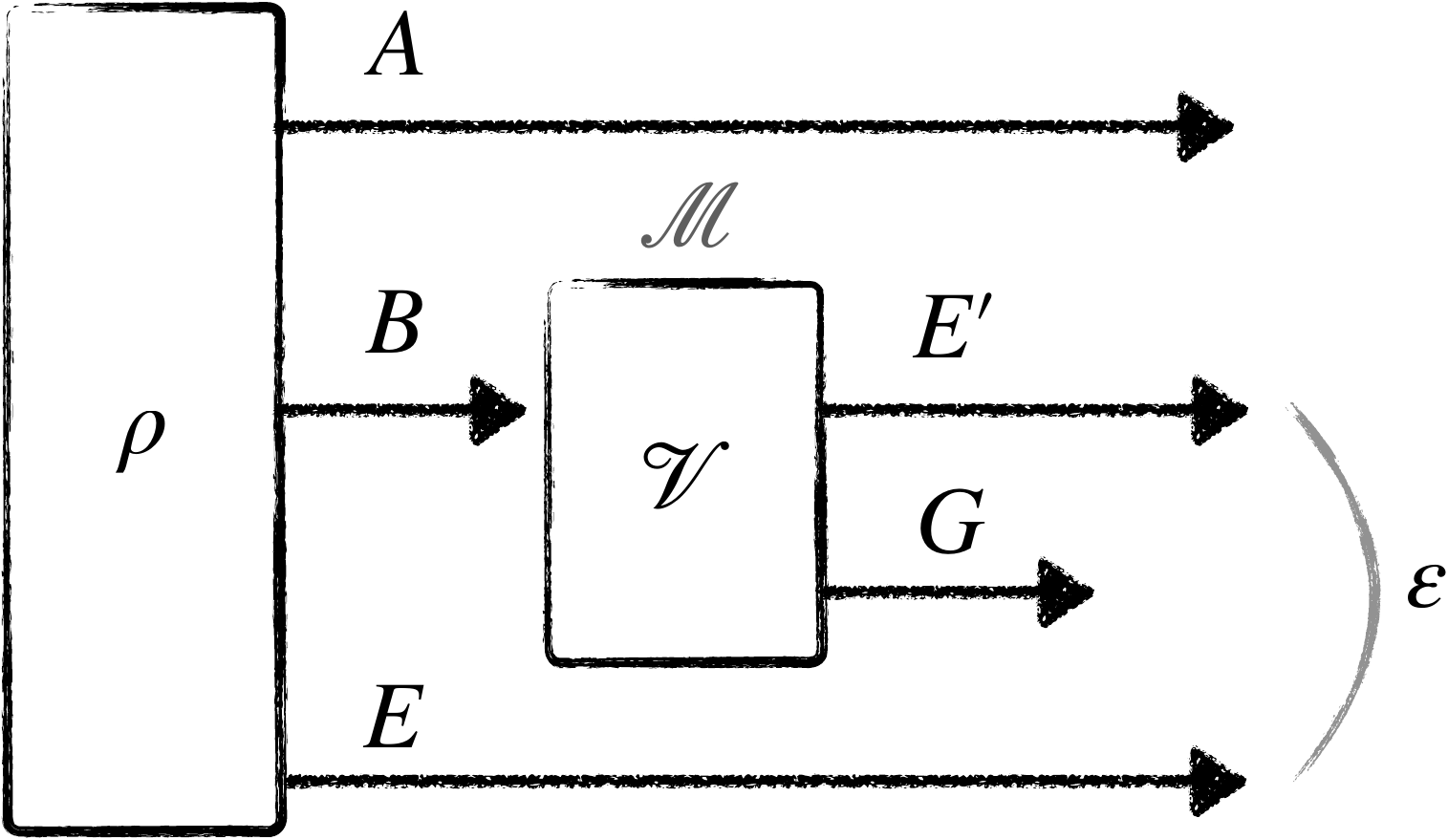} 
  \caption{Schematic illustration of an $\ve$-degradable state $\rho_{AB}$ and its $\ve$-degrading channel $\cM$, along with their respective Stinespring dilations $\cV(\cdot) = V(\cdot)V^\dag$.}
  \label{fig: ep state}
\end{figure}

\begin{definition}\label{def:Umrho}
Let $\rho \in \density(AB)$ be an $\ve$-degradable state with an $\ve$-degrading channel $\cM \in \CPTP(B:E')$. Let $V: \cH_B\to \cH_{E'}\ox \cH_G$ be a Stinespring isometry of $\cM$. Then we define 
\begin{align}
U_{\cM}(\rho_{A|B}) := H(G|E')_{\sigma} \quad \text{where} \quad \sigma_{AE'G} = V\rho_{AB}V^\dag.
\end{align}
\end{definition}

\begin{shaded}
\begin{theorem}[Improved upper bound on the one-way distillable entanglement] \label{thm:oneway_upperbound}
Let $\rho \in \density(AB)$ be an $\ve$-degradable quantum state with $\ve$-degrading channel $\cM\in \CPTP(B:E')$. Let $V: \cH_B\to \cH_{E'}\ox \cH_G$ be a Stinespring isometry of $\cM$ with $E' \cong E$. It satisfies that
\begin{align}\label{Eq:cont_upper_state}
D_{\to}(\rho_{A|B}) &\leq U_{\cM}(\rho_{A|B}) + \ve \log(|E|^2 - 1) + h(\ve).
\end{align}
\end{theorem}
\end{shaded}
\begin{proof} 
The argument here closely follows the proof of~\cite[Theorem 2.12]{Leditzky2018usefulsates}, leveraging Lemma~\ref{lem:better_cont_bound}. Let $\ket{\rho}_{ABE}$ be a purification of $\rho_{AB}$. Consider $n$ copies of the state $\rho_{AB}^{\ox n}$ and let $\cT_n: A^n \to A' M$ be an instrument with isometric extension
\begin{equation}
T_n: A^n \to A' M N, \quad T_n = \sum_m K_j \ox |m\rangle_M \ox |m\rangle_N,
\end{equation}
where $\{K_j\}$ are Kraus operators. For $t=1, \ldots, n$, define the pure states
\begin{align}
& \ket{\psi^t}_{A^n B_{t+1} \ldots B_n E_1' \ldots E_t' G_1 \ldots G_t E_1 \ldots E_n} := V_{B_1 \to E_1' G_1} \ox \cdots \ox V_{B_t \to E_t' G_t} \ket{\rho}_{ABE}^{\ox n}, \label{eq: distillation proof tmp1}\\
& \ket{\theta^t}_{A' M N B_{t+1} \ldots B_n E_1' \ldots E_t' G_1 \ldots G_t E_1 \ldots E_n} := T_n \ket{\psi^t},
\end{align}
where $V_{B_i \to E_i' G_i}$ are Stinespring isometries for the $\ve$-degrading channels.

\begin{figure}[H]
\centering
\includegraphics[width=0.55\textwidth]{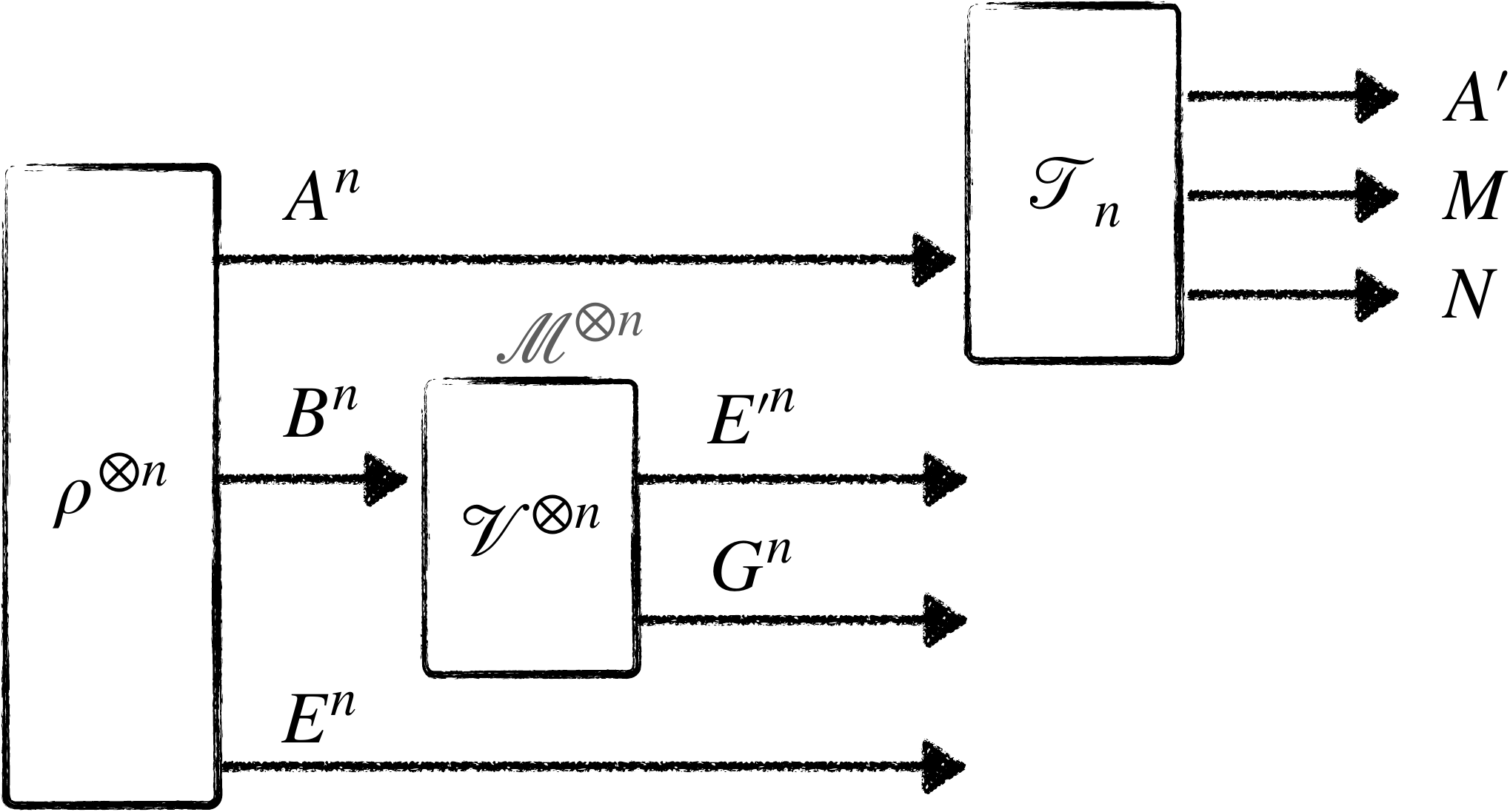}
\end{figure}

It was shown in~\cite[Eq.~(2.27)]{Leditzky2018usefulsates} that
\begin{equation}
I(A'\rangle MB^n)_{T_n\rho^{\ox n}T_n^\dagger} = H(G^n|ME^{\prime n})_{\theta^n} + \sum_{t=1}^n \left[ H(E'_t|ME'_{<t}E_{>t})_{\theta^t} - H(E_t|ME'_{<t}E_{>t})_{\theta^{t}} \right], \label{eq: LDS17 2.27}
\end{equation}
where $X_{<t} := X_1 \cdots X_{t-1}$ (with $X_{<1}$ the trivial system) and $X_{>t}$ is defined analogously. In~\cite{Leditzky2018usefulsates}, it is shown that the trace distance between $\theta^t_{ME'_1\ldots E'_{t-1}E_t \ldots E_n}$ and $\theta^t_{ME'_1 \ldots E'_tE_{t+1} \ldots E_n}$ is bounded as
\begin{align}
\frac{1}{2} \Big\|\theta^t_{ME'_1\ldots E'_{t-1}E_t \ldots E_n} - \theta^t_{ME'_1 \ldots E'_tE_{t+1} \ldots E_n}\Big\|_1 \leq \varepsilon.
\end{align}
Notice that 
$\tr_{E'_t} \theta^t_{ME'_1\ldots E'_{t}E_{t+1} \ldots E_n} = \tr_{E_t} \theta^t_{ME'_1\ldots E'_{t-1}E_{t+1}}$, i.e., the marginals on $ME'_{<t}E_{>t}$ coincide. Thus, by applying Lemma~\ref{lem:better_cont_bound}, the second term in Eq.~\eqref{eq: LDS17 2.27} can be bounded as
\begin{align}
\sum_{t=1}^n \left[ H(E'_t|ME'_{<t}E_{>t})_{\theta^t} - H(E_t|ME'_{<t}E_{>t})_{\theta^{t}} \right] \leq n \left( \varepsilon \log(|E|^2 - 1) + h(\varepsilon) \right).
\end{align}
It follows that
\begin{align}
I(A'\rangle MB^n)_{T_n\rho^{\ox n}T_n^\dagger}
&\leq H(G^n|ME^{\prime n})_{\theta^n} + n \left( \varepsilon \log(|E|^2 - 1) + h(\varepsilon) \right) \\
&\leq H(G^n|E^{\prime n})_{\psi^n} + n \left( \varepsilon \log(|E|^2 - 1) + h(\varepsilon) \right) \\
&= n H(G|E')_{\psi} + n \left( \varepsilon \log(|E|^2 - 1) + h(\varepsilon) \right),
\end{align}
where the second inequality uses the data-processing inequality for conditional entropy under partial trace, and the last equality uses the additivity of conditional entropy for tensor product states as in Eq.~\eqref{eq: distillation proof tmp1}. Since the above holds for any instrument $\cT_n$, optimizing over all such instruments and dividing both sides by $n$, then taking the limit $n \to \infty$ completes the proof. 
\end{proof}

\begin{remark}
Using the same idea to bound $H(G^n|E^{\prime n})_{\psi^n}$ through $I(A\rangle B)_{\rho^{\ox n}}$ as that in~\cite[Eq.~(2.30)]{Leditzky2018usefulsates}, we can also have
\begin{equation}\label{Eq:upperbound_coh_deg}
    D_{\to}(\rho_{A|B}) \leq I(A\rangle B)_\rho + 2\ve \log(|E|^2 - 1) + 2h(\ve).
\end{equation}
\end{remark}

Based on Theorem~\ref{thm:oneway_upperbound} and the framework of Riemannian optimization for state extensions (cf.~Section~\ref{sec:RiemanOpt}), we further have the following upper bound on the one-way distillable entanglement.

\begin{shaded}
\begin{corollary}\label{cor:state_upperbound_maniopt}
For any bipartite state $\rho_{AB}$, let $\ket{\phi}_{ABE} = \sum_i\sqrt{p_i}\ket{\psi_i}_{AB}\ox \ket{e_i}_{E}$ be the canonical purification of $\rho_{AB}$ where $\rho_{AB}=\sum_i p_i \ketbra{\psi_i}{\psi_i}_{AB}$ is the spectral decomposition. It holds that
\begin{align}\label{Eq:cont_rieman_state}
\begin{aligned}
D_{\to}(\rho_{A|B}) \leq \inf \; & \; U_{\cM}(\rho_{A|BF}) + \dg(\rho_{A|BF}) \log(|\widehat{E}|^2 - 1) + h(\dg(\rho_{A|BF})) \\ 
{\rm s.t.}  & \; V\in \mathrm{St}(|FR|,|E|),~|F|,|R|\in\NN_+, \\
&\; \rho_{ABF} = \tr_{R} (V \ketbra{\phi}{\phi} V^\dag),
\end{aligned}
\end{align}
where $\cM$ is the $\dg(\rho_{A|BF})$-degrading channel for $\rho_{A|BF}$ and $\widehat{E}$ is the auxiliary system for the purification of $\rho_{ABF}$.
\end{corollary}
\end{shaded}
\begin{proof} 
The bound directly follows from Fact~\ref{fact:state_ext} with Theorem~\ref{thm:oneway_upperbound} applying to $\rho_{A|BF}$. 
\end{proof}

As mentioned in Section~\ref{sec:RiemanOpt}, for any fixed $|F|,|R|\in\NN_+$, taking $\rho_{ABF} = \tr_{R} (V \ketbra{\phi}{\phi} V^\dag)$ into the objective in Eq.~\eqref{Eq:cont_rieman_state}, the problem reduces to
\begin{equation}\label{Eq:RiemanLoss}
    \min_{V\in \mathrm{St}(|FR|,|E|)} f(V) \coloneqq U_{\cM}(V) + \dg(V)\log(\eta^2-1) + h(\dg(V)).
\end{equation}
Here in Eq.~\eqref{Eq:RiemanLoss}, $\mathrm{St}(|FR|,|E|)$ is the complex Stiefel manifold, $U_{\cM}(V)$ and $\dg(V)$ are shorthand for $U_{\cM}(\rho_{A|BF}(V))$ and $\dg(\rho_{A|BF}(V))$, respectively, and $\eta = |ABF|$. This is because full-rank matrices are dense in $\density(AB)$, i.e., the probability of an iterative point landing on a full-rank point is 1 under the Lebesgue measure. The primary obstacle to solving this problem is that deriving an analytical expression for the Riemannian gradient of $f(V)$ is intractable. This difficulty arises because the degradability parameter, $\dg(\rho_{A|BF})$, and $\ve$-degrading channel $\cM$ themselves are the solution to an SDP~\cite[Lemma 2.10]{Leditzky2018usefulsates}, making Eq.~\eqref{Eq:RiemanLoss} a \textit{bilevel} optimization problem\cite{colson2007overview}. To obtain an analytical gradient, in principle, one may try to use implicit differentiation of the Karush-Kuhn-Tucker (KKT) conditions~\cite{Vandenberghe1996} of the inner SDP for $\dg(\rho_{A|BF})$, and analyze the gradient of purification function since we also need $\rho_{AE}$ for calculating $\dg(\rho_{A|BF})$. This approach is exceptionally complex to implement and requires solving a large, potentially ill-conditioned linear system at each iteration. For these practical reasons, we adopt a numerical finite difference scheme, which we will explain in detail in Section~\ref{sec:num_grad}.

\subsubsection{Numerical gradient approximation on the manifold}\label{sec:num_grad}

Since our primary goal is to establish a method that can be practically implemented to compute upper bounds for a wide range of quantum states and channels, we prioritize a pragmatic approach to optimization. We utilize an RGD algorithm but replace the intractable analytical gradient with a numerical approximation via a finite-difference scheme, a well-adopted approach in derivative-free optimization~\cite{mizumoto1973finite,Choi2000,kelley2011implicit,nesterov2017random,shi2021numerical}, and has been shown effective in some bilevel problems~\cite{Sow2022}. To apply gradient-based optimization to the objective function in Eq.~\eqref{Eq:RiemanLoss} (or Eq.~\eqref{Eq:upperbound_coh_deg}), we will show that the following objective function is actually~\textit{locally Lipschitz continuous}.
\begin{align}\label{Eq:fV_Ic}
    f(V) = I(A\rangle BF)_{\rho(V)} + 2\dg(V)\log(\eta^2-1) + 2h(\dg(V)).
\end{align}
For the first term $I(A\rangle B)_{\rho}$, we note that the von Neumann entropy $H(\rho)$ is continuously differentiable with respect to $\rho$ at any point where $\rho$ is full-rank, i.e., $\nabla H(\rho) = -(\log \rho + \1)$. Therefore, $I(A\rangle B)_{\rho}$ is also continuously differentiable on the full rank density matrices, which means that $\rho\mapsto I(A\rangle B)_{\rho}$ is locally Lipschitz continuous. The binary entropy term $h(\cdot)$ is also locally Lipschitz. For the second term, the degradability parameter, we have the following results.

\begin{shaded}
\begin{lemma}\label{lem:dg_continuity}
The degradability parameter $\dg(\cdot)$ is a H\"older continuous function on the set of bipartite quantum states $\mathscr{D}(AB)$ with respect to the trace distance. Specifically, for any two states $\rho_{AB}, \sigma_{AB} \in \mathscr{D}(AB)$ such that $\frac{1}{2}\|\rho_{AB}-\sigma_{AB}\|_1\leq \ve \leq 1$, it holds that
\begin{equation}
|\dg(\rho_{AB}) - \dg(\sigma_{AB})| \leq \sqrt{2\ve} + \ve.
\end{equation}
\end{lemma}
\end{shaded}

\begin{proof}
Let $\rho_{AB}$ and $\sigma_{AB}$ be two bipartite states on $\cH_A \ox \cH_B$. Choose $\ket{\phi^{\rho}}_{ABE}$ and $\ket{\phi^{\sigma}}_{ABE}$ to be purifications of $\rho_{AB}$ and $\sigma_{AB}$ on a common system $\cH_E$ with $|E| \ge \dim(\cH_A)\dim(\cH_B)$, such that they saturate the fidelity bound in Uhlmann's theorem, i.e.,
\begin{equation}\label{Eq:uhlthm}
F(\rho_{AB},\sigma_{AB}) = |\braket{\phi^{\rho}}{\phi^{\sigma}}|^2,
\end{equation}
where $F(\rho,\sigma)\coloneqq \|\sqrt{\rho}\sqrt{\sigma}\|_1^2$ is the fidelity between $\rho$ and $\sigma$. The complementary states are given by $\rho_{AE} = \tr_B \ketbra{\phi^{\rho}}{\phi^{\rho}}$ and $\sigma_{AE} = \tr_B \ketbra{\phi^{\sigma}}{\phi^{\sigma}}$. Let us define the function $f_{\cM}(\rho_{AB})$ as
\begin{equation}
f_{\cM}(\rho_{AB}) \coloneqq \frac{1}{2} \big\| \rho_{AE} - (\cI_A \ox \cM_{B\to E})(\rho_{AB}) \big\|_1.
\end{equation}
Consider that
\begin{align}
\big|f_{\cM}(\rho_{AB}) - f_{\cM}(\sigma_{AB})\big| &= \frac{1}{2} \Big|\big\| \rho_{AE} - (\cI_A \ox \cM)(\rho_{AB}) \big\|_1 - \big\| \sigma_{AE} - (\cI_A \ox \cM)(\sigma_{AB})\big\|_1 \Big|\nonumber\\
&\leq \frac{1}{2} \big\| \rho_{AE} - (\cI_A \ox \cM)(\rho_{AB}) - \sigma_{AE} + (\cI_A \ox \cM)(\sigma_{AB}) \big\|_1 \label{Eq:triangle1}\\
&\leq \frac{1}{2} \Big( \| \rho_{AE} - \sigma_{AE} \|_1 + \big\| (\cI_A \ox \cM)(\rho_{AB} - \sigma_{AB}) \big\|_1 \Big) \label{Eq:triangle2}\\
&\leq \frac{1}{2} \Big( \| \rho_{AE} - \sigma_{AE} \|_1 + \| \rho_{AB} - \sigma_{AB} \|_1 \Big). \label{Eq:dpi}
\end{align}
Here, in Eq.~\eqref{Eq:triangle1}, we used the reverse triangle inequality for the trace norm, i.e., $\|X\|_1 - \|Y\|_1 \leq \|X - Y\|_1$. We used $\|X - Y\|_1 \leq \|X\|_1 + \|Y\|_1$ in Eq.~\eqref{Eq:triangle2} and used $\| (\cI_A \ox \cM_{B\to E})(\rho_{AB} - \sigma_{AB}) \|_1 \leq \| \rho_{AB} - \sigma_{AB} \|_1$ in Eq.~\eqref{Eq:dpi}. Recall the Fuchs–van de Graaf inequalities
\begin{equation}\label{Eq:FdG_ineq}
    1-\sqrt{F(\rho_{AB},\sigma_{AB})} \leq \frac{1}{2}\|\rho_{AB}-\sigma_{AB}\|_1 \leq \sqrt{1-F(\rho_{AB},\sigma_{AB})}.
\end{equation}
We have that
\begin{align}\label{Eq:rhoAEsigAE}
    \|\rho_{AE} - \sigma_{AE} \|_1 \leq \|\ketbra{\phi^{\rho}}{\phi^{\rho}} - \ketbra{\phi^{\sigma}}{\phi^{\sigma}}\|_1 = 2\sqrt{1-F(\rho_{AB},\sigma_{AB})},
\end{align}
where the first inequality is due to the monotonicity of the trace distance under partial trace, the equality is thanks to Eq.~\eqref{Eq:uhlthm} and the fact that any pair of pure states saturates the upper bound in Eq.~\eqref{Eq:FdG_ineq}. Notice that
\begin{equation}
\begin{aligned}
\sqrt{1-F(\rho_{AB},\sigma_{AB})} &= \sqrt{(1+\sqrt{F(\rho_{AB},\sigma_{AB})})(1-\sqrt{F(\rho_{AB},\sigma_{AB})})}\\
&\leq \sqrt{2(1-\sqrt{F(\rho_{AB},\sigma_{AB})})}\\
&\leq \sqrt{\|\rho_{AB}-\sigma_{AB}\|_1},
\end{aligned}
\end{equation}
where the first inequality is because $F(\rho_{AB},\sigma_{AB}) \leq 1$ and the second inequality is followed from Eq.~\eqref{Eq:FdG_ineq}. Plugging into Eq.~\eqref{Eq:rhoAEsigAE}, we get 
\begin{equation}\label{Eq:rhoAE_rhoAB}
    \|\rho_{AE} - \sigma_{AE} \|_1 \leq 2\sqrt{\|\rho_{AB}-\sigma_{AB}\|_1}.
\end{equation}

Now, let $\cM_\rho$ be a CPTP map that achieves the minimum for $\dg(\rho_{AB})$, and let $\cM_\sigma$ be one that achieves the minimum for $\dg(\sigma_{AB})$. By definition, we have
\begin{align*}
\dg(\rho_{AB}) = f_{\cM_\rho}(\rho_{AB}), ~~\dg(\sigma_{AB}) = f_{\cM_\sigma}(\sigma_{AB}).
\end{align*}
Since $\cM_\rho$ is optimal for $\rho_{AB}$, we have $\dg(\rho_{AB}) \leq f_{\mathcal{D}}(\rho_{AB})$ for any map $\mathcal{D}$, including $\cM_\sigma$. This leads to
\begin{align}
\dg(\rho_{AB}) - \dg(\sigma_{AB}) &\leq f_{\cM_\sigma}(\rho_{AB}) - f_{\cM_\sigma}(\sigma_{AB}) \\
&\leq |f_{\cM_\sigma}(\rho_{AB}) - f_{\cM_\sigma}(\sigma_{AB})| \\
&\leq \frac{1}{2} \left( 2\sqrt{\|\rho_{AB}-\sigma_{AB}\|_1} + \|\rho_{AB} - \sigma_{AB}\|_1 \right),
\end{align}
where we used Eq.~\eqref{Eq:dpi} and Eq.~\eqref{Eq:rhoAE_rhoAB} for the last inequality. By swapping the roles of $\rho_{AB}$ and $\sigma_{AB}$, the same argument yields
\begin{equation}
\dg(\sigma_{AB}) - \dg(\rho_{AB}) \leq \frac{1}{2} \left( 2\sqrt{\|\rho_{AB}-\sigma_{AB}\|_1} + \|\sigma_{AB} - \rho_{AB}\|_1 \right).
\end{equation}
Combining these two inequalities gives
\begin{equation}
|\dg(\rho_{AB}) - \dg(\sigma_{AB})| \leq \frac{1}{2} \left( 2\sqrt{\|\rho_{AB}-\sigma_{AB}\|_1} + \|\rho_{AB} - \sigma_{AB}\|_1 \right).
\end{equation}
\end{proof}

\begin{shaded}
\begin{lemma}\label{lem:Vtorho}
For any bipartite state $\rho_{AB}$, let $\ket{\phi}_{ABE}$ be a purification of $\rho_{AB}$, $V:\cH_E\to \cH_R\ox \cH_F$ be an isometry and denote $\rho_{ABF}(V) = \tr_{R} (V \ketbra{\phi}{\phi} V^\dag)$. It holds that
\begin{align}
\|\rho_{ABF}(V) - \rho_{ABF}(V')\|_1 \leq 2\|V-V'\|_{\frob}.
\end{align}
\end{lemma}
\end{shaded}

\begin{proof}
Let $V, V' \in \mathrm{St}(|FR|,|E|)$ be two isometries and $\rho \equiv \rho_{ABF}(V),\rho' \equiv \rho_{ABF}(V')$. Consider that
\begin{equation}\label{Eq:diff_rho_to_V}
\begin{aligned}
\|\rho-\rho'\|_1 &= \big\|\tr_R\big(V\ketbra{\phi}{\phi}V^\dag - V'\ketbra{\phi}{\phi}V'^\dag\big)\big\|_1\\
&\leq \big\|V\ketbra{\phi}{\phi}V^\dag - V'\ketbra{\phi}{\phi}V'^\dag\big\|_1\\
&= \big\|(V-V')\ketbra{\phi}{\phi}V^\dag + V'\ketbra{\phi}{\phi}(V-V')^\dag\big\|_1\\
&\leq \big\|(V-V')\ketbra{\phi}{\phi}V^\dag\big\|_1 + \big\|V'\ketbra{\phi}{\phi}(V-V')^\dag\big\|_1\\
&\leq \|V-V'\|_{\infty} \cdot \|\ketbra{\phi}{\phi}\|_1 \cdot \|V^\dag\|_{\infty} + \|V'\|_{\infty} \cdot \|\ketbra{\phi}{\phi}\|_1 \cdot \|V-V^\dag\|_{\infty}\\
&= 2 \|V-V'\|_{\infty}\\
&\leq 2 \|V-V'\|_{\frob},
\end{aligned}
\end{equation}
where we used the data processing inequality in the second line, the triangle inequality in the fourth line, and the H\"older inequality twice in the fifth line, the fact that $\|V\|_{\infty} = \|V^\dag\|_{\infty}=1$ for an isometry in the second line from the bottom, and the fact that $\|X\|_{\infty}\leq \|X\|_{\frob}$ for any normal matrix $X$ in the last inequality.
\end{proof}

\begin{shaded}
\begin{corollary}\label{cor:deg_to_V}
For any bipartite state $\rho_{AB}$, let $\ket{\phi}_{ABE}$ be a purification of $\rho_{AB}$, $V:\cH_E\to \cH_R\ox \cH_F$ be an isometry and denote $\rho_{ABF}(V) = \tr_{R} (V \ketbra{\phi}{\phi} V^\dag)$. It holds that
\begin{equation}
    \big|\dg(\rho_{ABF}(V)) - \dg(\rho_{ABF}(V'))\big| \leq 2\|V-V'\|_{\frob}.
\end{equation}
\end{corollary}
\end{shaded}

\begin{proof}
According to the argument in Lemma~\ref{lem:dg_continuity}, we have that
\begin{equation*}
    \big|\dg(\rho_{A|BF}(V)) - \dg(\rho_{A|BF}(V'))\big| \leq \frac{1}{2} \Big( \| \rho_{AR}(V) - \rho_{AR}(V') \|_1 + \| \rho_{ABF}(V) - \rho_{ABF}(V') \|_1 \Big),
\end{equation*}
where 
\begin{equation}
    \rho_{AR}(V) = \tr_{BF}\Big[(\1_{AB}\ox V)\ketbra{\phi}{\phi}_{ABE} (\1_{AB}\ox V^\dag)\Big].
\end{equation}
Similar to Eq.~\eqref{Eq:diff_rho_to_V}, we have that
\begin{equation}
    \| \rho_{AR}(V) - \rho_{AR}(V') \|_1 \leq 2\|V-V'\|_{\frob},~\| \rho_{ABF}(V) - \rho_{ABF}(V') \|_1 \leq 2\|V-V'\|_{\frob}.
\end{equation}
Therefore, it follows that
\begin{equation}
    \big|\dg(\rho_{A|BF}(V)) - \dg(\rho_{A|BF}(V'))\big| \le 2\|V-V'\|_{\frob}.
\end{equation}
\end{proof}

Lemma~\ref{lem:dg_continuity} establishes the H\"older continuity of the degradability parameter $\dg(\cdot)$ with respect to states. More importantly, Lemma~\ref{lem:Vtorho} and Corollary~\ref{cor:deg_to_V} demonstrate that the degradability parameter $\dg(\cdot)$ is Lipschitz continuous with respect to isometries $V$. Consequently, the overall objective function $f(V)$ in Eq.~\eqref{Eq:fV_Ic} is locally Lipschitz continuous. This property is the theoretical cornerstone of our optimization approach. Specifically, although $f(V)$ may be nondifferentiable, its local Lipschitz nature justifies the use of concepts from nonsmooth optimization, namely the subgradient~\cite{Xiao2021}. In this context, our finite-difference scheme (cf.~Eqs.~\eqref{Eq:apprx_grad1}-\eqref{Eq:apprx_grad3}) serves as a reliable method to approximate a subgradient, as the local Lipschitz condition guarantees its truncation error is rigorously bounded linearly by $\cO(L h)$, where $L$ is the local Lipschitz constant and $h$ is the tunable step size~\cite{Michael2025}. Therefore, our algorithm acts as a practical implementation of a (inexact) Riemannian subgradient method, a principled approach known to converge to the set of Clarke stationary points~\cite{burke2005robust,bagirov2014introduction,Hosseini2017}. Based on the above, we explain the optimization algorithm as follows.

\paragraph{Riemannian gradient descent.} 
Our RGD algorithm, shown in Algorithm~\ref{alg:RGD_Doneway}, leverages standard tools from the field of Riemannian optimization. To ensure our presentation is self-contained, we briefly recall the essential definitions for the complex Stiefel manifold, which are well-established concepts. We refer the reader to~\cite{boumal2023intromanifolds} for a comprehensive treatment. 
The Riemannian gradient, denoted $\grad f(V)$, is a tangent vector in the tangent space $\tangent_V\mathscr{M}$ at a point $V$ on the manifold $\mathscr{M}$. It is defined by the relation
\begin{equation}
\big\langle \grad f(V), \dot{V}\big\rangle_V = \mathrm{D} f(V)[\dot{V}],~~\forall \dot{V} \in \tangent_V\mathscr{M},
\end{equation}
where $\dot{V}$ is any tangent vector, $\langle \cdot, \cdot \rangle_V$ is the Riemannian metric on the \textit{tangent space}, and $\mathrm{D} f(V)[\dot{V}]$ is the directional derivative of $f$ at $V$ along $\dot{V}$.

Specifically, for the \textit{complex Stiefel manifold} $\mathrm{St}(n,p) \coloneqq \{V\in\CC^{n\times p}: V^\dag V = \1\}$, its tangent space is given by
\begin{equation}\label{Eq:st_tangspace}
    \tangent_V\mathrm{St}(n,p) = \big\{VA:A\in\CC^{n\times n},~A^\dag+A = 0\big\}.
\end{equation}
The Riemannian metric on $\tangent_V\mathrm{St}(n,p)$ we use is given by
\begin{equation}
    \langle \cdot,\cdot \rangle_{V}: \tangent_V\mathrm{St}(n,p) \times \tangent_V\mathrm{St}(n,p) \to \RR, \qquad \langle X,Y \rangle_{V} = \tr (X^\dag Y).
\end{equation}
The projection onto the tangent space at any point $V\in\mathrm{St}(n,p)$ reads
\begin{equation}\label{Eq:st_proj}
    \mathrm{Proj}_V(X) = (\1 - VV^\dag) X + V\frac{V^\dag X - X^\dag V}{2}.
\end{equation}
The Riemannian gradient is then given by 
\begin{equation}
    \grad f(V) = \mathrm{Proj}_V\big(\nabla \bar{f}(V)\big),
\end{equation}
where $\nabla \bar{f}(V)$ is the gradient of an extension $\bar{f}$ of $f$ on Euclidean space. When doing an optimization on a manifold, in order to move in the direction of a tangent vector while staying on the manifold, we also need a retraction mapping~\cite[Section 4]{absil2008optimization}. There are different retractions for the Stiefel manifold, a standard one, and the one we used in Section~\ref{sec:lower_qcap} is the retraction based on the QR decomposition, i.e.,
\begin{equation}\label{Eq:retr_qr}
    \mathrm{R}_V(X) \coloneqq \mathrm{qf}(V+X),
\end{equation}
where $\mathrm{qf}(X)$ denotes the $Q$ factor of the decomposition of $X$ as $X=QR$ where $Q\in\mathrm{St}(n,p)$ and $R$ is an $p\times p$ upper triangular with nonnegative diagonal entries.

\begin{algorithm}[t]
\caption{Riemannian Gradient Descent (RGD) for upper bounds on $D_{\to}(\cdot)$}\label{alg:RGD_Doneway}
\begin{algorithmic}[1]
    \REQUIRE A bipartite state $\rho_{AB}$ and a purification $\ket{\phi}_{ABE}$; Initial guess $V^{(0)} \in \mathrm{St}(n, p)$.
    \WHILE{the stopping criteria are not satisfied}
        \STATE 
        Approximate $\dot{V}^{(t)} = -\grad f(V_k)$ using the finite-difference scheme as described in Eqs.~\eqref{Eq:apprx_grad1}-\eqref{Eq:apprx_grad3}.
        \STATE Compute a stepsize $s^{(t)}$.
        \STATE Update $V^{(t+1)}= \mathrm{R}_{V^{(t)}} \big(s^{(t)}\dot{V}^{(t)}\big)$ by Eq.~\eqref{Eq:retr_qr}; $t= t+1$.
    \ENDWHILE
    \STATE Compute the upper bound $d_{\mathrm{upp}}$ in Eq.~\eqref{Eq:cont_rieman_state} using the optimal $V^*$.
    \ENSURE $V^*,d_{\mathrm{upp}}$.
\end{algorithmic}
\end{algorithm}

To optimize Eq.~\eqref{Eq:fV_Ic}, since we cannot compute the gradient analytically, we approximate the directional derivative using a first-order finite difference. For a small step $t$, the directional derivative is approximated as
\begin{equation}\label{Eq:apprx_grad1}
    \mathrm{D} f(V)[\dot{V}] \approx \frac{f(\mathrm{R}_V(t\dot{V})) - f(V)}{t}.
\end{equation}
To compute the full gradient vector $\grad f(V)$, we can compute the following steps:
\begin{itemize}
\item[i).] Define an orthonormal basis $\{\zeta_i\}_{i=1}^d$ for the tangent space $\tangent_V\mathscr{M}$, where $d = \dim(\tangent_V\mathscr{M})$. 
\item[ii).] For each basis vector $\zeta_i$, compute the approximate directional derivative using the finite-difference formula
\begin{equation}\label{Eq:apprx_grad2}
    c_i = \mathrm{D} f(V)[\zeta_i] \approx \frac{f(\mathrm{R}_V(t\zeta_i)) - f(V)}{t}.
\end{equation}
\item[iii).] The Riemannian gradient is then constructed by combining these components
\begin{equation}\label{Eq:apprx_grad3}
    \grad f(V) = \sum_{i=1}^d c_i\zeta_i.
\end{equation}
\end{itemize}
This process requires $d+1$ function evaluations to compute a single gradient. While computationally intensive, it bypasses the need for any analytical derivatives.

\subsection{Upper bounds on the quantum capacity}\label{sec:upper_qcapacity}
In this section, we apply the Riemannian optimization methods for searching extensions of a target channel to obtain upper bounds on the quantum capacity. The Lloyd–Shor–Devetak theorem shows that the quantum capacity of a channel $\cN$ is equal to the regularized channel coherent information~\cite{Schumacher1996a,Lloyd1997,Barnum2000,Shor2002a,Devetak2005a},
\begin{align}\label{eq: quantum channel coding theorem}
  Q(\cN) = \lim_{n\to \infty} \frac{1}{n} Q^{(1)}(\cN^{\ox n}),
\end{align}
where the \textit{channel coherent information} is defined as $Q^{(1)}(\cN) \coloneqq \max_{\rho \in \density} I_c(\rho,\cN)$ (also denoted as $I_c(\cN)$) with 
\begin{equation}
    I_c(\rho,\cN) \coloneqq H(\cN(\rho)) - H(\cN^c(\rho)).
\end{equation}

A quantum channel $\cN_{A\to B}$ is said to be \textit{degradable} if there exists a channel $\cM \in \CPTP(B : E)$ such that $\cN^c = \cM \circ \cN$, where $\cN^c$ is the complementary channel of $\cN$. The primary utility of this property is that the coherent information of a degradable channel is additive. Consequently, the regularization in Eq.~\eqref{eq: quantum channel coding theorem} becomes unnecessary, leading to $Q(\cN) = Q^{(1)}(\cN)$. However, non-degradable channels generally exhibit non-additive coherent information\cite{DiVincenzo1998a,Felix2018}, making the regularization essential. This non-additivity can be so pronounced that an unbounded number of channel uses may be required to detect capacity~\cite{Cubitt2015}.

This sharp dichotomy between degradable and non-degradable channels motivates a more nuanced approach for channels that, while not strictly degradable, are \textit{close} to being so. To formalize this, Sutter \textit{et al.} introduced the \textit{$\ve$-degradable channel}~\cite{Sutter2014approxepChannel}, defined as follows.

\begin{definition}
Let $\cN \in \CPTP(A:B)$ be a quantum channel with complementary channel $\cN^{c} \in \CPTP(A:E)$, where $E$ denotes the environment system. We say that $\cN$ is \emph{$\ve$-degradable} if there exists a quantum channel $\cM \in \CPTP(B:E')$ such that $E' \cong E$ and 
\begin{align}
\frac{1}{2} \big\|\cN^{c} - \cM \circ \cN \big\|_\diamond \leq \ve,
\end{align}
where $\|\cdot\|_\diamond$ denotes the diamond norm. The channel $\cM$ is referred to as an $\ve$-degrading channel. The \emph{degradability parameter} of $\cN$ is defined as\footnote{Compared with the definition in~\cite{Sutter2014approxepChannel}, the definition here includes an additional factor of $\frac{1}{2}$ to unify the results in the subsequent text.}
\begin{align}
\dg(\cN) \coloneqq \inf_{\cM \in \CPTP(B:E)} \frac{1}{2} \big\|\cN^c - \cM \circ \cN\big\|_\diamond.
\end{align}
\end{definition}
This framework provides bounds on the quantum capacity of a channel based on the approximate degradability~\cite[Theorem 3.4]{Sutter2014approxepChannel}, and the degradability parameter can be computed via an SDP: 
\begin{align}\label{eq: epsilon degradable SDP}
  \dg (\cN) = \inf \|\tr_E Z\|_\infty \quad \suchthat\quad Z \geq J_{\cN^c} - J_{\cM\circ \cN},~Z \geq 0,~J_{\cM} \geq 0,~\tr_E J_{\cM} = \1_B.
\end{align}

Recently, building upon the improved continuity bound on conditional entropy, the quantum capacity upper bound based on the approximate degradability has been improved in \cite[Proposition 10]{Mario2025} as follows.

\begin{proposition}[Improved quantum capacity upper bound \cite{Mario2025}]\label{prop:refined_upperbound_qcapacity}
Let $\cN \in \CPTP(A : B)$ be an $\ve$-degradable channel with an $\ve$-degrading channel $\cM \in \CPTP(B:E')$. 
It satisfies that
\begin{align}\label{Eq:cont_upper_channel}
Q(\cN) \leq U_{\cM}(\cN) + \ve \log(\left| E \right|^2 - 1) + h(\ve).
\end{align}
Here, $U_{\cM}(\cN)$ is defined as
\begin{align}\label{eq: U N M definition}
U_{\cM}(\cN) \coloneqq  \max\Big\{ H(G|E')_{\sigma}: \sigma_{E' G} = V \cN(\rho) V^\dag,\,\rho \in \density(A)\Big\},
\end{align}
where $V : \cH_B\to \cH_{E'}\ox \cH_G$ is the Stinespring isometry of $\cM$.
\end{proposition}

Similarly to Corollary~\ref{cor:state_upperbound_maniopt} for the state case, by combining Fact~\ref{fact:channel_ext} and Proposition~\ref{prop:refined_upperbound_qcapacity}, we have the following upper bound on the quantum capacity through channel extensions. 

\begin{shaded}
\begin{corollary} \label{cor: capacity_upperbound_maniopt}
Let $\cN \in \CPTP(A : B)$ be a quantum channel with a Stinespring isometry $U:\cH_A\to \cH_B\ox \cH_E$ and $V: \cH_E \to \cH_R\ox \cH_F$ be an arbitrary isometry. Consider the channel $\widehat{\cN}\in \CPTP(A : BF)$ whose Stinespring isometry is given by $W = (\1_B \ox V)U$. Then it holds that
\begin{align}\label{Eq:cont_rieman_channel}
\begin{aligned}
Q(\cN) \leq \inf \; & \; U_{\cM}(\widehat{\cN}) + \dg(\widehat{\cN}) \log(|E'|^2 - 1) + h(\dg(\widehat{\cN}))\\
{\rm s.t.} &\; V \in \mathrm{St}(|FR|,|E|),~|F|,|R|\in\NN_+,\\
&\; \widehat{\cN}(\cdot) = \tr_{R} (\1_B \ox V)U(\cdot)U^\dag(\1_B \ox V^\dag).
\end{aligned}
\end{align}
\end{corollary}
\end{shaded}

\begin{remark}
In principle, our state and channel extension methods can be combined with any upper bound on the target state's one-way distillable entanglement or channel's capacity. However, the method provides no improvement for `meta-converse' bounds, e.g., the geometric-Rains information~\cite[Theorem 12]{fangGeometricRenyiDivergence2021}. Such bounds are always defined via a minimization of a divergence $\mathbf{D}$ over a set of \textit{free} channels $\FF$, i.e., $\min_{\cM\in\FF(A:B)} \mathbf{D}(\cN_{A\to B}\|\cM_{A\to B})$. By the data-processing inequality, if $\FF$ is closed under partial trace on the output, i.e., $\cM_{A\to BF}\in\FF(A:BF) \implies \cM_{A\to B}\in\FF(A:B)$, the value of the bound is invariant under channel extensions, i.e.,  \begin{equation}
\min_{\cM\in\FF(A:B)} \mathbf{D}(\cN_{A\to B}\|\cM_{A\to B}) = \min_{\widehat{\cM}\in\FF(A:BF)} \mathbf{D}(\widehat{\cN}_{A\to BF}\|\widehat{\cM}_{A\to BF}).    
\end{equation}
In contrast, the non-monotonicity of the continuity bound in Eq.~\eqref{Eq:cont_upper_state} and Eq.~\eqref{Eq:cont_upper_channel} makes them an ideal testbed for the extension method. Moreover, in Eq.~\eqref{Eq:cont_rieman_state} and Eq.~\eqref{Eq:cont_rieman_channel}, we are essentially searching for a degradable extension with the tail terms in the continuity bound serving as a proper regularization term.
\end{remark}

\subsection{Improved upper bounds on isotropic states and depolarizing channels}\label{sec:improve_iso}
In this section, we establish improved upper bounds on the one-way distillable entanglement of isotropic states using our proposed methods. As a consequence, we obtain the SOTA upper bounds on the quantum capacity of qubit depolarizing channels. 

For a local dimension $d$, the \textit{isotropic state} is defined by
\begin{align}
I_d(f)\coloneqq f \Phi_+ + \frac{1-f}{d^2-1}(\1_{d^2}-\Phi_+), 
\end{align}
where $\Phi_+$ is the $d\times d$ maximally entangled state and $f\in[0,1]$. It is also known as the Choi state of the qudit depolarizing channel
\begin{equation}
\cD_p(\cdot) = (1-p)(\cdot) + \frac{p}{d^2-1} \sum_{\substack{0\leq i,j\leq d-1\\(i,j)\neq (0,0)}} X^iZ^j(\cdot)(X^iZ^j)^\dag,
\end{equation}
where $p=1-f$ and $X,Z$ are the generalized Pauli operators defined by
\begin{equation}
    X\ket{k} \coloneqq \ket{k+1(\mathrm{mod}~d)},~~~Z\ket{k}\coloneqq \omega^k\ket{k},~~~\omega = \exp(2\pi i/d).
\end{equation}

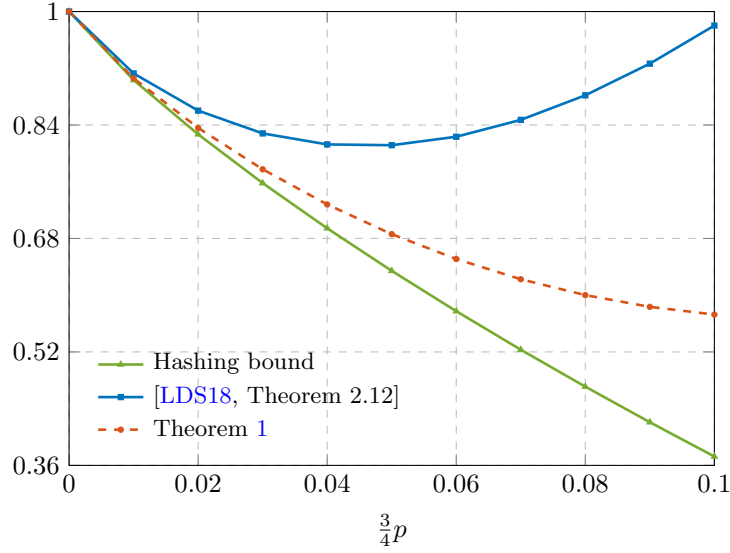
\begin{figure}[H]
\centering
\begin{tikzpicture}
    \node[anchor=south west,inner sep=0] (image) at (0,0) {\includegraphics{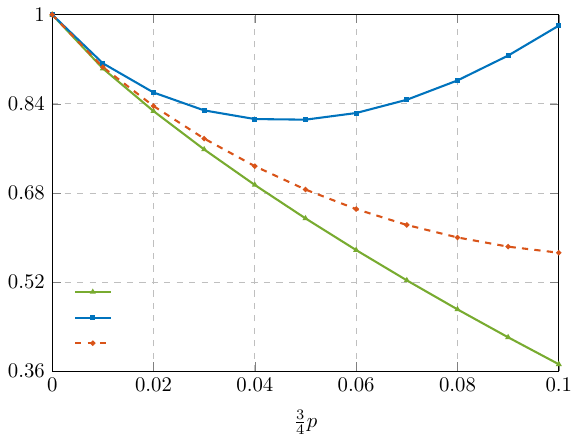}};
    \node[anchor=south west,inner sep=0] at (2,2.42){\footnotesize Hashing bound};
    \node[anchor=south west,inner sep=0] at (2,1.95){\footnotesize \!\!\cite[Theorem 2.12]{Leditzky2018usefulsates}};
    \node[anchor=south west,inner sep=0] at (2,1.6){\footnotesize Theorem~\ref{thm:oneway_upperbound}}; 
    \end{tikzpicture}
    \caption{Bounds on the one-way distillable entanglement of the qubit isotropic states. The $x$-axis is the parameter $3/4p=1-f \in [0,0.1]$ of the state $I_2(f)$.
    }
    \label{fig:distillable entanglement dp choi state}
\end{figure}

First, we directly compare our bound in Theorem~\ref{thm:oneway_upperbound} with that given in \cite[Theorem 2.12]{Leditzky2018usefulsates}. The numerical results are shown in Figure~\ref{fig:distillable entanglement dp choi state}. The green solid line is the coherent information of states, which gives a lower bound on $D_{\to}(I_d(f))$. The red solid line and the blue solid line are the upper bound given by Theorem~\ref{thm:oneway_upperbound} and~\cite[Theorem 2.12]{Leditzky2018usefulsates}, respectively. We can observe that a considerably tighter upper bound can be obtained by refining the continuity estimate via Lemma~\ref{lem:better_cont_bound} and by removing an additional approximation step through the additivity of conditional entropy. In particular, the refined continuity bound exhibits a more favorable scaling behavior, maintaining its effectiveness especially for relatively large degradability parameters (when $p$ goes large), where previous bounds tend to become very loose.

\begin{figure}[H]
\centering
\begin{tikzpicture}
    \node[anchor=south west,inner sep=0] (image) at (0,0) {\includegraphics{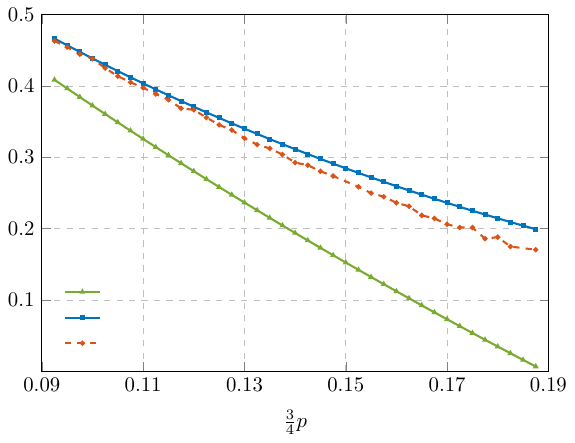}};
    \node[anchor=south west,inner sep=0] at (1.8,2.42){\footnotesize Hashing bound};
    \node[anchor=south west,inner sep=0] at (1.8,1.95){\footnotesize \!\!\cite[Proposition 5.1]{kianvashBoundingQuantumCapacity2022}};
    \node[anchor=south west,inner sep=0] at (1.8,1.52){\footnotesize Corollary~\ref{cor:state_upperbound_maniopt}}; 
    \end{tikzpicture}
    \caption{Bounds on the quantum capacity of the qubit depolarizing channel. The $x$-axis is the parameter $3/4p=1-f \in [0.0925, 0.1875]$ of the qubit depolarizing channel $\cD_p$.}
\label{fig:depo_sota}
\end{figure}
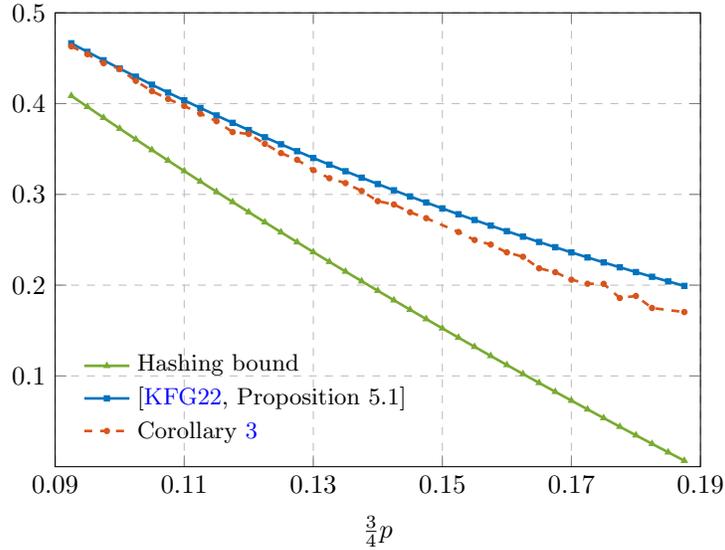

Notably, the quantum capacity of a teleportation-simulable channel is equal to the one-way distillable entanglement of its Choi state~\cite{Bennett1993a,bennettMixedstateEntanglementQuantum1996,Braunstein1998}, i.e., $Q(\cD_{1-f}) = D_{\to}(I_d(f))$. Therefore, the upper bound on $D_{\to}(I_d(f))$ also gives an upper bound on the quantum capacity of depolarizing channels, $Q(\cD_{1-f})$. The previously best-known upper bound on $Q(\cD_p)$ is given by the flagged extension method~\cite[Proposition 5.1 and Figure~1]{kianvashBoundingQuantumCapacity2022}. Here, based on Corollary~\ref{cor:state_upperbound_maniopt} and the numerical optimization method on a Stiefel manifold as described in Section~\ref{sec:num_grad}, we obtained SOTA upper bounds on the quantum capacity of qubit depolarizing channels. Specifically, for qubit isotropic states, we have $|E|=4$ and choose $|R|=4, |F|=2$ in our numerical experiments. Thereby, we are performing optimization over $\mathrm{St}(8,4)$ for the extensions of isotropic states $\rho_{ABF}$, with a qubit extension system $F$. The comparison of our bound with the best-known ones is presented in Figure~\ref{fig:depo_sota}. The red dashed line is the upper bound given by the optimized extensions of the isotropic states, according to Corollary~\ref{cor:state_upperbound_maniopt}. The blue solid line corresponds to the upper bound given by~\cite[Proposition 5.1]{kianvashBoundingQuantumCapacity2022}, using flag extensions. We can observe a notable improvement in the given noise range. 

We emphasize that all our numerical experiments leverage high-precision solvers where the primal-dual gap for all semidefinite programs is consistently driven below $10^{-14}$, and all other entropy function calculations are performed at standard double-precision machine epsilon ($\approx 10^{-16}$). This high degree of accuracy ensures that the reported gains are significant and reliable. All state extensions as well as corresponding isometries (cf.~Fact~\ref{fact:state_ext}) we have obtained for the qubit isotropic states, which yield the SOTA upper bounds, can be found in~\cite{coderepo}.

\subsection{Further examples}\label{sec:upper_eg}
To demonstrate the efficacy of our method, we conduct numerical experiments on more general bipartite states and quantum channels to estimate upper bounds on their one-way distillable entanglement and quantum capacity, respectively.

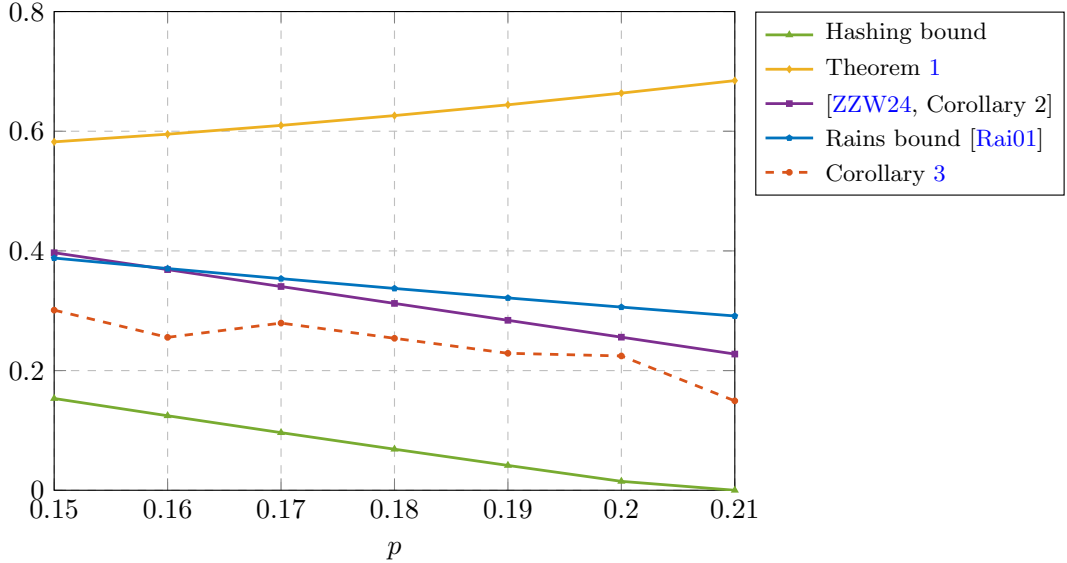
\begin{figure}[t]
\centering
\begin{tikzpicture}
    \node[anchor=south west,inner sep=0] (image) at (0,0) {\includegraphics[width=0.86\linewidth]{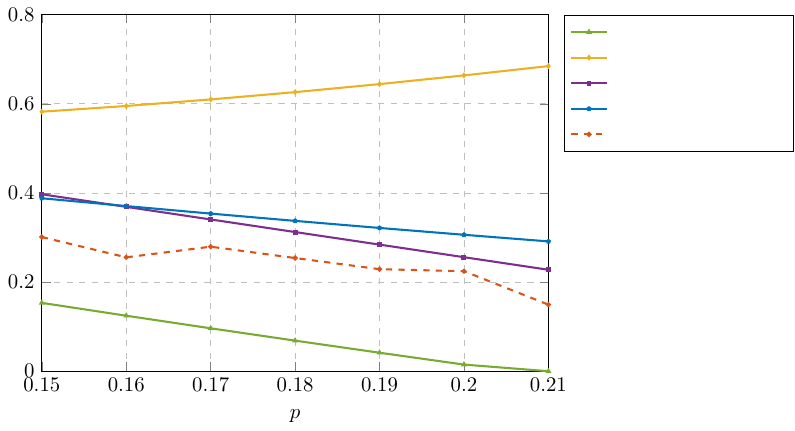}};
    \def\legendX{11}
    \node[anchor=south west,inner sep=0] at (\legendX,7.0){\footnotesize Hashing bound};
    \node[anchor=south west,inner sep=0] at (\legendX,6.58){\footnotesize Theorem~\ref{thm:oneway_upperbound}}; 
    \node[anchor=south west,inner sep=0] at (\legendX,6.02){\footnotesize \!\!\cite[Corollary 2]{Zhu_2024}};
    \node[anchor=south west,inner sep=0] at (\legendX,5.56){\footnotesize Rains bound~\cite{Rains2001}};
    \node[anchor=south west,inner sep=0] at (\legendX,5.11){\footnotesize Corollary~\ref{cor:state_upperbound_maniopt}} ;
\end{tikzpicture}
\caption{Bounds on the one-way distillable entanglement of maximum entangled states affected by $\cD_p$ and $\cA_{0.1}$. Select $p \in [0.15,0.22]$ as the plot range. The result includes a comparison between Theorem~\ref{thm:oneway_upperbound}, Ref.~\cite[Corollary 2]{Zhu_2024}, Rains bound~\cite{Rains2001}, Corollary~\ref{cor:state_upperbound_maniopt} and the hashing bound~\cite{devetakDistillationSecretKey2005} given by the state's coherent information.}
\label{fig:distillable entanglement DPAD}
\end{figure}

\paragraph{Quantum states.} Suppose Alice and Bob share some maximally entangled states affected by local noises, i.e.,
\begin{equation}\label{Eq:noisyMES}
    \rho_{A'B'} \coloneqq \cN_{A\to A'} \ox \cM_{B\to B'}(\Phi_{AB}).
\end{equation}
A canonical model for physical processes such as spontaneous emission is the \textit{amplitude damping channel} (AD) $\cA_p(\cdot)$~\cite{nielsen2010quantum}, which is defined by
\begin{equation}
    \cA_p(\cdot) = K_1 (\cdot) K_1 ^\dagger + K_2 (\cdot) K_2 ^\dagger,
\end{equation}
where $K_1 = \ketbra{0}{0}+ \sqrt{1-p}\ketbra{1}{1}, ~K_2 = \sqrt{p}\ketbra{0}{1}$ are Kraus operators. Set $\cN_{A\to A'}$ as the qubit amplitude damping channel $\cA_{0.1}$ and $\cM_{B\to B'}$ as the qubit depolarizing channel $\cD_{p}$. In Figure~\ref{fig:distillable entanglement DPAD}, we compare our continuity bound in Theorem~\ref{thm:oneway_upperbound} and the bound optimized through Corollary~\ref{cor:state_upperbound_maniopt} with previous bounds. The green line represents the Rains bound~\cite{Rains2001}, and the blue line represents the bound in~\cite[Corollary 2]{Zhu_2024} based on decomposing the state into degradable and antidegradable parts~\cite {Leditzky2018usefulsates}, which was reported as the best-known bound for this family of noisy states. We can observe that the optimized upper bound following Corollary~\ref{cor:state_upperbound_maniopt} outperforms all previous ones and can provide tighter estimations on such noisy states.

\paragraph{Quantum channels.} 
For channels, to numerically demonstrate the efficacy of Proposition~\ref{prop:refined_upperbound_qcapacity} and Corollary~\ref{cor: capacity_upperbound_maniopt}, we conduct numerical experiments on several examples. Specifically, we consider \textit{damping-dephasing channel} and \textit{damping-erasure channel}, which are compositions of an AD channel $\cA_p(\cdot)$ with a dephasing channel $\cZ_p(\cdot)$, or an erasure channel $\cE_p(\cdot)$, respectively. The dephasing channel and the erasure channel are defined as
\begin{align}
  \cZ_p(\cdot) = (1-p)(\cdot) + pZ(\cdot) Z ^\dagger,~\cE_p(\cdot) = (1-p)(\cdot) + p\ketbra{e}{e},
\end{align}
where $\ket{e}$ is the quantum state orthogonal to original quantum system, $Z=\mathrm{diag}(1,-1)$ is the Pauli $Z$ matrix. We will abbreviate
\begin{equation}
    \cZ\cA_{g,p}(\cdot) \coloneqq \cZ_p\circ\cA_g(\cdot),~~\text{and}~~\cE\cA_{g,p}(\cdot) \coloneqq \cE_p\circ\cA_g(\cdot)
\end{equation}
as the damping-dephasing and the damping-erasure channel, respectively, where the first parameter in the subscript denotes the noise parameter of the first composited channel. In particular, the damping-dephasing channel is argued as a natural and important noise model in~\cite{ghosh2012}. To use Proposition~\ref{prop:refined_upperbound_qcapacity} and Corollary~\ref{cor: capacity_upperbound_maniopt}, a central challenge is to calculate $U_\cM(\widehat{\cN})$ for any parameterized extended channel in the optimization iterations. Notice that after calculating $\dg(\widehat{\cN})$ and $\ve$-degrading channel $\cM$ via Eq.~\eqref{eq: epsilon degradable SDP}, the overall channel $\cM\circ\widehat{\cN}$ is fixed, and the quantity $U_\cM(\cN)$, i.e., maximizing the conditional entropy with a specified channel, can be computed via an SDP~\cite{fawzi2018efficient}. However, the SDP formulation~\cite{fawzi2018efficient} is still time-consuming, considering our need to estimate the gradient through a finite-difference scheme. Therefore, we opt to employ an alternative numerical technique during our implementation as follows.

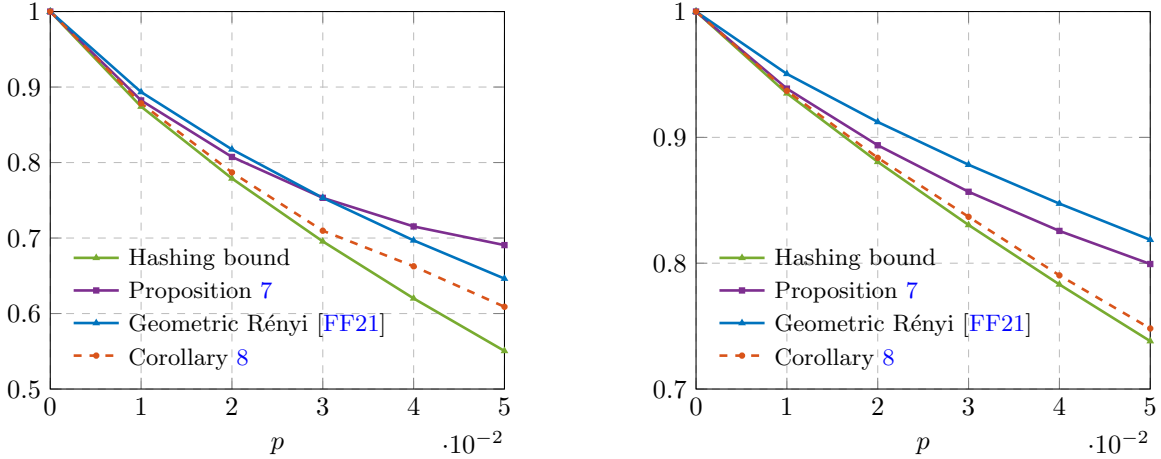
\begin{figure}[t]
\centering
\begin{subfigure}[t]{0.48\textwidth}
\centering
\begin{tikzpicture}
    \node[anchor=south west,inner sep=0] (image) at (0,0) {\includegraphics{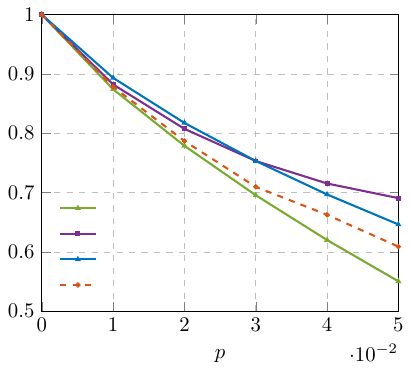}};
    \node[anchor=south west,inner sep=0] at (1.75,2.6){\footnotesize Hashing bound};
    \node[anchor=south west,inner sep=0] at (1.75,2.15){\footnotesize Proposition~\ref{prop:refined_upperbound_qcapacity}};
    \node[anchor=south west,inner sep=0] at (1.75,1.7){\footnotesize Geometric R\'enyi \cite{fangGeometricRenyiDivergence2021}};
    \node[anchor=south west,inner sep=0] at (1.75,1.27){\footnotesize  Corollary~\ref{cor: capacity_upperbound_maniopt}};
\end{tikzpicture}
\caption{Bounds on the quantum capacity of the damping-dephasing channel $\cZ\cA_{p,p}(\cdot)$.}
\label{fig:QC compare dephrasure3p}
\end{subfigure}
\hfill
\begin{subfigure}[t]{0.48\textwidth}
\centering
    \begin{tikzpicture}
    \node[anchor=south west,inner sep=0] (image) at (0,0) {\includegraphics{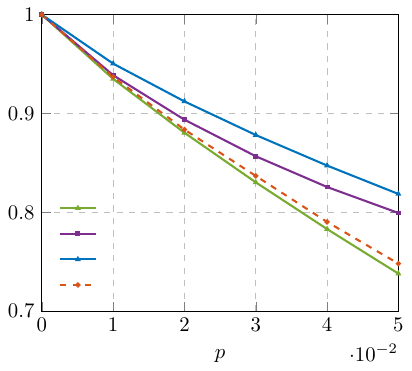}};
    \node[anchor=south west,inner sep=0] at (1.75,2.6){\footnotesize Hashing bound}; 
     \node[anchor=south west,inner sep=0] at (1.75,2.15){\footnotesize Proposition~\ref{prop:refined_upperbound_qcapacity}};
    \node[anchor=south west,inner sep=0] at (1.75,1.7){\footnotesize Geometric R\'enyi \cite{fangGeometricRenyiDivergence2021}};
    \node[anchor=south west,inner sep=0] at (1.75,1.27){\footnotesize  Corollary~\ref{cor: capacity_upperbound_maniopt}};        
\end{tikzpicture}
\caption{Bounds on the quantum capacity of the damping-erasure channel $\cE\cA_{p,p}(\cdot)$.}
\label{fig:QC_DamEra}
\end{subfigure}
\caption{Bounds on the quantum capacity of damping-dephasing channels and damping-erasure channels. The $x$-axis represents the parameter $p \in [0,0.05]$ for all channels. The results include a comparison between Proposition~\ref{prop:refined_upperbound_qcapacity}, Ref.~\cite[Theorem 12]{fangGeometricRenyiDivergence2021} and Corollary~\ref{cor: capacity_upperbound_maniopt}. The green solid line represents the coherent information of the channel's Choi state.
}
\label{fig:qcap_composited_channel}
\end{figure}

Note that the Choi matrix of the above channel is real. Thus, if we set the transformation $V$ to be restricted to a real unitary matrix, then the Choi matrix of the extended channel is also real. We argue that the Choi matrix of the optimal $\ve$-degrading channel $\cM$ in Eq.~\eqref{eq: epsilon degradable SDP} can be taken as a real matrix by convexity. That is, if $\sigma$ is a maximizing input state, then its conjugate $\sigma^*$ is also a maximizing state due to symmetry. Therefore $\rho = \frac{1}{2}(\sigma + \sigma^*)$ is also a maximizing state, leading to $\rho$ is real. Finally, we can argue that the optimal solution of $U_{\cM}(\cN)$ can be restricted to real matrices when the Choi matrix of $\cN$ is real. Thus to find the optimal value $U_{\cM}(\cN)$ we can simply scan all possible feasible solutions 
\begin{equation}
    \rho = \begin{bmatrix} a & b \\ b & 1-a \end{bmatrix},~~\text{where}~~ a \in [0,1], ~b \in \big[-\sqrt{a(1-a)},\sqrt{a(1-a)}\big].
\end{equation}
The accuracy can be controlled by adjusting the scanning step size. Typically, a step size of 0.005 is chosen, which provides sufficient precision compared to convex optimization results while reducing computation time. After the optimization is terminated, we can then calculate the upper bound in Proposition~\ref{prop:refined_upperbound_qcapacity} with the optimized extended channel.

The numerical results are shown in Figure~\ref{fig:qcap_composited_channel}. The red dashed line corresponds to the upper bound provided by Corollary~\ref{cor: capacity_upperbound_maniopt} combined with our optimization method, where we choose $|F| = 2$ and $|R| =|E|$. The purple solid line corresponds to the upper bound given by $R_\alpha(\widehat{\cN})$~\cite[Theorem 12]{fangGeometricRenyiDivergence2021}, which is the best-known SDP computable upper bound for general quantum channels, according to the best of our knowledge. Specifically, $R_\alpha(\widehat{\cN})$ is computed by an SDP~\cite[Proposition 13]{fangGeometricRenyiDivergence2021} and we set the parameter to $\alpha = 1 + 2^{-10}$. The green solid line shows the coherent information of the channel's Choi state, which serves as the hashing lower bound on its one-way distillable entanglement and is therefore a lower bound on the quantum capacity. We can observe that our method can achieve significantly tighter upper bounds than previous ones. Notably, in Figure~\ref{fig:QC_DamEra} for the damping-erasure channel, our upper bound nearly meets the lower bound, providing a sharp estimate of the channel's quantum capacity.

\section{Lower bounds on quantum communication}\label{sec:lower}
In this section, we develop Riemannian optimization methods for estimating lower bounds on the one-way distillable entanglement and the quantum capacity.

\subsection{Lower bounds on the one-way distillable entanglement}\label{sec:lower_oneway_DE}
Recall that the regularized formula of the one-way distillable entanglement is
\begin{align}
D_{\to}(\rho_{AB}) = \lim_{n\rightarrow \infty} \frac{1}{n} D^{(1)}_{\to}(\rho_{AB}^{\ox n}) = \lim_{n\rightarrow \infty} \frac{1}{n}\max_{\cT} I(A'\rangle B^nM)_{\cT(\rho_{AB}^{\ox n})},
\end{align}
where the maximization is over all quantum instruments $\cT:A\to A'M$ on Alice’s system. It is defined by $\cT(\cdot) = \sum_{j} T_j (\cdot) \ox \ketbra{j}{j}$, where $\{\ket{j}\}_{j=0}^{m-1}$ is an orthonormal basis for the classical register, each $T_j$ is a completely positive map such that $\sum_{j}T_j$ is a CPTP. Note that it suffices to consider instrument $T$ where each map $T_j$ consists of only one Kraus operator~\cite[Section 13.2.5]{khatri2024}, i.e.,
\begin{equation}\label{Eq:onekraus_T}
    \cT(\cdot) = \sum_{j=0}^{m-1} K_j (\cdot) K_j^\dagger \ox \ketbra{j}{j}.
\end{equation}
The one-way distillable entanglement is an operational quantity of great interest in entanglement theory. Due to the superadditivity of $D_{\to}^{(1)}(\cdot)$, to obtain a lower bound on $D_{\to}(\rho_{AB})$, a direct approach is to tackle the calculation of $D_{\to}^{(1)}(\rho_{AB}^{\ox n})$ for as large $n$ as possible. To this end, we formulate the problem as a Riemannian optimization over the \textit{unitary manifold} as follows. For any fixed $|M|=m$ and any $j=0,1,...,m-1$, construct
\begin{equation}\label{Eq:K_to_U}
    K_j \coloneqq (\1_A \ox \bra{j}_M)U_{AM}(\1_A \ox \ket{0}_M),
\end{equation}
where $U_{AM}$ is a unitary operator. It lies in $\cU(|AM|) = \mathrm{St}(|AM|,|AM|)$, the unitary manifold, a special case of the Stiefel manifold. This parameterization is nothing but the Stinespring dilation theorem. One can check that 
\begin{equation*}
\begin{aligned}
    \sum_{j=0}^{m-1} K_j^\dagger K_j &= \sum_{j=0}^{m-1}(\1_A \ox \bra{0})U_{AM}^\dagger(\1_A \ox \ketbra{j}{j})U_{AM}(\1_A \ox \ket{0})\\
    &=(\1_A \ox \bra{0}_M)U_{AM}^\dagger U_{AM}(\1_A \ox \ket{0}_M) =\1_A,
\end{aligned}
\end{equation*}
which shows $\{K_j\}$ is a set of Kraus operators. Taking Eq.~\eqref{Eq:K_to_U} into Eq.~\eqref{Eq:onekraus_T}, for any fixed $n,m\in\NN_+$, For a quantum state $\rho_{A|B}$, let us define the \textit{coherent information cost function $\rho_{A|B}$} as 
\begin{equation}
\mathrm{coh}_{\rho_{A|B}}(U_{AM}) = - I(A'\rangle BM)_{\sigma},
\end{equation}
where 
\begin{equation}
    \sigma_{A|BM} = \sum_{j=0}^{m-1} (\1_{AB} \ox \bra{j}_M)(\1_B\ox U_{AM}) (\rho_{AB} \ox \ketbra{0}{0}_M) (\1_B\ox U^{\dag}_{AM})(\1_{AB} \ox \ket{j}_M) \ox \ketbra{j}{j}_M.
\end{equation}

\begin{shaded}
\begin{lemma}\label{lem:state_coh_u_grad}
For a bipartite quantum state $\rho_{AB}$ and a system $M$ with $|M|=m\in\NN_+$, the Euclidean gradient of $\mathrm{coh}_{\rho_{AB}}(U_{AM})$ with respect to $U_{AM}$ is given by
\begin{equation}
    \nabla \mathrm{coh}_{\rho_{AB}} = 2\tr_B\bigg[\sum_{j=0}^{m-1}  \Big(\big(\1_A\ox \log \sigma_{B}^{(j)} - \log \sigma_{AB}^{(j)} \big)\ox \ketbra{j}{j}_M \Big)(\1_B\ox U_{AM}) (\rho_{AB}\ox \ketbra{0}{0}_M) \bigg],
\end{equation}
where $\sigma_{AB}^{(j)} = (\1_{AB}\ox \bra{j}_M)(\1_B\ox U_{AM}) (\rho_{AB}\ox \ketbra{0}{0}_M) (\1_B\ox U^{\dag}_{AM}) (\1_{AB}\ox \ket{j}_M)$ and $\sigma_{B}^{(j)} = \tr_A \sigma_{AB}^{(j)}$.
\end{lemma}
\end{shaded}

\begin{proof}
For any unitary $U_{AM}$, let's construct $K_j = (\1_A \ox \bra{j}_M)U_{AM}(\1_A \ox \ket{0}_M)$ as that in Eq.~\eqref{Eq:K_to_U}. We introduce some notations
\begin{equation}
    \rho^U_{ABM} = (\1_B\ox U_{AM}) (\rho_{AB}\ox \ketbra{0}{0}_M) (\1_B\ox U^{\dag}_{AM})
\end{equation}
and
\begin{equation}
    \sigma_{AB}^{(j)} = (\1_{AB}\ox \bra{j}_M)\rho^U_{ABM} (\1_{AB}\ox \ket{j}_M).
\end{equation}
Then the output state $\cT(\rho) = \sum_{j=0}^{m-1}K_j\rho_{AB} K_j^{\dag}\ox \ketbra{j}{j}_{M}$ and its marginal can be written as
\begin{equation}
    \sigma_{ABM} = \sum_{j=0}^{m-1} \sigma_{AB}^{(j)}\ox \ketbra{j}{j}_M~~\text{and}~~\sigma_{BM} = \sum_{j=0}^{m-1} \sigma_{B}^{(j)}\ox \ketbra{j}{j}_M,
\end{equation}
respectively. Now, we express the coherent information cost function $\rho_{A|B}$ as
\begin{equation}
    \mathrm{coh}_{\rho_{A|B}}(U_{AM}) = \tr \big(\sigma_{BM} \log \sigma_{BM}\big) - \tr \big(\sigma_{ABM}\log \sigma_{ABM}\big),
\end{equation}
and will calculate the gradient of it with respect to $U_{AM}$, i.e., $d \mathrm{coh}_{\rho_{A|B}} = \mathrm{Re}\big(\tr[ (\nabla_U \mathrm{coh}_{\rho_{A|B}})^\dag dU ]\big)$. Notice that for block-diagonal matrices, we have that
\begin{equation}
    \log\sigma_{ABM} = \sum_{j=0}^{m-1} \Big(\log \sigma_{AB}^{(j)}\Big)\ox \ketbra{j}{j}_M,~~\log \sigma_{BM} = \sum_{j=0}^{m-1} \Big(\log\sigma_{B}^{(j)}\Big)\ox \ketbra{j}{j}_M.
\end{equation}
It follows that
\begin{equation}
    \tr \big(\sigma_{ABM}\log \sigma_{ABM}\big) = \sum_{j=0}^{m-1} \tr \Big(\sigma_{AB}^{(j)}\log \sigma_{AB}^{(j)}\Big),~~\tr \big(\sigma_{BM}\log \sigma_{BM}\big) = \sum_{j=0}^{m-1} \tr \Big(\sigma_{B}^{(j)}\log \sigma_{B}^{(j)}\Big).
\end{equation}
and
\begin{equation}
\mathrm{coh}_{\rho_{A|B}}(U_{AM}) = \sum_{j=0}^{m-1} \tr \Big(\sigma_{B}^{(j)}\log \sigma_{B}^{(j)}\Big) - \tr \Big(\sigma_{AB}^{(j)}\log \sigma_{AB}^{(j)}\Big).
\end{equation}
Using the fact that $d \tr (X\log X) = \tr[(\log X + \1) dX]$ and $d\sigma_{AB}^{(j)} = (\1_{AB}\ox \bra{j}_M)d\rho^U_{ABM} (\1_{AB}\ox \ket{j}_M)$, we can deduce that
\begin{equation}\label{Eq:dcoh1}
\begin{aligned}
d \mathrm{coh}_{\rho_{A|B}} &= \sum_{j=0}^{m-1} d\tr \Big(\sigma_{B}^{(j)}\log \sigma_{B}^{(j)}\Big) - d\tr \Big(\sigma_{AB}^{(j)}\log \sigma_{AB}^{(j)}\Big) \\
&=\sum_{j=0}^{m-1} \tr \Big[\big(\log \sigma_{B}^{(j)} + \1_B\big)d\sigma_{B}^{(j)}\Big] - \tr \Big[\big(\log \sigma_{AB}^{(j)} + \1_{AB}\big)d\sigma_{AB}^{(j)}\Big] \\
&=\sum_{j=0}^{m-1} \tr \Big[\big(\1_A\ox \log \sigma_{B}^{(j)} - \log \sigma_{AB}^{(j)} \big)(\1_{AB}\ox \bra{j}_M)d\rho^U_{ABM} (\1_{AB}\ox \ket{j}_M)\Big]\\
&= \tr \Big(Y_{ABM} d\rho^U_{ABM}\Big),
\end{aligned}
\end{equation}
where in the last equality we defined
\begin{equation}
    Y_{ABM} \coloneqq \sum_{j=0}^{m-1}  \big(\1_A\ox \log \sigma_{B}^{(j)} - \log \sigma_{AB}^{(j)} \big)\ox \ketbra{j}{j}_M.
\end{equation}
Note the differential of $\rho^U_{ABM}$ is given by
\begin{equation}
    d\rho^U_{ABM} = (\1_B\ox dU_{AM}) (\rho_{AB}\ox \ketbra{0}{0}_M) (\1_B\ox U^{\dag}_{AM}) + (\1_B\ox U_{AM}) (\rho_{AB}\ox \ketbra{0}{0}_M) (\1_B\ox dU^{\dag}_{AM}).
\end{equation}
We can plug it into Eq.~\eqref{Eq:dcoh1} to obtain
\begin{equation}
\begin{aligned}
d\mathrm{coh}_{\rho_{A|B}} &= \tr \Big[Y_{ABM}(\1_B\ox dU_{AM}) (\rho_{AB}\ox \ketbra{0}{0}_M) (\1_B\ox U^{\dag}_{AM})\Big]\\
&\quad + \tr \Big[Y_{ABM}(\1_B\ox U_{AM}) (\rho_{AB}\ox \ketbra{0}{0}_M) (\1_B\ox dU^{\dag}_{AM})\Big]\\
&= \tr \Big[ \tr_B \Big((\rho_{AB}\ox \ketbra{0}{0}_M) (\1_B\ox U^{\dag}_{AM})Y_{ABM}\Big) dU_{AM}\Big]\\
&\quad + \tr \Big[\tr_B\Big(Y_{ABM}(\1_B\ox U_{AM}) (\rho_{AB}\ox \ketbra{0}{0}_M) \Big)dU^{\dag}_{AM}\Big].
\end{aligned}
\end{equation}
Therefore, we can identify the gradient
\begin{equation}
    \nabla \mathrm{coh}_{\rho_{AB}} = 2\tr_B\bigg[\sum_{j=0}^{m-1}  \Big(\big(\1_A\ox \log \sigma_{B}^{(j)} - \log \sigma_{AB}^{(j)} \big)\ox \ketbra{j}{j}_M \Big)(\1_B\ox U_{AM}) (\rho_{AB}\ox \ketbra{0}{0}_M) \bigg].
\end{equation}
\end{proof}

The Riemannian gradient is then given as follows.

\begin{shaded}
\begin{corollary}
For a bipartite quantum state $\rho_{AB}$ and a system $M$ with $|M|=m\in\NN_+$, the Riemannian gradient of $\mathrm{coh}_{\rho_{AB}}(U_{AM})$ with respect to $U_{AM}$ is given by
\begin{equation}\label{Eq:Rieman_grad_UAM}
\begin{aligned}
\grad &\mathrm{coh}_{\rho_{AB}}(U_{AM}) = \tr_B\bigg[\sum_{j=0}^{m-1}  \Big(\big(\1_A\ox \log \sigma_{B}^{(j)} - \log \sigma_{AB}^{(j)} \big)\ox \ketbra{j}{j}_M \Big)(\1_B\ox U_{AM}) (\rho_{AB}\ox \ketbra{0}{0}_M) \bigg] \\
& -U_{AM} \tr_B\bigg[(\rho_{AB}\ox \ketbra{0}{0}_M)(\1_B\ox U_{AM}^\dag) \sum_{j=0}^{m-1}  \Big(\big(\1_A\ox \log \sigma_{B}^{(j)} - \log \sigma_{AB}^{(j)} \big)\ox \ketbra{j}{j}_M \Big)\bigg] U_{AM}.
\end{aligned}
\end{equation}
where $\sigma_{AB}^{(j)} = (\1_{AB}\ox \bra{j}_M)(\1_B\ox U_{AM}) (\rho_{AB}\ox \ketbra{0}{0}_M) (\1_B\ox U^{\dag}_{AM}) (\1_{AB}\ox \ket{j}_M)$ and $\sigma_{B}^{(j)} = \tr_A \sigma_{AB}^{(j)}$.
\end{corollary}
\end{shaded}

\begin{proof}
For the unitary manifold, according to Eq.~\eqref{Eq:st_proj}, the projection onto the tangent space is given by
\begin{equation}
    \mathrm{Proj}_V(X) = \frac{1}{2} (X-VX^\dag V).
\end{equation}
Therefore, by Lemma~\ref{lem:state_coh_u_grad}, we have that
\begin{equation}
\begin{aligned}
\grad &\mathrm{coh}_{\rho_{AB}}(U_{AM}) = \tr_B\bigg[\sum_{j=0}^{m-1}  \Big(\big(\1_A\ox \log \sigma_{B}^{(j)} - \log \sigma_{AB}^{(j)} \big)\ox \ketbra{j}{j}_M \Big)(\1_B\ox U_{AM}) (\rho_{AB}\ox \ketbra{0}{0}_M) \bigg] \\
& -U_{AM} \tr_B\bigg[(\rho_{AB}\ox \ketbra{0}{0}_M)(\1_B\ox U_{AM}^\dag) \sum_{j=0}^{m-1}  \Big(\big(\1_A\ox \log \sigma_{B}^{(j)} - \log \sigma_{AB}^{(j)} \big)\ox \ketbra{j}{j}_M \Big)\bigg] U_{AM}.
\end{aligned}
\end{equation}
\end{proof}

Equipped with the Riemannian gradient, we can thereby apply the standard RGD algorithm as shown in Algorithm~\ref{alg:RGD_Dlowerbound} to estimate $D_{\to}^{(1)}(\rho_{AB}^{\ox n})$ for a fixed bipartite state $\rho_{AB}$ and a fixed $n\in\NN_+$.

\begin{algorithm}[t]
\caption{RGD algorithm for lower bounds on the one-way distillable entanglement}\label{alg:RGD_Dlowerbound}
\begin{algorithmic}[1]
      \REQUIRE A quantum state $\rho_{AB}$, $m\in\NN_+$; Initial guess $U^{(1)}\in \cU(|AM|)$. 
      \WHILE{the stopping criteria are not satisfied}
          \STATE Compute $\dot{U}^{(t)} = -\grad \mathrm{coh}_{\rho_{AB}}(U^{(t)})$ by Eq.~\eqref{Eq:Rieman_grad_UAM}.
          \STATE Compute a stepsize $s^{(t)}$.
          \STATE Update $U^{(t+1)}= \mathrm{R}_{U^{(t)}} \big(s^{(t)}\dot{U}^{(t)}\big)$ by Eq.~\eqref{Eq:retr_qr}; $t= t+1$.
      \ENDWHILE
      \STATE Compute $I_{\mathrm{low}} = -\frac{1}{n}\mathrm{coh}_{\rho_{AB}}(U^*)$
      \ENSURE $U^*,~I_{\mathrm{low}}$.
\end{algorithmic}
\end{algorithm}

\subsection{Lower bounds on the quantum capacity}\label{sec:lower_qcap}

A key challenge in determining a channel's quantum capacity, $Q(\cN)$, is the superadditivity of its underlying coherent information, i.e., entangled input state across multiple channel uses $\cN^{\ox n}$ can strictly improve the coherent information compared with strategies that use the channels independently. Consequently, the standard approach to establishing lower bounds is to compute $\frac{1}{n}Q^{(1)}(\cN^{\ox n})$ for large $n$, a quantity that becomes progressively tighter as the blocklength $n$ increases. Recall that the $n$-shot channel coherent information is given by
\begin{equation}\label{Eq:Q1n_opt}
    Q^{(1)}(\cN^{\ox n}) \coloneqq \max_{\psi_{RA^n}} I(R\rangle B)_{\cN^{\ox n}_{A\to B}(\psi_{RA^n})},
\end{equation}
where the optimization is with respect to all \textit{code states} $\ket{\psi}_{RA^n}$ and we abbreviate $\ketbra{\psi}{\psi}$ as $\psi$. One key strategy to obtaining a tight lower bound is to restrict the search to code state families with high symmetry~\cite{Johannes2021,bhalerao2025}. This approach can render the entropy computation manageable even for large blocklengths, enabling a more accurate assessment of superadditive gains. A notable recent success in this method is the use of permutation-invariant codes~\cite{bhalerao2025}, which have led to improved capacity thresholds for several important channel models. Another strategy moves away from rigid symmetries and instead utilizes highly expressive, general-purpose ansatz. A key example is to utilize a neural network state ansatz to estimate lower bounds on quantum capacity~\cite{Bausch_2020}.

In fact, it is straightforward to see that Eq.~\eqref{Eq:Q1n_opt} is an optimization on the \textit{complex unit sphere} of the corresponding Hilbert space, a canonical example of a Riemannian manifold. Thereby, we can apply Riemannian optimization algorithms to compute Eq.~\eqref{Eq:Q1n_opt}. The tangent space of a complex unit sphere $\mathrm{S}^{|A|}$ is given by $\tangent_{\ket{\psi}}\mathrm{S}^{|A|} = \{\ket{\dot{\psi}}\in\CC^{|A|}: \braket{\psi}{\dot{\psi}} = 0\}$, the projection onto which is
\begin{equation}\label{Eq:sphere_proj}
\mathrm{Proj}_{\ket{\psi}}(\ket{v}) = (\1-\ketbra{\psi}{\psi})\ket{v},~\forall\ket{v}\in\CC^{|A|}.
\end{equation}
Then we can use the RGD algorithm to compute Eq.~\eqref{Eq:Q1n_opt} once we can obtain the Riemannian gradient. To this end, we first calculate the Euclidean gradient of the objective function as follows.

\begin{shaded}
\begin{lemma}\label{lem:grad_psi}
For a quantum channel $\cN_{A\to B}$ and a system $R$, let $f_{\cN}(\psi_{RA}) = - I(R\rangle B)_{\cN_{A\to B}(\psi_{RA})}$. The Euclidean gradient of $f_{\cN}(\psi_{RA})$ at $\ket{\psi}_{RA}$ is given by
\begin{equation}
\nabla_{\ket{\psi}} f_{\cN} = 2 \left\{\1_R \ox \cN^{\dagger}_{B\to A} \Big[\log \tr_R\cN_{A\to B}(\psi_{RA}) \Big] -\cN^{\dagger}_{B\to A} \Big[\log \cN_{A\to B}(\psi_{RA})\Big]\right\}\ket{\psi}_{RA}.
\end{equation}
\end{lemma}
\end{shaded}

\begin{proof}
Since $\tr_R X_{RA} = \sum_j(\bra{j}_R\ox\1_A)(X_{RA})(\ket{j}_R\ox \1_A)$, by denoting the Kraus operators of $\cN_{A\to B}$ as $\{K_k\}_k$, we can express the function $f_{\cN}(\psi_{RA}) \coloneq I(R\rangle B)_{\cN_{A\to B}(\psi_{RA})}$ as
\begin{equation}\label{Eq:coh_psi}
\begin{aligned}
    f_{\cN}(\psi_{RA}) =&\tr \Big[\sum_{k,j} (\bra{j}_{R}\ox K_{k})\ketbra{\psi}{\psi}(\ket{j}_{R}\ox K_{k}^{\dagger}) \log \Big(\sum_{k,j} (\bra{j}_{R}\ox K_{k})\ketbra{\psi}{\psi}(\ket{j}_{R}\ox K_{k}^{\dagger})\Big)\Big]\\
    &\quad - \sum_{k} \bra{\psi}(\1_{R}\ox K_{k}^{\dagger}) \log \Big(\sum_{k} (\1_{R}\ox K_{k})\ketbra{\psi}{\psi}(\1_{R}\ox K_{k}^{\dagger})\Big)(\1_{R}\ox K_{k})\ket{\psi}.
\end{aligned}
\end{equation}
In order to compute the derivatives, we introduce the following argument
\begin{equation}\label{Eq:dir_de}
    \frac{\partial}{\partial \ket{\ell}} \bra{\psi}\mathsf{W}(\ket{\psi})\ket{\psi} =\bra{\psi}\frac{\partial}{\partial \ket{\ell}} \mathsf{W}(\ket{\psi}) \ket{\psi} + 2\bra{\ell} \mathsf{W}(\ket{\psi})\ket{\psi},
\end{equation}
that computes the directional derivative of the function $\bra{\psi}\mathsf{W}(\ket{\psi})\ket{\psi}$ along a vector $\ket{\ell}\in\mathbb{C}^{n}$, where $\mathsf{W}:\mathbb{C}^{n}\to\mathbb{C}^{n\times n}$ is a mapping. 
Denote the first term and the second term in Eq.~\eqref{Eq:coh_psi} as
\begin{equation*}
\begin{aligned}
    &f_{\cN}^{(1)} = \bra{\psi} \Big[\sum_{k_1,j_1}(\ket{j_1}_{R}\ox K_{k_1}^{\dagger}) \log \Big(\sum_{k_2,j_2} (\bra{j_2}_{R}\ox K_{k_2})\ketbra{\psi}{\psi}(\ket{j_2}_{R}\ox K_{k_2}^{\dagger})\Big) (\bra{j_1}_{R}\ox K_{k_1})\Big] \ket{\psi}\\
    &f_{\cN}^{(2)} = \bra{\psi} \Big[\sum_{k_1} (\1_{R}\ox K_{k_1}^{\dagger}) \log \Big(\sum_{k_2} (\1_{R}\ox K_{k_2})\ketbra{\psi}{\psi}(\1_{R}\ox K_{k_2}^{\dagger})\Big)(\1_{R}\ox K_{k_1})\Big]\ket{\psi}.
\end{aligned}
\end{equation*}
For $f_{\cN}^{(1)}$, according to Eq.~\eqref{Eq:dir_de}, we can calculate the directional derivative along the state $\ket{\ell}$ as
\begin{equation}\label{Eq:deri_f1}
\begin{aligned}
    \langle\nabla_{\ket{\psi}} f_{\cN}^{(1)}\ket{\ell} &= 2\bra{\ell} \sum_{k_1,j_1}(\ket{j_1}\ox K_{k_1}^{\dagger}) \log \Big(\sum_{k_2,j_2} (\bra{j_2}\ox K_{k_2})\ketbra{\psi}{\psi}(\ket{j_2}\ox K_{k_2}^{\dagger})\Big) (\bra{j_1}\ox K_{k_1}) \ket{\psi}\\
    +&\bra{\psi} \frac{\partial}{\partial \ket{\ell}} \sum_{k_1,j_1}(\ket{j_1}\ox K_{k_1}^{\dagger}) \log \Big(\sum_{k_2,j_2} (\bra{j_2}\ox K_{k_2})\ketbra{\psi}{\psi}(\ket{j_2}\ox K_{k_2}^{\dagger})\Big) (\bra{j_1}\ox K_{k_1}) \ket{\psi}.
\end{aligned}
\end{equation}
The second term can be calculated by
\begin{equation}\label{Eq:deri_f12}
\begin{aligned}
    &\bra{\psi} \frac{\partial}{\partial \ket{\ell}} \sum_{k_1,j_1}(\ket{j_1}\ox K_{k_1}^{\dagger}) \log \Big(\sum_{k_2,j_2} (\bra{j_2}\ox K_{k_2})\ketbra{\psi}{\psi}(\ket{j_2}\ox K_{k_2}^{\dagger})\Big) (\bra{j_1}\ox K_{k_1}) \ket{\psi}\\
    &= \bra{\psi} \sum_{k_1,j_1} (\ket{j_1}\ox K_{k_1}^{\dagger}) \Big( \sum_{k_2,j_2} (\bra{j_2}\ox K_{k_2})(\ketbra{\ell}{\psi}+\ketbra{\psi}{\ell}) (\ket{j_2} \ox K_{k_2}^\dagger)\Big) \sigma^{-1} (\bra{j_1} \ox K_{k_1})\ket{\psi}\\
    &= \tr\sum_{k_1,j_1} \Big[\Big( \sum_{k_2,j_2} (\bra{j_2} \ox K_{k_2})(\ketbra{\ell}{\psi}+\ketbra{\psi}{\ell}) (\ket{j_2} \ox K_{k_2}^\dagger)\Big) \sigma^{-1} (\bra{j_1}\ox K_{k_1})\ketbra{\psi}{\psi}(\ket{j_1}\ox K_{k_1}^\dagger)\Big]\\
    &= \tr \Big( \sum_{k_2,j_2} (\bra{j_2} \ox K_{k_2})(\ketbra{\ell}{\psi}+\ketbra{\psi}{\ell}) (\ket{j_2} \ox K_{k_2}^\dagger)\Big)\\
    &= \tr \Big( \sum_{k_2,j_2} (\ketbra{\ell}{\psi}+\ketbra{\psi}{\ell}) (\ketbra{j_2}{j_2} \ox K_{k_2}^\dagger K_{k_2})\Big)\\
    &= 2\braket{\ell}{\psi}.
\end{aligned}
\end{equation}
Similarly, for $f_{\cN}^{(2)}$, we have
\begin{equation}\label{Eq:deri_f2}
\begin{aligned}
    \langle\nabla_{\ket{\psi}} f_{\cN}^{(2)}\ket{\ell} &= 2\bra{\ell} \sum_{k_1} (\1\ox K_{k_1}^{\dagger}) \log \Big(\sum_{k_2} (\1\ox K_{k_2})\ketbra{\psi}{\psi}(\1\ox K_{k_2}^{\dagger})\Big)(\1\ox K_{k_1})\ket{\psi}\\
    +&\bra{\psi} \frac{\partial}{\partial \ket{\ell}}\sum_{k_1} (\1\ox K_{k_1}^{\dagger}) \log \Big(\sum_{k_2} (\1\ox K_{k_2})\ketbra{\psi}{\psi}(\1_{R}\ox K_{k_2}^{\dagger})\Big)(\1_{R}\ox K_{k_1})\ket{\psi}.
\end{aligned}
\end{equation}
The second term can be calculated by
\begin{equation}\label{Eq:deri_f22}
\begin{aligned}
    &\bra{\psi} \frac{\partial}{\partial \ket{\ell}}\sum_{k_1} (\1\ox K_{k_1}^{\dagger}) \log \Big(\sum_{k_2} (\1\ox K_{k_2})\ketbra{\psi}{\psi}(\1\ox K_{k_2}^{\dagger})\Big)(\1\ox K_{k_1})\ket{\psi}\\
    &= \bra{\psi} \sum_{k_1} (\1\ox K_{k_1}^\dagger) \Big( \sum_{k_2} (\1\ox K_{k_2})(\ketbra{\ell}{\psi}+\ketbra{\psi}{\ell}) (\1\ox K_{k_2}^\dagger)\Big) \sigma^{-1} (\1\ox K_{k_1})\ket{\psi}\\
    &= \tr\sum_{k_1} \Big[\Big( \sum_{k_2} (\1\ox K_{k_2})(\ketbra{\ell}{\psi}+\ketbra{\psi}{\ell}) (\1\ox K_{k_2}^\dagger)\Big) \sigma^{-1} (\1\ox K_{k_1})\ketbra{\psi}{\psi}(\1\ox K_{k_1}^\dagger)\Big]\\
    &= \tr\sum_{k_2} \Big[ (\ketbra{\ell}{\psi}+\ketbra{\psi}{\ell}) (\1\ox K_{k_2}^\dagger K_{k_2})\Big]\\
    &= \tr \big[\ketbra{\ell}{\psi} + \ketbra{\psi}{\ell}\big]\\
    &= 2\braket{\ell}{\psi}.
\end{aligned}
\end{equation}
Combining Eqs.~\eqref{Eq:deri_f1}~\eqref{Eq:deri_f12}~\eqref{Eq:deri_f2}~\eqref{Eq:deri_f22}, we have that
\begin{equation}
\begin{aligned}
    \nabla_{\ket{\psi}} f_{\cN} &= \nabla_{\ket{\psi}} f_{\cN}^{(1)} - \nabla_{\ket{\psi}} f_{\cN}^{(2)}\\
    &= 2 \Big\{\sum_{k_1,j_1}(\ket{j_1}\ox K_{k_1}^{\dagger}) \log \big[\tr_R\cN_{A\to B}(\psi)\big] (\bra{j_1}\ox K_{k_1}) -\cN^{\dagger}_{B\to A}\Big[\log \cN_{A\to B}(\psi)\Big]\Big\}\ket{\psi}\\
    &= 2 \left\{\1 \ox \cN^{\dagger}_{B\to A} \Big[\log \tr_R\cN_{A\to B}(\psi) \Big] -\cN^{\dagger}_{B\to A} \Big[\log \cN_{A\to B}(\psi)\Big]\right\}\ket{\psi}.
\end{aligned}
\end{equation}
\end{proof}

Based on Lemma~\ref{lem:grad_psi} and the projection given in Eq.~\eqref{Eq:sphere_proj}, we can obtain the Riemannian gradient and use the standard RGD method to tackle the problem in Eq.~\eqref{Eq:Q1n_opt}. However, this formulation faces a significant scalability challenge. As the blocklength $n$ increases, the dimension of this sphere grows exponentially, quickly rendering the problem computationally intractable. To circumvent this curse of dimensionality, we reformulate the problem as an optimization over a \textit{product of unitary manifolds} in the manner described below.

\begin{figure}[t]
    \centering
    \includegraphics[width=0.75\linewidth]{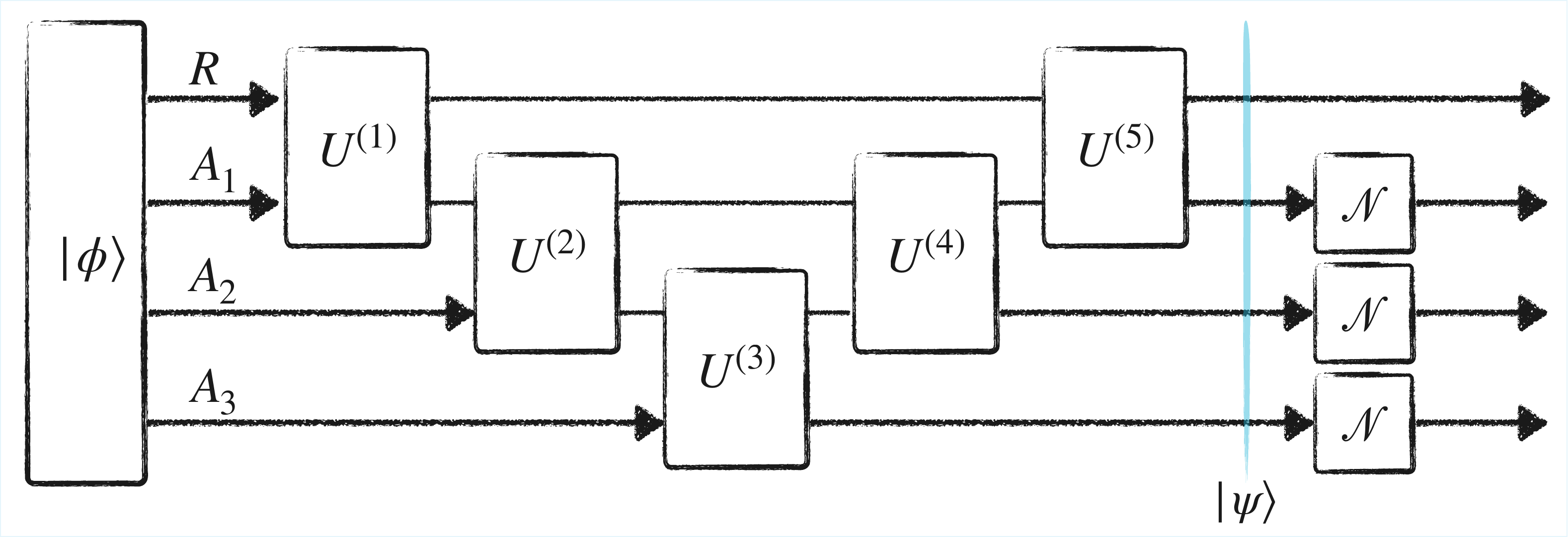}
    \caption{An example of estimating $\frac{1}{3}Q^{(1)}(\cN_{A\to B}^{\ox 3})$ using a product of unitary manifolds. Five local unitaries are interleaved to parameterize an input code state $\ket{\psi}_{RA_1A_2A_3}$ for 3 copies of the channel.}
    \label{fig:interleave_U}
\end{figure}

To illustrate our method, we consider an example of computing the average 3-shot coherent information, $\frac{1}{3}Q^{(1)}(\cN_{A\to B}^{\ox 3})$. Let the three input systems be denoted by $A_1, A_2, A_3$, where $A_i \cong A$, and let their joint system be $A^3 \coloneqq A_1A_2A_3$. We parameterize the input state $|\psi\rangle_{RA^3}$ with an auxiliary system $R$ as
\begin{equation}\label{Eq:psi_RA3}
    \ket{\psi}_{RA^3} = (U^{(5)}_{RA_1}\ox \1_{A_2A_3}) (U^{(4)}_{A_1A_2}\ox \1_{RA_3})(U^{(3)}_{A_2A_3}\ox \1_{RA_1}) (U^{(2)}_{RA_2}\ox \1_{A_1A_3})(U^{(1)}_{RA_1}\ox \1_{A_2A_3})\ket{\phi}_{RA^3},
\end{equation}
where $\ket{\phi}_{RA^3}$ is a fixed initial state, e.g., $\ket{0}^{\ox |R||A|^3}$, and each $U^{(i)}$ is a unitary operator acting on a specific subsystem, creating an interleaved structure as depicted in Figure~\ref{fig:interleave_U}. We call such a circuit an \textit{interleaved local unitary ansatz}. Similar ideas have been adopted to optimize quantum circuits for Hamiltonian simulation, e.g., using a brick wall circuit ansatz and then applying Riemannian optimization~\cite{Kotil_2024,Le2025riemannianquantum}. Then the optimization is over a product of unitary manifolds
\begin{equation}
\cU(|RA|) \times \cU(|A|^2) \times \cdots \times \cU(|A|^2)\times \cU(|RA|).
\end{equation}
The key insight of this method is to interleave local unitaries to avoid the curse of dimensionality while maintaining expressivity of the parameterized input state. If $|R|$ is a fixed constant, then each constituent manifold has a fixed dimension that does not scale with $n$, ensuring the problem remains manageable even for large blocklengths. Even if we want to make $|R|$ increase dependent on $n$, e.g., $|R|=|A|^n$, we can alternatively divide $R$ into local subsystems and construct a similar manageable local structure.

Formally, for $n$-copy uses of a quantum channel $\cN$, let us define the \textit{coherent information cost function of $n$-shot $\cN$} as 
\begin{equation}
\mathrm{coh}_{\cN, n}(\mathbf{U}_n) = - I(R\rangle B)_{\cN^{\ox n}_{A\to B}(\psi(\mathbf{U}_n))},
\end{equation}
where 
\begin{equation}\label{Eq:U_n}
    \mathbf{U}_n = (U^{(1)}, U^{(2)}, \cdots, U^{(2n-1)})\in \underbrace{\cU(|RA|) \times \cU(|A|^2) \times \cdots \times \cU(|RA|)}_{2n-1},
\end{equation}
and
\begin{equation}
    \ket{\psi(\mathbf{U})}_{RA^n} = {\overleftarrow{\prod}}_{k=n+1}^{2n-1} \Big(U^{(k)}_{S_{2n-1-k}S_{2n-k}}\ox \1_{S^{(k)}_c}\Big) {\overleftarrow{\prod}}_{k=1}^{n} \Big(U^{(k)}_{S_{k-1}S_k}\ox \1_{S^{(k)}_c}\Big)\ket{0}_{RA^n}.
\end{equation}
Here, the notation $\overleftarrow{\prod}$ denotes the ordered product of operators, arranged from left to right in descending order of the index. We set $S_0=R,S_j=A_j$ and denote the corresponding complementary subsystem by 
\begin{equation}
S_c^{(k)} \coloneqq \left\{
\begin{aligned}
& S_1S_2\cdots S_n \setminus S_{k-1}S_k,\quad  &&\text{for}~ 1\leq k\leq n,\\
& S_1S_2\cdots S_n \setminus S_{2n-1-k}S_{2n-k},\quad &&\text{for}~ n< k\leq 2n-1.
\end{aligned}\right.
\end{equation}
The tangent space of the product manifold is given by
\begin{equation}
    \tangent_{U^{(1)}} \cU(|RA|) \times \tangent_{U^{(2)}} \cU(|A|^2) \times \cdots \tangent_{U^{(2n-1)}} \cU(|RA|),
\end{equation}
where each local tangent space is defined in Eq.~\eqref{Eq:st_tangspace}. The Riemannian gradient is given by
\begin{equation}\label{Eq:Rieman_grad_prod}
    \grad \mathrm{coh}_{\cN,n}(\mathbf{U}) = \Big(\mathrm{Proj}_{U^{(1)}}\big(\partial_{U^{(1)}} \widebar{\mathrm{coh}}(\mathbf{U})\big), \cdots, \mathrm{Proj}_{U^{(2n-1)}}\big(\partial_{U^{(2n-1)}} \widebar{\mathrm{coh}}(\mathbf{U})\big)\Big),
\end{equation}
where $\mathrm{Proj}_{U^{k}}(\cdot)$ is the projection onto the tangent space as defined in Eq.~\eqref{Eq:st_proj} and $\widebar{\mathrm{coh}}: \CC^{|RA|\times |RA|}\times \CC^{|A|^2\times |A|^2}\times \cdots \CC^{|RA|\times |RA|} \to \RR$ is an extension function of $\mathrm{coh}$ that coincide with the later on the product of unitary manifolds. The retraction map is given in Eq.~\eqref{Eq:retr_qr}.

Since such an ansatz (cf.~Eq.~\eqref{Eq:U_n} and Figure~\ref{fig:interleave_U}) is not guaranteed to express all possible input states, we have the relation
\begin{equation}\label{Eq:min_coh_opt}
    Q^{(1)}(\cN^{\ox n}) \geq -\min_{\mathbf{U}_n} \mathrm{coh}_{\cN,n}(\mathbf{U}_n),
\end{equation}
to give a valid lower bound on $Q^{(1)}(\cN^{\ox n})$, thus a lower bound on $Q(\cN)$. Then Eq.~\eqref{Eq:min_coh_opt} is the main optimization we are concerned with, which we utilize the RGD algorithm to solve. The Riemannian gradient can be calculated as follows.

\begin{shaded}
\begin{lemma}\label{lem:grad_local_U}
For a quantum channel $\cN_{A\to B}$ and its $n$-shot coherent information cost function $\mathrm{coh}_{\cN,n}(\mathbf{U})$, denote $V^{(k)}=U^{(k)}\ox \1_{S_c^{(k)}}, \ket{\phi_k} = {\overleftarrow{\prod}}_{j=1}^k V^{(j)}\ket{0}_{RA^n}$, and $\ket{\psi}_{RA^n} = \ket{\phi_{2n-1}}$. Then the partial derivative of $\mathrm{coh}_{\cN,n}(\mathbf{U})$ 
at $U^{(k)}$ is given by
\begin{equation}
\partial_{U^{(k)}} \mathrm{coh}_{\cN,n} = \tr_{S_c^{(k)}} \Big[\big(V^{(2n-1)} V^{(2n-2)}\cdots V^{(k+1)} \big)^\dag\ketbra{G_{\psi}}{\phi_{k-1}}\Big],
\end{equation}
for $k=1,2,\cdots, 2n-1$, where $\ket{G_{\psi}} = \partial_{\ket{\psi}}f_{\cN}$ is given in Lemma~\ref{lem:grad_psi}.
\end{lemma}
\end{shaded}

\begin{proof}
Based on Lemma~\ref{lem:grad_psi}, we can use the chain rule to obtain the desired partial derivative. Denote the gradient vector with respect to a state $\ket{\psi}_{RA^n}$ as $\ket{G_{\psi}}$. We have 
\begin{equation}\label{Eq:df_dcoh_def}
    d f_{\cN} = \mathrm{Re}\big(\bra{G_{\psi}} d\ket{\psi}\big),~~d \mathrm{coh}_{\cN,n} = \mathrm{Re}\Big(\tr\big[(\nabla_{\mathbf{U}} \mathrm{coh}_{\cN,n})^\dag d\mathbf{U}\big]\Big).
\end{equation}
We will calculate the gradient with respect to a single, arbitrary unitary $U^{(k)}$. To this end, we consider the change $d \mathrm{coh}_{\cN,n}$ that results only from an infinitesimal change $d U^{(k)}$, while all other $U^{(j)}$ for $j \neq k$ are held constant. This is equivalent to taking a partial derivative. In this case, $dV^{(j)} = 0$ for all $j \neq k$. Using the fact that
\begin{equation}
    d\ket{\psi}|_{U^{(k)}} = V^{(2n-1)} V^{(2n-2)}\cdots V^{(k+1)} dV^{(k)}\ket{\phi_{k-1}},
\end{equation}
we can deduce
\begin{equation}
\begin{aligned}
    d \mathrm{coh}_{\cN,n} &= \mathrm{Re}\Big( \tr\Big[ \ketbra{\phi_{k-1}}{G_{\psi}} V^{(2n-1)} V^{(2n-2)}\cdots V^{(k+1)} dV^{(k)}\Big]\Big)\\
    &= \mathrm{Re}\Big( \tr\Big[ \ketbra{\phi_{k-1}}{G_{\psi}} V^{(2n-1)} V^{(2n-2)}\cdots V^{(k+1)} (dU^{(k)}\ox \1_{S_k})\Big]\Big)\\
    &= \mathrm{Re}\Big( \tr \Big[ \tr_{S_k} \big(\ketbra{\phi_{k-1}}{G_{\psi}} V^{(2n-1)} V^{(2n-2)}\cdots V^{(k+1)} \big)dU^{(k)}\Big]\Big).
\end{aligned}
\end{equation}
Compared with Eq.~\eqref{Eq:df_dcoh_def}, we find that
\begin{equation}
    \nabla_{U^{(k)}} \mathrm{coh}_{\cN,n} = \tr_{S_k} \Big[\big(V^{(2n-1)} V^{(2n-2)}\cdots V^{(k+1)} \big)^\dag\ketbra{G_{\psi}}{\phi_{k-1}}\Big].
\end{equation}
\end{proof}

The Riemannian gradient is then given as follows.

\begin{shaded}
\begin{corollary}\label{cor:local_U_Riemangrad}
For a quantum channel $\cN_{A\to B}$ and a unitary ensemble $\{U^{(k)}\}_{k=1}^{2n-1}$, the Riemannian gradient of the $n$-shot coherent information cost function at $\mathbf{U} = (U^{(1)}, U^{(2)}, \cdots, U^{(2n-1)})$ is given by
\begin{equation}\label{Eq:Rieman_grad_localU}
    \big[\grad \mathrm{coh}_{\cN,n}(\mathbf{U})\big]_k = \frac{1}{2} \tr_{S_c^{(k)}} \bigg[\Big({\overleftarrow{\prod}}_{j=k+1}^{2n-1} V^{(j)}\Big)^\dag\ketbra{G_{\psi}}{\phi_{k-1}} - \ketbra{\phi_{k}}{G_{\psi}}{\overleftarrow{\prod}}_{j=k}^{2n-1} V^{(j)}\bigg]
\end{equation}
for $k=1,2,\cdots, 2n-1$.
\end{corollary}
\end{shaded}

\begin{proof}
For the unitary manifold, according to Eq.~\eqref{Eq:st_proj}, the projection onto the tangent space is given by
\begin{equation}\label{Eq:uni_proj}
    \mathrm{Proj}_V(X) = \frac{1}{2} (X-VX^\dag V).
\end{equation}
Therefore, by Lemma~\ref{lem:grad_local_U} and Eq.~\eqref{Eq:Rieman_grad_prod}, we have that
\begin{equation*}
\begin{aligned}
    \big[\grad \mathrm{coh}_{\cN,n}(\mathbf{U})\big]_k &= \frac{1}{2} \bigg(\tr_{S_c^{(k)}} \bigg[\Big({\overleftarrow{\prod}}_{j=k+1}^{2n-1} V^{(j)}\Big)^\dag\ketbra{G_{\psi}}{\phi_{k-1}}\bigg]\\
    &\quad - \tr_{S_c^{(k)}} \bigg[U^{(k)}\ketbra{\phi_{k-1}}{G_{\psi}}\Big({\overleftarrow{\prod}}_{j=k+1}^{2n-1} V^{(j)}\Big)U^{(k)}\bigg]\bigg)\\
    &= \frac{1}{2} \tr_{S_c^{(k)}} \bigg[\Big({\overleftarrow{\prod}}_{j=k+1}^{2n-1} V^{(j)}\Big)^\dag\ketbra{G_{\psi}}{\phi_{k-1}} - \ketbra{\phi_{k}}{G_{\psi}}{\overleftarrow{\prod}}_{j=k}^{2n-1} V^{(j)}\bigg].
\end{aligned}
\end{equation*}
\end{proof}

Equipped with the Riemannian gradient, we have Algorithm~\ref{alg:RGD_Qlowerbound} to optimize a code state for a given channel $\cN$ and a positive integer $n$, thereby establishing a lower bound on the quantum capacity. 
\begin{remark}
Regarding Algorithm~\ref{alg:RGD_Dlowerbound} for computing lower bounds on the one-way distillable entanglement, we can also formulate the problem on a product of unitary manifolds by constructing a local unitary ansatz like Figure~\ref{fig:interleave_U}. It should be noted, however, that while our ansatz circumvents the curse of dimensionality in the \textit{parameter} space, the evaluation of the coherent information objective function remains a computational bottleneck. This challenge arises from the need to compute the logarithm of density matrices in a Hilbert space whose dimension grows exponentially with $n$.
\end{remark}

\begin{algorithm}[t]
\caption{RGD algorithm for lower bounds on the quantum capacity}\label{alg:RGD_Qlowerbound}
\begin{algorithmic}[1]
      \REQUIRE A quantum channel $\cN_{A\to B}$, $|R|,n\in\NN_+$; Initial guess $(U^{(1)}, U^{(2)}, \cdots, U^{(2n-1)})$. 
      \WHILE{the stopping criteria are not satisfied}
          \STATE Compute $\dot{\mathbf{U}}^{(t)} = -\grad \mathrm{coh}_{\cN,n}(\mathbf{U}^{(t)})$ by Eq.~\eqref{Eq:Rieman_grad_localU}.
          \STATE Compute a stepsize $s^{(t)}$.
          \STATE Update $\mathbf{U}^{(t+1)}= \mathrm{R}_{\mathbf{U}^{(t)}} \big(s^{(t)}\dot{\mathbf{U}}^{(t)}\big)$ by Eq.~\eqref{Eq:retr_qr}; $t= t+1$.
      \ENDWHILE
      \STATE Compute $q_{\mathrm{low}} = -\frac{1}{n}\mathrm{coh}_{\cN,n}(\mathbf{U}_n)$
      \ENSURE $\ket{\psi(\mathbf{U}_n)},~q_{\mathrm{low}}$.
\end{algorithmic}
\end{algorithm}

\subsection{Superadditivity in one-way entanglement distillation}\label{sec:superadd_state}

To demonstrate the efficacy of our method on computing lower bounds on the one-way distillable entanglement, we apply Algorithm~\ref{alg:RGD_Dlowerbound} to a class of noisy bipartite states. 

\paragraph{Generalized amplitude damping channel.}
Similar to Eq.~\eqref{Eq:noisyMES}, we consider a pair of Bell state $\ket{\Phi_+}=\frac{1}{\sqrt{2}}(\ket{00}+\ket{11})$ shared between Alice and Bob is affected by a local \textit{generalized amplitude damping channel} (GADC)~\cite{Khatri_2020} channel. As a generalization of the AD channel, it is defined by 
\begin{equation}\label{Eq:GADC}
    \cA_{\gamma,N} (\cdot) = \sum_{i=1}^4 A_i(\cdot)A_i^\dag,
\end{equation}
where the Kraus operators $\{A_i\}_{i=1}^4$ are given by
\begin{align}
    A_1 &= \sqrt{1-N} \big( \ketbra{0}{0} + \sqrt{1-\gamma} \ketbra{1}{1} \big)\\
    A_2 &= \sqrt{\gamma(1-N)} \ketbra{0}{1} \\
    A_3 &= \sqrt{N} \big( \sqrt{1-\gamma} \ketbra{0}{0} + \ketbra{1}{1} \big)\\
    A_4 &= \sqrt{\gamma N} \ketbra{1}{0},
\end{align}
with $\gamma,N\in[0,1]$. When $N=0$, it reduces to the AD channel. Thereby, we are interested in the Choi state of the GADC, i.e., $\rho_{AB'} = (\cI\ox\cA_{\gamma, N})(\Phi_+)$. For a fixed number of copies $n=2,4$, we compute lower bounds on $D_{\to}^{(1)}(\rho_{AB}^{\ox n})$ by optimizing a feasible one-way LOCC protocol using Algorithm~\ref{alg:RGD_Dlowerbound}, with a fixed dimension of the classical register to $|M|=2$. The optimization is terminated when the norm of the Riemannian gradient falls below $10^{-7}$. To mitigate convergence to local optima, each data point is the result of 200 independent optimization runs with random initial unitaries.

\begin{figure}[H]
\centering
\begin{subfigure}[t]{0.48\textwidth}
\centering
\begin{tikzpicture}
    \node[anchor=south west,inner sep=0] (image) at (0,0) {\includegraphics{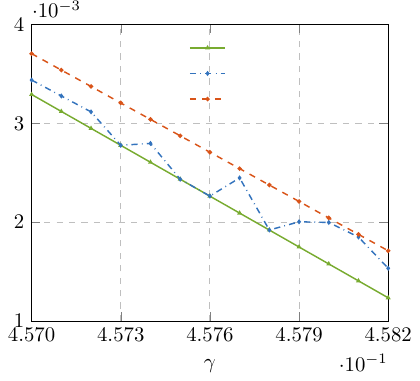}};
    \node[anchor=south west,inner sep=0] at (4,5.5){\footnotesize Hashing bound};
    \node[anchor=south west,inner sep=0] at (4,5.05){\footnotesize 2-copy average};
    \node[anchor=south west,inner sep=0] at (4,4.6){\footnotesize 4-copy average};
\end{tikzpicture}
\caption{$\cA_{\gamma,0.05}$}
\label{fig: GAD choi 0.05 EDnt lowerbound}
\end{subfigure}
\hfill
\begin{subfigure}[t]{0.48\textwidth}
\centering
\begin{tikzpicture}
    \node[anchor=south west,inner sep=0] (image) at (0,0) {\includegraphics{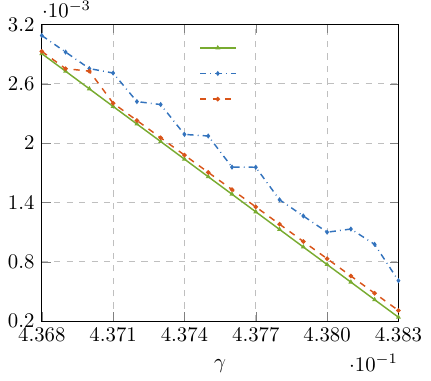}};
    \node[anchor=south west,inner sep=0] at (4.1,5.5){\footnotesize Hashing bound};
    \node[anchor=south west,inner sep=0] at ((4.1,5.05){\footnotesize 2-copy average};
    \node[anchor=south west,inner sep=0] at ((4.1,4.6){\footnotesize 4-copy average};
\end{tikzpicture}
\caption{$\cA_{\gamma,0.1}$}
\label{fig:GADchoi_DEbound0.1}
\end{subfigure}
\caption{Distillable entanglement lower bound on the Choi state of GADC by optimizing $\frac{1}{n} I(A'\rangle B^n M)_{\cT(\rho_{AB}^{\ox n})}$.}
\label{fig:GADchoi_DEbound}
\end{figure}
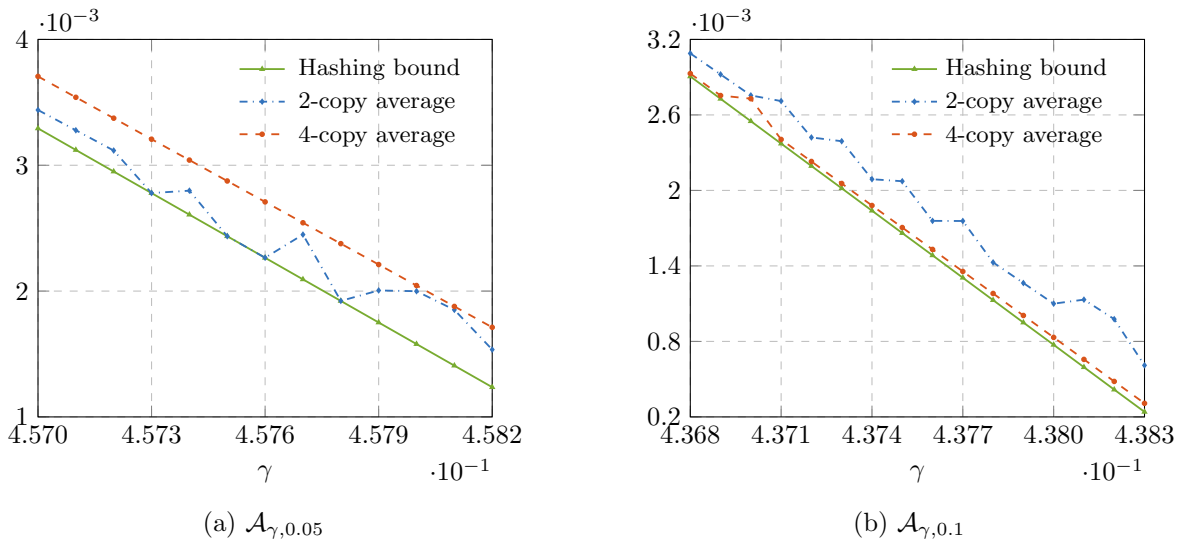

The results of our numerical experiments are presented in Figure~\ref{fig:GADchoi_DEbound}. The green solid line represents the coherent information as a lower bound. The blue and red dashed lines show our improved multi-copy bounds for $n=2, 4$, respectively, which are given by the quantity
\begin{equation}
\frac{1}{n} I(A'\rangle B^n M)_{\cT^*(\rho_{AB}^{\ox n})}.
\end{equation}
These bounds were calculated using the instrument $\cT^*$, which was optimized via Algorithm~\ref{alg:RGD_Dlowerbound} through the parameterization of the output unitary $U^*$. It demonstrates that for all tested noise parameters, the calculated lower bounds for multi-copy cases are strictly greater than the single-shot coherent information, which confirms that multi-copy protocols unlock additional distillable entanglement. The fluctuations observed in the dashed lines are attributed to the nature of the numerical optimization Algorithm~\ref{alg:RGD_Dlowerbound}. As the channel parameter is varied, the algorithm may settle into different local solutions, causing small variations in the resulting lower bound. Because we set $|M|=2$ for all experiments, and the issue of local minima, there may be cases where a 4-copy average rate is smaller than a 2-copy average rate in Figure~\ref{fig:GADchoi_DEbound0.1}. We can adjust the dimension of the classical system $M$ to further optimize the lower bounds. Overall, the results highlight the effectiveness of our Riemannian optimization framework in discovering the non-trivial quantum instruments required to reveal these superadditive effects, even with a small measurement dimension.

\subsection{Superadditivity in channel coherent information}\label{sec:superadd_channel}
To demonstrate the efficacy of our method on computing lower bounds on the quantum capacity, we apply Algorithm~\ref{alg:RGD_Qlowerbound} to different channels of interest.

\paragraph{Generalized amplitude damping channel.}
First, we consider the GADC as defined in Eq.~\eqref{Eq:GADC}. We compute lower bounds on the quantum capacity, $\frac{1}{n}Q^{(1)}(\cA_{\gamma,N}^{\ox n})$, for blocklength $n=3, 4, 5$ using our interleaved unitary ansatz with a reference system of dimension $|R|=2$, and choose the same parameters as that in~\cite[Appendix A.2]{Bausch_2020}. For each channel with fixed parameters, we execute Algorithm~\ref{alg:RGD_Qlowerbound} 200 times from random initial points to thoroughly explore the optimization landscape, selecting the highest coherent information rate achieved. The algorithm is terminated when the norm of the Riemannian gradient falls below $10^{-7}$.

The resulting rates for the optimized code states, $\frac{1}{n}Q^{(1)}(\psi(\mathbf{U}_n), \cA_{\gamma,N}^{\ox n})$, are then compared to the results in Ref.\cite{Bausch_2020}. Finding a rate that exceeds the single-shot coherent information demonstrates the superadditive nature of the channel, where a larger value signifies a more pronounced superadditive effect for that blocklength. In Table~\ref{tab:44035} to~\ref{tab:39169}, we present the results for the GADC with different parameters $(\gamma, N)$. We can see that our method achieves an improvement over the results in Ref.~\cite{Bausch_2020} even with a small, 2- or 3-dimensional auxiliary system $R$. This shows that our method is highly effective at discovering complex entangled code states, allowing us to reveal stronger superadditivity in the channel coherent information.

\paragraph{Dephrasure channel.}
The \textit{dephrasure channel} is defined by the composition of a dephasing channel and an erasure channel, i.e.,
\begin{equation}
    \cE\cZ_{p,q}(\cdot) \coloneqq (1-q)\big((1-p)(\cdot) + pZ(\cdot)Z\big) + q \tr(\cdot) \ketbra{e}{e},
\end{equation}
where $\ket{e}$ is an erasure flag orthogonal to the input space. In Table~\ref{tab: dephrasure 0.32 0.1} to~\ref{tab: dephrasure 0.08, 0.4}, we present our numerical results for the dephrasure channels with different parameters $(p,q)$ using Algorithm~\ref{alg:RGD_Qlowerbound}. We compare the coherent information yielded by our optimized code states with the ones given by the permutation-invariant code states in~\cite[Table 4]{bhalerao2025}. We observe that our code states can achieve higher rates with only 3 or 4 copy uses, compared with the 9-copy uses in Ref.~\cite{bhalerao2025}.

\paragraph{Damping-dephasing channel.}
Another channel of particular interest is the \textit{damping-dephasing} channel $\cZ\cA_{g,p}(\cdot)$, as introduced in Section~\ref{sec:upper_eg}, which has been shown to have a non-additive coherent information~\cite{Siddhu2024}. It has Kraus operators
\begin{align}
    O_1 &= \sqrt{1-p} \big( \ketbra{0}{0} + \sqrt{1-g} \ketbra{1}{1} \big)\\
    O_2 &= \sqrt{g} \ketbra{0}{1} \\
    O_3 &= \sqrt{p} \big( \ketbra{0}{0} -\sqrt{1-g} \ketbra{1}{1} \big),
\end{align}
where $p,g\in[0,1]$.
The optimized rates $\frac{1}{n}Q^{(1)}(\psi, \cZ\cA_{g,p}^{\ox n})$ for $n=2,3,4,5$ are shown in Table~\ref{tab:p016g020}, where they are compared with the results from~\cite[Table 3]{bhalerao2025}. For an auxiliary system of dimension $|R|=2$ or $|R|=3$, our optimized code states can provide higher rates, indicating a stronger superadditivity. It is worth noting that the primary focus of Ref.~\cite{bhalerao2025} is on improving capacity thresholds in the high-noise regime using permutation-invariant codes, an approach that is not necessarily optimized for maximizing rates at small blocklengths. Nevertheless, this comparison serves to highlight the superadditive gains that our method can unlock even with very few channel uses and small auxiliary dimensions.

\begin{table}[htbp]
\centering
\begin{tabular}{cccc}
& & \hspace{-10mm} $(\gamma, N) = (0.44035, 0.1)$& \\
\toprule
\multirow{2}{*}{$n$ copy} & \multicolumn{2}{c}{This work (cf.~Alg.~\ref{alg:RGD_Qlowerbound})} & Ref.~\cite{Bausch_2020} \\
\cmidrule(lr){2-3} \cmidrule(lr){4-4}
& $|R|=2$ & $|R|=3$ & $|R|=2^n$ \\
\midrule
3 & $1.7515 \cdot 10^{-3}$ & $1.7515 \cdot 10^{-3}$& $5.7598 \cdot 10^{-4}$ \\
4 & $2.0657 \cdot 10^{-3}$ & $2.0657 \cdot 10^{-3}$ & $1.2683 \cdot 10^{-3}$ \\
5 & $1.9124 \cdot 10^{-3}$ & $1.9116 \cdot 10^{-3}$  & $9.1537 \cdot 10^{-4}$ \\
\bottomrule
\end{tabular}
\caption{Lower bounds on the $n$-shot coherent information, $\frac{1}{n}Q^{(1)}(\psi, \cA_{\gamma,N}^{\ox n})$ with an optimized code state $\ket{\psi}$, for a GADC with $(\gamma, N) = (0.44035, 0.1)$. The increasing value as a function of $n$ demonstrates strong superadditivity of the coherent information.}
\label{tab:44035}
\end{table}

\begin{table}[htbp]
\centering
\begin{tabular}{cccc}
& & \hspace{-10mm} $(\gamma, N) = (0.41488, 0.2)$& \\
\toprule
\multirow{2}{*}{$n$ copy} & \multicolumn{2}{c}{This work (cf.~Alg.~\ref{alg:RGD_Qlowerbound})} & Ref.~\cite{Bausch_2020} \\

\cmidrule(lr){2-3} \cmidrule(lr){4-4}
& $|R|=2$ & $|R|=3$ & $|R|=2^n$ \\
\midrule
3 & $2.4565\cdot 10^{-3}$ & $2.4565\cdot 10^{-3}$ & $1.6923 \cdot 10^{-3}$ \\
4 & $2.4633 \cdot 10^{-3}$ & $2.4188 \cdot 10^{-3}$ & $1.4132 \cdot 10^{-3}$ \\
5 & $2.5366\cdot 10^{-3}$ & $2.5245\cdot 10^{-3}$ & $9.8025 \cdot 10^{-4}$ \\
\bottomrule
\end{tabular}
\caption{Lower bounds on the $n$-shot coherent information, $\frac{1}{n}Q^{(1)}(\psi, \cA_{\gamma,N}^{\ox n})$ with an optimized code state $\ket{\psi}$, for a GADC with $(\gamma, N) = (0.41488, 0.2)$.}
\label{tab:41488}
\end{table}

\begin{table}[htbp]
\centering
\begin{tabular}{cccc}
& & \hspace{-10mm} $(\gamma, N) = (0.40102, 0.3)$& \\
\toprule
\multirow{2}{*}{$n$ copy} & \multicolumn{2}{c}{This work (cf.~Alg.~\ref{alg:RGD_Qlowerbound})} & Ref.~\cite{Bausch_2020} \\

\cmidrule(lr){2-3} \cmidrule(lr){4-4}
& $|R|=2$ & $|R|=3$ & $|R|=2^n$ \\
\midrule
3 & $2.8213 \cdot 10^{-3}$ & $2.8213 \cdot 10^{-3}$ & $2.1889 \cdot 10^{-3}$ \\
4 & $2.7973\cdot 10^{-3}$ & $2.7921\cdot 10^{-3}$ & $7.3635 \cdot 10^{-4}$ \\
5 & $2.8460\cdot 10^{-3}$ & $2.7500\cdot 10^{-3}$ & - \\
\bottomrule
\end{tabular}
\caption{Lower bounds on the $n$-shot coherent information, $\frac{1}{n}Q^{(1)}(\psi, \cA_{\gamma,N}^{\ox n})$ with an optimized code state $\ket{\psi}$, for a GADC with $(\gamma, N) = (0.40102, 0.3)$.}
\label{tab:40102}
\end{table}

\begin{table}[htbp]
\centering
\begin{tabular}{cccc}
& & \hspace{-10mm} $(\gamma, N) = (0.39392, 0.4)$& \\
\toprule
\multirow{2}{*}{$n$ copy} & \multicolumn{2}{c}{This work (cf.~Alg.~\ref{alg:RGD_Qlowerbound})} & Ref.~\cite{Bausch_2020} \\

\cmidrule(lr){2-3} \cmidrule(lr){4-4}
& $|R|=2$ & $|R|=3$ & $|R|=2^n$ \\
\midrule
3 & $2.8517\cdot 10^{-3}$ & $2.8517\cdot 10^{-3}$ & $2.3456 \cdot 10^{-3}$ \\
4 & $2.8377\cdot 10^{-3}$ & $2.8340\cdot 10^{-3}$ & $1.7592 \cdot 10^{-3}$ \\
5 & $2.8458\cdot 10^{-3}$ & $2.7590\cdot 10^{-3}$ & - \\
\bottomrule
\end{tabular}
\caption{Lower bounds on the $n$-shot coherent information, $\frac{1}{n}Q^{(1)}(\psi, \cA_{\gamma,N}^{\ox n})$ with an optimized code state $\ket{\psi}$, for a GADC with $(\gamma, N) = (0.39392, 0.4)$.}
\label{tab:39392}
\end{table}

\begin{table}[htbp]
\centering
\begin{tabular}{cccc}
& & \hspace{-10mm} $(\gamma, N) = (0.39169, 0.5)$& \\
\toprule
\multirow{2}{*}{$n$ copy} & \multicolumn{2}{c}{This work (cf.~Alg.~\ref{alg:RGD_Qlowerbound})} & Ref.~\cite{Bausch_2020} \\

\cmidrule(lr){2-3} \cmidrule(lr){4-4}
& $|R|=2$ & $|R|=3$ & $|R|=2^n$ \\
\midrule
3 & $2.7901\cdot 10^{-3}$ & $2.7901\cdot 10^{-3}$ & $2.3948 \cdot 10^{-3}$ \\
4 & $2.7491\cdot 10^{-3}$ & $2.7365\cdot 10^{-3}$ & $1.7913 \cdot 10^{-3}$ \\
5 & $2.7222\cdot 10^{-3}$ &$2.1595\cdot 10^{-3}$  & $1.3393 \cdot 10^{-3}$ \\
\bottomrule
\end{tabular}
\caption{Lower bounds on the $n$-shot coherent information, $\frac{1}{n}Q^{(1)}(\psi, \cA_{\gamma,N}^{\ox n})$ with an optimized code state $\ket{\psi}$, for a GADC with $(\gamma, N) = (0.39169, 0.5)$.}
\label{tab:39169}
\end{table}


\begin{table}[htbp]
\centering
\begin{tabular}{cccc}
& & \hspace{-10mm} $(p, q) = (0.32, 0.1)$& \\
\toprule
\multirow{2}{*}{$n$ copy} & \multicolumn{2}{c}{This work (cf.~Alg.~\ref{alg:RGD_Qlowerbound})} & Ref.~\cite{bhalerao2025} \\
\cmidrule(lr){2-3} \cmidrule(lr){4-4}
& $|R|=2$ & $|R|=3$ & $|R|=2$ \\
\midrule
3 & $7.4634 \cdot 10^{-5}$ & $1.1178 \cdot 10^{-4}$& - \\
4 & $8.6365 \cdot 10^{-5}$ & $7.9767 \cdot 10^{-5}$ & - \\
9& - & - &$5.2223 \cdot 10^{-5}$\\
\bottomrule
\end{tabular}
\caption{Lower bounds on the $n$-shot coherent information, $\frac{1}{n}Q^{(1)}(\psi, \cE\cZ_{p,q}^{\ox n})$ with an optimized code state $\ket{\psi}$, for a dephrasure channel with $(p, q) = (0.32, 0.1)$. The increasing value as a function of $n$ demonstrates strong superadditivity of the coherent information.}
\label{tab: dephrasure 0.32 0.1}
\end{table}

\begin{table}[htbp]
\centering
\begin{tabular}{cccc}
& & \hspace{-10mm} $(p, q) = (0.24, 0.2)$& \\
\toprule
\multirow{2}{*}{$n$ copy} & \multicolumn{2}{c}{This work (cf.~Alg.~\ref{alg:RGD_Qlowerbound})} & Ref.~\cite{bhalerao2025} \\
\cmidrule(lr){2-3} \cmidrule(lr){4-4}
& $|R|=2$ & $|R|=3$ & $|R|=2$ \\
\midrule
3 & $6.1854 \cdot 10^{-6}$ & $4.5631 \cdot 10^{-6}$& -\\
4 & $1.2144 \cdot 10^{-5}$ & $3.4223 \cdot 10^{-6}$ & - \\
9 & - & - & $1.3181 \cdot 10^{-6}$\\
\bottomrule
\end{tabular}
\caption{Lower bounds on the $n$-shot coherent information, $\frac{1}{n}Q^{(1)}(\psi, \cE\cZ_{p,q}^{\ox n})$ with an optimized code state $\ket{\psi}$, for a dephrasure channel with $(p, q) = (0.24, 0.2)$.}
\label{tab: dephrasure 0.24 0.2}
\end{table}

\begin{table}[htbp]
\centering
\begin{tabular}{cccc}
& & \hspace{-10mm} $(p, q) = (0.16, 0.3)$& \\
\toprule
\multirow{2}{*}{$n$ copy} & \multicolumn{2}{c}{This work (cf.~Alg.~\ref{alg:RGD_Qlowerbound})} & Ref.~\cite{bhalerao2025} \\
\cmidrule(lr){2-3} \cmidrule(lr){4-4}
& $|R|=2$ & $|R|=3$ & $|R|=2$ \\
\midrule
3 & $3.9674 \cdot 10^{-5}$ & $3.3592 \cdot 10^{-5}$& - \\
4 & $1.5026 \cdot 10^{-5}$ & $2.5206 \cdot 10^{-5}$ & - \\
9 & - & - & $2.3103 \cdot 10^{-5}$\\
\bottomrule
\end{tabular}
\caption{Lower bounds on the $n$-shot coherent information, $\frac{1}{n}Q^{(1)}(\psi, \cE\cZ_{p,q}^{\ox n})$ with an optimized code state $\ket{\psi}$, for a dephrasure channel with $(p, q) = (0.16, 0.3)$.}
\label{tab: dephrasure 0.16, 0.3}
\end{table}

\begin{table}[htbp]
\centering
\begin{tabular}{cccc}
& & \hspace{-10mm} $(p, q) = (0.08, 0.4)$& \\
\toprule
\multirow{2}{*}{$n$ copy} & \multicolumn{2}{c}{This work (cf.~Alg.~\ref{alg:RGD_Qlowerbound})} & Ref.~\cite{bhalerao2025} \\
\cmidrule(lr){2-3} \cmidrule(lr){4-4}
& $|R|=2$ & $|R|=3$ & $|R|=2$ \\
\midrule
3 & $4.7900 \cdot 10^{-5}$ & $1.3663 \cdot 10^{-6}$& - \\
4 & $8.1678 \cdot 10^{-7}$ & $7.5609 \cdot 10^{-5}$ & - \\
9 & - & - & $5.4524 \cdot 10^{-5}$\\
\bottomrule
\end{tabular}
\caption{Lower bounds on the $n$-shot coherent information, $\frac{1}{n}Q^{(1)}(\psi, \cE\cZ_{p,q}^{\ox n})$ with an optimized code state $\ket{\psi}$, for a dephrasure channel with $(p, q) = (0.08, 0.4)$.}
\label{tab: dephrasure 0.08, 0.4}
\end{table}


\begin{table}[htbp]
\centering
\begin{tabular}{ccccc}
& & \hspace{-10mm} $(p, g) = (0.16, 0.20)$& \\
\toprule
\multirow{2}{*}{$n$ copy} & \multicolumn{2}{c}{This work (cf.~Alg.~\ref{alg:RGD_Qlowerbound})} & \multicolumn{2}{c}{Ref.~\cite{bhalerao2025}} \\

\cmidrule(lr){2-3} \cmidrule(lr){4-5}
& $|R|=2$ & $|R|=3$ & $|R|=2$ & $|R|=3$ \\
\midrule
2 & $1.9987\cdot 10^{-2}$ & $2.0105\cdot 10^{-2}$ & - & - \\
3 & $2.0978\cdot 10^{-2}$ & $2.1637\cdot 10^{-2}$ & - & - \\
4 & $2.0083\cdot 10^{-2}$ & $2.2073\cdot 10^{-2}$ & - & - \\
5 & $1.8714\cdot 10^{-2}$ & $2.1816\cdot 10^{-2}$ & $1.4707\cdot 10^{-2}$ & $1.9899 \cdot 10^{-2}$ \\
\bottomrule
\end{tabular}
\caption{Lower bounds on the $n$-shot coherent information, $\frac{1}{n}Q^{(1)}(\psi, \cZ\cA_{g,p}^{\ox n})$ with an optimized code state $\ket{\psi}$, for a damping-dephasing channel with $(p, g) = (0.16, 0.20)$.}
\label{tab:p016g020}
\end{table}

\newpage
\subsection{Amortization does not enhance channel coherent information}\label{sec:amort}

It is known that the channel coherent information, and thus the quantum channel capacity, can be expressed in terms of a quantum channel divergence and its regularization~\cite[Section 5]{fang_towards_2025}. One natural idea for obtaining potentially tighter lower bounds is to consider the amortized channel divergence~\cite{Wilde_2020}, where different input states are allowed for the two channels being distinguished. We introduce the notion of \textit{amortized channel coherent information} and show that it is sandwiched between the channel coherent information and the regularized channel coherent information, suggesting a possible improvement over the single-letter lower bound by the channel coherent information. However, we subsequently prove that the amortized channel coherent information always coincides with the channel coherent information, thereby ruling out the potential enhancement of the quantum capacity lower bound via amortization.

\begin{definition}\label{def:amor_unstab_c_diver}
Let $\mathbf{D}$ be a general quantum divergence. For any $\cN \in \CPTP(A:B)$ and $\cM \in \CP(A:B)$, the unstabilized channel divergence is defined by
\begin{align}
\underline{\mathbf{D}}(\cN \| \cM) \coloneqq \sup _{\rho\in \density(A)} \mathbf{D}\big(\cN_{A\to B}(\rho) \| \cM_{A\to B}(\rho)\big),
\end{align}
where the supremum is taken over all quantum states $\rho$ on system $A$. The regularized and unstabilized channel divergence is defined by
\begin{align}
\underline{\mathbf{D}}^{\mathrm{reg}}(\cN \| \cM):=\lim _{n \to \infty} \frac{1}{n} \underline{\mathbf{D}}\left(\cN^{\ox n} \| \cM^{\ox n}\right)=\sup _{n \in \mathbb{N}} \frac{1}{n} \underline{\mathbf{D}}\left(\cN^{\ox n} \| \cM^{\ox n}\right),
\end{align}
where the second equality is followed by the super-additivity of unstabilized channel divergence $\underline{\mathbf{D}}$. The amortized and unstabilized channel divergence is defined by
\begin{align}\label{Eq:_D^A}
\underline{\mathbf{D}}^A(\cN \| \cM) \coloneqq \sup _{\rho, \sigma \in \density(A)} \big[\mathbf{D}\left(\cN_{A\to B}(\rho) \| \cM_{A\to B}(\sigma)\right)-\mathbf{D}(\rho \| \sigma)\big],
\end{align}
where the supremum is taken over all quantum states $\rho$ and $\sigma$ on system $A$.
\end{definition}

See the definition of the amortized channel divergence in~\cite[Definition 3]{Wilde_2020}. A common choice of $\mathbf{D}(\cdot\|\cdot)$ is the Umegaki relative entropy (also called quantum relative entropy)~\cite{Umegaki1962}, defined by
\begin{align}
    D(\rho\|\sigma) \coloneqq 
    \begin{cases}
        \tr [\rho(\log \rho - \log \sigma)]~~&\text{if}~ \supp(\rho)\subseteq \supp(\sigma),\\
        +\infty~~&\text{otherwise}.
    \end{cases}
\end{align}
for any quantum states $\rho$ and semidefinite operator $\sigma$. It was shown in~\cite[Theorem 34]{fang_towards_2025} that the coherent information of a quantum channel can be expressed as the unstabilized channel relative entropy as follows.

\begin{lemma}[\!\!\cite{fang_towards_2025}]\label{lem:I_cN}
For any $\cN \in \CPTP(A : B)$, it holds that
\begin{align}
I_c(\cN) & =\underline{D}\left(\cI_E \ox \cN_{A\to B} \| \cR_E^{\1} \ox \cN_{A\to B}\right),  \\
Q(\cN)   & =\underline{D}^{\mathrm{reg}}\left(\cI_E \ox \cN_{A\to B} \| \cR_E^{\1} \ox \cN_{A\to B}\right),
\end{align}
where $E$ is isomorphic to $A$ and $\cR_E^{\1}(\cdot)=\tr_E(\cdot) \1_E$ is a CP map that always outputs the identity operator $\1_E$.
\end{lemma}

Then, we introduce the amortized channel coherent information as follows.

\begin{definition}[Amortized channel coherent information]
For any $\cN \in \CPTP(A:B)$, the amortized channel coherent information $I_c^A(\cN)$ is defined by
\begin{align}\label{Eq:I_c^A}
    I_c^A(\cN)\coloneqq \underline{D}^A\left(\cI_E \ox \cN_{A\to B} \| \cR_E^{\1} \ox \cN_{A\to B}\right),
\end{align}
where $\underline{D}^A(\cdot)$ is the amortized unstabilized channel relative entropy.
\end{definition}

\begin{remark}
Following the convention of defining an amortized entanglement measure for quantum channels, e.g., the amortized entanglement~\cite[Section 2]{Kaur_2017}, one could also define another variant of the amortized channel coherent information as
\begin{equation}
    I_c^A(\cN) = \max_{\rho_{A'AB'}} I(A'\rangle BB')_{\cN_{A\to B}(\rho_{A'AB'})} - I(A'A\rangle B)_{\rho},
\end{equation}
which is different from our consideration here.
\end{remark}

Followed by Lemma~\ref{lem:I_cN}, we have the following chain inequalities for any quantum channel $\cN$.

\begin{shaded}
\begin{lemma}\label{lem:Ic_IcA_Q}
For any $\cN \in \CPTP(A:B)$ it holds that $I_c(\cN) \leq I_c^A(\cN) \leq Q(\cN)$.
\end{lemma}
\end{shaded}
\begin{proof}
Since $\underline{D}(\cdot\|\cdot) \leq \underline{D}^A(\cdot\|\cdot)$, we have $I_c(\cN) \leq I_c^A(\cN)$. On the other hand, by the chain rule of quantum relative entropy~\cite[Theorem 2]{Fang2020chainrule}, we have
\begin{align}
I_c^A(\cN) & = \underline{D}^A\left(\cI_E\ox \cN_{A\to B}\big\|\cR_E^\1\ox \cN_{A\to B}\right)\\
& \leq \underline{D}^{\mathrm{reg}}\left(\cI_E\ox \cN_{A\to B}\big\|\cR_E^\1\ox \cN_{A\to B}\right)\\
& = Q(\cN),
\end{align}
where we used Lemma~\ref{lem:I_cN} for the second equality.
\end{proof}

Lemma~\ref{lem:Ic_IcA_Q} indicates that the amortized channel coherent information could potentially provide a tighter lower bound on quantum capacity. However, we show that the amortized channel coherent information in fact equals the channel coherent information for any channel $\cN$.

\begin{shaded}
\begin{theorem}\label{thm:Ic_amort}
For any $\cN \in \CPTP(A : B)$ it holds that $I_c^A(\cN) = I_c(\cN)$.
\end{theorem}
\end{shaded}

\begin{proof}
Combining (\ref{Eq:I_c^A}) and (\ref{Eq:_D^A}), we have
\begin{align}
I_c^A(\cN) = \sup _{\rho, \sigma \in \density(EA)}\Big[ D\Big(\cI_E \ox \cN_{A\to B}(\rho) \| \cR_E^{\1} \ox \cN_{A\to B}(\sigma)\Big)-D(\rho \| \sigma) \Big].
\end{align}
By denoting $\rho_{EB} = \cI_E \ox \cN_{A\to B}(\rho)$ and $\rho_B = \tr_{E}\rho_{EB}$, we can consider
\begin{equation}
\begin{aligned}
I_c^A(\cN) = & \sup _{\rho, \sigma \in \density(EA)}\Big[ D\Big(\cI_E \ox \cN_{A\to B}(\rho) \| \cR_E^{\1} \ox \cN_{A\to B}(\sigma)\Big)- D(\rho \| \sigma) \Big]\\
=& \sup _{\rho, \sigma \in \density(EA)}\Big[ \tr(\rho_{EB}\log\rho_{EB}) - \tr[\rho_{EB}\log(\1_E\ox \tr_{E}\sigma_{EB})] - D(\rho \| \sigma)
\Big] \\
= & \sup_{\rho, \sigma \in \density(EA)} \Big[ \tr(\rho_{EB}\log\rho_{EB}) - \tr[\rho_{EB}\log(\1_E\ox \tr_{E}\rho_{EB})] \\
& \qquad\qquad + \tr[\rho_{EB}\log(\1_E\ox \tr_{E}\rho_{EB})] - \tr[\rho_{EB}\log(\1_E\ox \tr_{E}\sigma_{EB})] - D(\rho \| \sigma)\Big]    \\
=& \sup _{\rho, \sigma \in \density(EA)}\Big[ D\Big(\cI_E \ox \cN_{A\to B}\left(\rho\right) \| \cR_E^{\1} \ox \cN_{A\to B}(\rho)\Big)             \\
& \qquad\qquad + \tr[\rho_B \log\rho_B] - \tr[\rho_B \log\sigma_B] - D(\rho \| \sigma) \Big] \\
=  & \sup _{\rho, \sigma \in \density(EA)}\Big[ D\Big(\cI_E \ox \cN_{A\to B}\left(\rho\right) \| \cR_E^{\1} \ox \cN_{A\to B}(\rho)\Big) + D\left(\rho_B \| \sigma_B\right) - D(\rho \| \sigma) \Big], \label{Eq:rela_entr_B-rela_entr}
\end{aligned}
\end{equation}
where the second equality follows from the definition of quantum relative entropy, in the third equality we add and subtract the term $\tr[\rho_{EB}\log(\1_E\ox \tr_{E}\rho_{EB})]$, and the fourth equality follows from the fact that $\log(A\ox B) =\log A\ox \1 + \1\ox \log B$. Notice that
\begin{align}
D\left(\rho_B \| \sigma_B\right) =  D\left(\tr_E\ox \cN_{A\to B}(\rho) \|\tr_E\ox \cN_{A\to B}(\sigma)\right) \leq D\left(\rho \| \sigma\right),
\end{align}
where $\tr_E(\cdot)$ is the partial trace over the system $E$, and the inequality follows from the monotonicity of the quantum relative entropy under CPTP maps. Then we have that
\begin{align}
(\ref{Eq:rela_entr_B-rela_entr})\leq \sup _{\rho \in \density(EA)}\Big[ D\Big(\cI_E \ox \cN_{A\to B}\left(\rho\right) \| \cR_E^{\1} \ox \cN_{A\to B}(\rho)\Big)\Big] = I_c(\cN),
\end{align}
where the equality follows from Lemma~\ref{lem:I_cN}.
\end{proof}

\section{Concluding remarks}\label{sec:conclu}

In this work, we have developed a comprehensive framework based on Riemannian optimization to derive improved, computable two-sided bounds on the one-way distillable entanglement and the quantum capacity. 

Our main technical contributions are twofold. First, for upper bounds, we transformed the search for state and channel extensions that can minimize information-theoretic bounds into a tractable optimization over the Stiefel manifold. This method, when combined with a refined continuity bound for conditional entropy, delivers strictly tighter upper bounds than previously known records for large values of the depolarizing parameter~\cite{kianvashBoundingQuantumCapacity2022}. We can also obtain improved upper bounds for different states and channels of interest. Second, for lower bounds, we transform the problem to optimizations over unitary manifolds. Our algorithm for states and channels can be flexibly used to explore the superadditivity in the one-way entanglement distillation rate and the channel coherent information. We demonstrate improved lower bounds on both quantities for various states and channels of interest, compared to previous results. Our methods may also serve as a guide for the design of entanglement distillation protocols and quantum error correction codes. Furthermore, we proved that amortization of the underlying channel divergence does not enhance the channel coherent information. This result excludes the amortization as a pathway to obtain a tighter lower bound on the quantum capacity, and can be of independent interest.

The methods presented here open several avenues for future research. For the upper-bound framework, it would be of great interest to see if the structure of the optimal isometries found via optimization can inspire new analytical insights or closed-form bounds for specific channel families. The computational cost of the finite-difference gradient approximation could be reduced by exploring more advanced non-smooth or bilevel optimization techniques tailored to the Stiefel manifold~\cite{Sow2022,dutta2024riemannianbileveloptimization,Andi2025}. For the lower-bound approach, the key bottleneck of computing the entropy function remains unsolved in our methods, which limits the scalability of our algorithms. It would be interesting to explore new manifolds with a symmetry structure that can aid in the calculation of the entropy function, combined with the idea of permutation-invariant codes~\cite{bhalerao2025}.  It would also be interesting and potentially impactful to explore geometric optimization for other quantum resource theories~\cite{Chitambar2018,Lami2023,Kaur2018,Gallego2015,WWS19,Wang2018e,Lami2018c,Regula2017c,Wang2019,Takagi2019,Wang2023b} and quantum information–processing tasks~\cite{Datta2016,Datta2013b,Zhu2024,Campbell2017b,Chitambar2014,Zhao2021}, both to establish fundamental limits and to discover better protocols or lower bounds of achievable rates.

\paragraph{Acknowledgments.} 
C. Z. thanks Bin Gao, Renfeng Peng, and Yan Yang for insightful discussion and consultation on Riemannian optimization. K. F. thanks Haonan Zhang for bringing his attention to the improved continuity bound during his visit at CUHK-Shenzhen.  X. W. and C. Z. acknowledge the support from the National Key R\&D Program of China (Grant No.~2024YFB4504004), the National Natural Science Foundation of China (Grant. No.~12447107), the Guangdong Provincial Quantum Science Strategic Initiative (Grant No.~GDZX2403008, GDZX2403001), the Guangdong Provincial Key Lab of Integrated Communication, Sensing and Computation for Ubiquitous Internet of Things (Grant No.~2023B1212010007), the Quantum Science Center of Guangdong-Hong Kong-Macao Greater Bay Area, and the Education Bureau of Guangzhou Municipality. K. F. is supported by the National Natural Science Foundation of China (grant No. 92470113 and 12404569), the Shenzhen Science and Technology Program (grant No. JCYJ20240813113519025), the Shenzhen Fundamental Research Program (grant No. JCYJ20241202124023031), the 1+1+1 CUHK-CUHK(SZ)-GDST Joint Collaboration Fund (grant No. GRD\ P2025-022), and the University Development Fund (grant No. UDF01003565).  

\bibliographystyle{alpha}
\bibliography{main}

\newcommand{\etalchar}[1]{$^{#1}$}
\begin{thebibliography}{KDWW19}

\bibitem[ABD{\etalchar{+}}24]{Koenraad2024}
Koenraad Audenaert, Bjarne Bergh, Nilanjana Datta, Michael~G. Jabbour, {\'A}ngela Capel, and Paul Gondolf.
\newblock Continuity bounds for quantum entropies arising from a fundamental entropic inequality, December 2024.

\bibitem[AF04]{Alicki_2004}
R~Alicki and M~Fannes.
\newblock Continuity of quantum conditional information.
\newblock {\em Journal of Physics A: Mathematical and General}, 37(5):L55–L57, January 2004.

\bibitem[AMS08]{absil2008optimization}
P-A Absil, Robert Mahony, and Rodolphe Sepulchre.
\newblock {\em Optimization algorithms on matrix manifolds}.
\newblock Princeton University Press, 2008.

\bibitem[AT60]{atiyah1960complex}
MF~Atiyah and JA~Todd.
\newblock On complex stiefel manifolds.
\newblock In {\em Mathematical Proceedings of the Cambridge Philosophical Society}, volume~56, pages 342--353. Cambridge University Press, 1960.

\bibitem[BBC{\etalchar{+}}93]{Bennett1993a}
Charles~H. Bennett, Gilles Brassard, Claude Cr\'epeau, Richard Jozsa, Asher Peres, and William~K. Wootters.
\newblock Teleporting an unknown quantum state via dual classical and einstein-podolsky-rosen channels.
\newblock {\em Physical Review Letters}, 70:1895--1899, Mar 1993.

\bibitem[BBP{\etalchar{+}}96]{Bennett1996}
Charles~H Bennett, Gilles Brassard, Sandu Popescu, Benjamin Schumacher, John~A Smolin, and William~K Wootters.
\newblock {Purification of Noisy Entanglement and Faithful Teleportation via Noisy Channels}.
\newblock {\em Physical Review Letters}, 76(5):722--725, jan 1996.

\bibitem[BBPS96]{Bennett1996b}
Charles~H Bennett, Herbert~J Bernstein, Sandu Popescu, and Benjamin Schumacher.
\newblock {Concentrating partial entanglement by local operations}.
\newblock {\em Physical Review A}, 53(4):2046--2052, apr 1996.

\bibitem[BDSW96]{bennettMixedstateEntanglementQuantum1996}
Charles~H. Bennett, David~P. DiVincenzo, John~A. Smolin, and William~K. Wootters.
\newblock Mixed-state entanglement and quantum error correction.
\newblock {\em Physical Review A}, 54(5):3824--3851, November 1996.

\bibitem[BK98]{Braunstein1998}
Samuel~L. Braunstein and H.~J. Kimble.
\newblock Teleportation of continuous quantum variables.
\newblock {\em Physical Review Letters}, 80:869--872, Jan 1998.

\bibitem[BKM14]{bagirov2014introduction}
Adil Bagirov, Napsu Karmitsa, and Marko~M M{\"a}kel{\"a}.
\newblock {\em Introduction to Nonsmooth Optimization: theory, practice and software}, volume~12.
\newblock Springer, 2014.

\bibitem[BKN00]{Barnum2000}
Howard Barnum, Emanuel Knill, and M.A. Nielsen.
\newblock {On quantum fidelities and channel capacities}.
\newblock {\em IEEE Transactions on Information Theory}, 46(4):1317--1329, jul 2000.

\bibitem[BL20]{Bausch_2020}
Johannes Bausch and Felix Leditzky.
\newblock Quantum codes from neural networks.
\newblock {\em New Journal of Physics}, 22(2):023005, feb 2020.

\bibitem[BL21]{Johannes2021}
Johannes Bausch and Felix Leditzky.
\newblock Error thresholds for arbitrary pauli noise.
\newblock {\em SIAM Journal on Computing}, 50(4):1410--1460, 2021.

\bibitem[BL25]{bhalerao2025}
Sujeet Bhalerao and Felix Leditzky.
\newblock Improving quantum communication rates with permutation-invariant codes, 2025.

\bibitem[BLO05]{burke2005robust}
James~V Burke, Adrian~S Lewis, and Michael~L Overton.
\newblock A robust gradient sampling algorithm for nonsmooth, nonconvex optimization.
\newblock {\em SIAM Journal on Optimization}, 15(3):751--779, 2005.

\bibitem[BLT25]{Mario2025}
Mario Berta, Ludovico Lami, and Marco Tomamichel.
\newblock Continuity of entropies via integral representations.
\newblock {\em IEEE Transactions on Information Theory}, 71(3):1896--1908, March 2025.

\bibitem[BMAS14]{boumal2014manopt}
Nicolas Boumal, Bamdev Mishra, P-A Absil, and Rodolphe Sepulchre.
\newblock Manopt, a {Matlab} toolbox for optimization on manifolds.
\newblock {\em The Journal of Machine Learning Research}, 15(1):1455--1459, 2014.

\bibitem[Bon13]{Bonnabel_2013}
Silvere Bonnabel.
\newblock Stochastic gradient descent on riemannian manifolds.
\newblock {\em IEEE Transactions on Automatic Control}, 58(9):2217–2229, September 2013.

\bibitem[Bou23]{boumal2023intromanifolds}
Nicolas Boumal.
\newblock {\em An introduction to optimization on smooth manifolds}.
\newblock Cambridge University Press, 2023.

\bibitem[BSST02]{Bennett2002}
C.H. Bennett, P.W. Shor, J.A. Smolin, and A.V. Thapliyal.
\newblock {Entanglement-assisted capacity of a quantum channel and the reverse Shannon theorem}.
\newblock {\em IEEE Transactions on Information Theory}, 48(10):2637--2655, oct 2002.

\bibitem[BW92]{Bennett1992}
Charles~H Bennett and Stephen~J Wiesner.
\newblock {Communication via one- and two-particle operators on Einstein-Podolsky-Rosen states}.
\newblock {\em Physical Review Letters}, 69(20):2881--2884, nov 1992.

\bibitem[CEM{\etalchar{+}}15]{Cubitt2015}
Toby Cubitt, David Elkouss, William Matthews, Maris Ozols, David P{{e}}rez-Garc{{i}}a, and Sergii Strelchuk.
\newblock {Unbounded number of channel uses may be required to detect quantum capacity}.
\newblock {\em Nature Communications}, 6(1):6739, dec 2015.

\bibitem[CG19]{Chitambar2018}
Eric Chitambar and Gilad Gour.
\newblock {Quantum resource theories}.
\newblock {\em Reviews of Modern Physics}, 91(2):025001, apr 2019.

\bibitem[CH17]{Campbell2017b}
Earl~T. Campbell and Mark Howard.
\newblock {Unified framework for magic state distillation and multiqubit gate synthesis with reduced resource cost}.
\newblock {\em Physical Review A}, 95(2):022316, feb 2017.

\bibitem[CK00]{Choi2000}
T.~D. Choi and C.~T. Kelley.
\newblock Superlinear convergence and implicit filtering.
\newblock {\em SIAM Journal on Optimization}, 10(4):1149--1162, 2000.

\bibitem[CLM{\etalchar{+}}14]{Chitambar2014}
Eric Chitambar, Debbie Leung, Laura Man{\v{c}}inska, Maris Ozols, and Andreas Winter.
\newblock {Everything You Always Wanted to Know About LOCC (But Were Afraid to Ask)}.
\newblock {\em Communications in Mathematical Physics}, 328(1):303--326, may 2014.

\bibitem[CMS07]{colson2007overview}
Beno{\^\i}t Colson, Patrice Marcotte, and Gilles Savard.
\newblock An overview of bilevel optimization.
\newblock {\em Annals of operations research}, 153(1):235--256, 2007.

\bibitem[COT24]{Casanova2024}
Miguel Casanova, Kentaro Ohki, and Francesco Ticozzi.
\newblock Finding quantum codes via riemannian optimization, 2024.

\bibitem[DCS24]{dutta2024riemannianbileveloptimization}
Sanchayan Dutta, Xiang Cheng, and Suvrit Sra.
\newblock Riemannian bilevel optimization, 2024.

\bibitem[Dev05]{Devetak2005a}
Igor Devetak.
\newblock {The Private Classical Capacity and Quantum Capacity of a Quantum Channel}.
\newblock {\em IEEE Transactions on Information Theory}, 51(1):44--55, jan 2005.

\bibitem[DMHB13]{Datta2013b}
Nilanjana Datta, Milan Mosonyi, Minhsiu Min-Hsiu Hsieh, and Fernando G S~L Brandao.
\newblock {A smooth entropy approach to quantum hypothesis testing and the classical capacity of quantum channels}.
\newblock {\em IEEE Transactions on Information Theory}, 59(12):8014--8026, dec 2013.

\bibitem[DS05]{Devetak2005}
Igor Devetak and Peter~W Shor.
\newblock {The Capacity of a Quantum Channel for Simultaneous Transmission of Classical and Quantum Information}.
\newblock {\em Communications in Mathematical Physics}, 256(2):287--303, jun 2005.

\bibitem[DSS98]{DiVincenzo1998a}
David~P. DiVincenzo, Peter~W. Shor, and John~A. Smolin.
\newblock {Quantum-channel capacity of very noisy channels}.
\newblock {\em Physical Review A}, 57(2):830--839, feb 1998.

\bibitem[DTW16]{Datta2016}
Nilanjana Datta, Marco Tomamichel, and Mark~M Wilde.
\newblock {On the second-order asymptotics for entanglement-assisted communication}.
\newblock {\em Quantum Information Processing}, 15(6):2569--2591, jun 2016.

\bibitem[DW05]{devetakDistillationSecretKey2005}
Igor Devetak and Andreas Winter.
\newblock Distillation of secret key and entanglement from quantum states.
\newblock {\em Proceedings of the Royal Society A: Mathematical, Physical and Engineering Sciences}, 461(2053):207--235, January 2005.

\bibitem[Eke91]{Ekert1991}
Artur~K. Ekert.
\newblock Quantum cryptography based on bell's theorem.
\newblock {\em Physical Review Letters}, 67:661--663, Aug 1991.

\bibitem[FF18]{fawzi2018efficient}
Hamza Fawzi and Omar Fawzi.
\newblock Efficient optimization of the quantum relative entropy.
\newblock {\em Journal of Physics A: Mathematical and Theoretical}, 51(15):154003, 2018.

\bibitem[FF21]{fangGeometricRenyiDivergence2021}
Kun Fang and Hamza Fawzi.
\newblock Geometric {{R{\'e}nyi Divergence}} and its {{Applications}} in {{Quantum Channel Capacities}}.
\newblock {\em Communications in Mathematical Physics}, 384(3):1615--1677, June 2021.

\bibitem[FFRS20]{Fang2020chainrule}
Kun Fang, Omar Fawzi, Renato Renner, and David Sutter.
\newblock Chain rule for the quantum relative entropy.
\newblock {\em Physical Review Letters}, 124:100501, Mar 2020.

\bibitem[FGW25]{fang_towards_2025}
Kun Fang, Gilad Gour, and Xin Wang.
\newblock Towards the ultimate limits of quantum channel discrimination and quantum communication.
\newblock {\em SCIENCE CHINA Information Sciences}, 68(8), July 2025.

\bibitem[FKG20]{fanizzaQuantumFlagsNew2020}
Marco Fanizza, Farzad Kianvash, and Vittorio Giovannetti.
\newblock Quantum flags, and new bounds on the quantum capacity of the depolarizing channel.
\newblock {\em Physical Review Letters}, 125(2):020503, July 2020.

\bibitem[FW08]{Fern2008}
Jesse Fern and K.~Birgitta Whaley.
\newblock {Lower bounds on the nonzero capacity of Pauli channels}.
\newblock {\em Physical Review A}, 78(6):062335, dec 2008.

\bibitem[GA15]{Gallego2015}
Rodrigo Gallego and Leandro Aolita.
\newblock {Resource Theory of Steering}.
\newblock {\em Physical Review X}, 5(4):41008, 2015.

\bibitem[GB14]{cvx}
Michael Grant and Stephen Boyd.
\newblock {CVX}: Matlab software for disciplined convex programming, version 2.1.
\newblock \url{http://cvxr.com/cvx}, March 2014.

\bibitem[GFG12]{ghosh2012}
Joydip Ghosh, Austin~G. Fowler, and Michael~R. Geller.
\newblock Surface code with decoherence: An analysis of three superconducting architectures.
\newblock {\em Physical Review A}, 86:062318, December 2012.

\bibitem[HHHH09]{Horodecki_2009}
Ryszard Horodecki, Paweł Horodecki, Michał Horodecki, and Karol Horodecki.
\newblock Quantum entanglement.
\newblock {\em Reviews of Modern Physics}, 81(2):865–942, June 2009.

\bibitem[HKY{\etalchar{+}}24]{Hsu2024}
Ming-Chien Hsu, En-Jui Kuo, Wei-Hsuan Yu, Jian-Feng Cai, and Min-Hsiu Hsieh.
\newblock Quantum state tomography via nonconvex riemannian gradient descent.
\newblock {\em Physical Review Letters}, 132:240804, Jun 2024.

\bibitem[HMJT25]{Andi2025}
Andi Han, Bamdev Mishra, Pratik Jawanpuria, and Akiko Takeda.
\newblock A framework for bilevel optimization on riemannian manifolds.
\newblock In {\em Proceedings of the 38th International Conference on Neural Information Processing Systems}, NIPS '24, Red Hook, NY, USA, 2025. Curran Associates Inc.

\bibitem[HU17]{Hosseini2017}
Seyedehsomayeh Hosseini and Andr\'{e} Uschmajew.
\newblock A riemannian gradient sampling algorithm for nonsmooth optimization on manifolds.
\newblock {\em SIAM Journal on Optimization}, 27(1):173--189, 2017.

\bibitem[Inc22]{MATLAB}
The~MathWorks Inc.
\newblock Matlab version: 9.13.0 (r2022b), 2022.

\bibitem[JD24]{jabbour2024tightening}
Michael~G Jabbour and Nilanjana Datta.
\newblock Tightening continuity bounds for entropies and bounds on quantum capacities.
\newblock {\em IEEE Journal on Selected Areas in Information Theory}, 2024.

\bibitem[JKL{\etalchar{+}}25]{Michael2025}
Michael~I. Jordan, Guy Kornowski, Tianyi Lin, Ohad Shamir, and Manolis Zampetakis.
\newblock Deterministic nonsmooth nonconvex optimization, 2025.

\bibitem[Joh16]{qetlab}
Nathaniel Johnston.
\newblock {QETLAB}: A {MATLAB} toolbox for quantum entanglement, version 0.9.
\newblock \url{https://qetlab.com}, Jan 2016.

\bibitem[KBHM24]{Kotil_2024}
Ayse Kotil, Rahul Banerjee, Qunsheng Huang, and Christian~B Mendl.
\newblock Riemannian quantum circuit optimization for hamiltonian simulation.
\newblock {\em Journal of Physics A: Mathematical and Theoretical}, 57(13):135303, March 2024.

\bibitem[KDWW19]{Kaur2018}
Eneet Kaur, Siddhartha Das, Mark~M. Wilde, and Andreas Winter.
\newblock {Extendibility Limits the Performance of Quantum Processors}.
\newblock {\em Physical Review Letters}, 123(7):070502, aug 2019.

\bibitem[Kel11]{kelley2011implicit}
Carl~T Kelley.
\newblock {\em Implicit filtering}.
\newblock SIAM, 2011.

\bibitem[KFG22]{kianvashBoundingQuantumCapacity2022}
Farzad Kianvash, Marco Fanizza, and Vittorio Giovannetti.
\newblock Bounding the quantum capacity with flagged extensions.
\newblock {\em Quantum}, 6:647, February 2022.

\bibitem[KSW20]{Khatri_2020}
Sumeet Khatri, Kunal Sharma, and Mark~M. Wilde.
\newblock Information-theoretic aspects of the generalized amplitude-damping channel.
\newblock {\em Physical Review A}, 102(1), July 2020.

\bibitem[KW17]{Kaur_2017}
Eneet Kaur and Mark~M Wilde.
\newblock Amortized entanglement of a quantum channel and approximately teleportation-simulable channels.
\newblock {\em Journal of Physics A: Mathematical and Theoretical}, 51(3):035303, December 2017.

\bibitem[KW24]{khatri2024}
Sumeet Khatri and Mark~M. Wilde.
\newblock Principles of quantum communication theory: A modern approach, 2024.

\bibitem[LCD{\etalchar{+}}21]{Xiao2021}
Xiao Li, Shixiang Chen, Zengde Deng, Qing Qu, Zhihui Zhu, and Anthony Man-Cho~So.
\newblock Weakly convex optimization over stiefel manifold using riemannian subgradient-type methods.
\newblock {\em SIAM Journal on Optimization}, 31(3):1605--1634, 2021.

\bibitem[LDS18]{Leditzky2018usefulsates}
Felix Leditzky, Nilanjana Datta, and Graeme Smith.
\newblock Useful states and entanglement distillation.
\newblock {\em IEEE Transactions on Information Theory}, 64(7):4689–4708, July 2018.

\bibitem[LHZ{\etalchar{+}}25]{Li2025}
Ze-Tong Li, Xin-Lin He, Cong-Cong Zheng, Yu-Qian Dong, Tian Luan, Xu-Tao Yu, and Zai-Chen Zhang.
\newblock Quantum comb tomography via learning isometries on stiefel manifold.
\newblock {\em Physical Review Letters}, 134:010803, January 2025.

\bibitem[LL18]{Youngrong2018}
Youngrong Lim and Soojoon Lee.
\newblock Activation of the quantum capacity of gaussian channels.
\newblock {\em Physical Review A}, 98:012326, Jul 2018.

\bibitem[Llo97]{Lloyd1997}
Seth Lloyd.
\newblock {Capacity of the noisy quantum channel}.
\newblock {\em Physical Review A}, 55(3):1613--1622, March 1997.

\bibitem[LLS18]{Felix2018}
Felix Leditzky, Debbie Leung, and Graeme Smith.
\newblock Dephrasure channel and superadditivity of coherent information.
\newblock {\em Physical Review Letters}, 121:160501, Oct 2018.

\bibitem[LLS{\etalchar{+}}23]{Felix2023}
Felix Leditzky, Debbie Leung, Vikesh Siddhu, Graeme Smith, and John~A. Smolin.
\newblock Generic nonadditivity of quantum capacity in simple channels.
\newblock {\em Physical Review Letters}, 130:200801, May 2023.

\bibitem[LRFO21]{Luchnikov_2021}
Ilia Luchnikov, Alexander Ryzhov, Sergey Filippov, and Henni Ouerdane.
\newblock Qgopt: Riemannian optimization for quantum technologies.
\newblock {\em SciPost Physics}, 10(3), March 2021.

\bibitem[LRW{\etalchar{+}}18]{Lami2018c}
Ludovico Lami, Bartosz Regula, Xin Wang, Rosanna Nichols, Andreas Winter, and Gerardo Adesso.
\newblock {Gaussian quantum resource theories}.
\newblock {\em Physical Review A}, 98(2):022335, aug 2018.

\bibitem[LRWW23]{Lami2023}
Ludovico Lami, Bartosz Regula, Xin Wang, and Mark~M. Wilde.
\newblock {Upper bounds on the distillable randomness of bipartite quantum states}.
\newblock In {\em 2023 IEEE Information Theory Workshop (ITW)}, pages 203--208. IEEE, apr 2023.

\bibitem[LSM25]{Le2025riemannianquantum}
Isabel Nha~Minh Le, Shuo Sun, and Christian~B. Mendl.
\newblock Riemannian quantum circuit optimization based on matrix product operators.
\newblock {\em {Quantum}}, 9:1833, August 2025.

\bibitem[LTAL19]{Youngrong2019}
Youngrong Lim, Ryuji Takagi, Gerardo Adesso, and Soojoon Lee.
\newblock Activation and superactivation of single-mode gaussian quantum channels.
\newblock {\em Physical Review A}, 99:032337, Mar 2019.

\bibitem[Miz73]{mizumoto1973finite}
Hisao Mizumoto.
\newblock A finite-difference method on a riemann surface.
\newblock {\em Hiroshima Mathematical Journal}, 3(2):277--332, 1973.

\bibitem[NC10]{nielsen2010quantum}
Michael~A Nielsen and Isaac~L Chuang.
\newblock {\em Quantum computation and quantum information}.
\newblock Cambridge university press, 2010.

\bibitem[NPJ20]{Noh_2020}
Kyungjoo Noh, Stefano Pirandola, and Liang Jiang.
\newblock Enhanced energy-constrained quantum communication over bosonic gaussian channels.
\newblock {\em Nature Communications}, 11(1), January 2020.

\bibitem[NS17]{nesterov2017random}
Yurii Nesterov and Vladimir Spokoiny.
\newblock Random gradient-free minimization of convex functions.
\newblock {\em Foundations of Computational Mathematics}, 17(2):527--566, 2017.

\bibitem[Rai01]{Rains2001}
E.M. Rains.
\newblock {A semidefinite program for distillable entanglement}.
\newblock {\em IEEE Transactions on Information Theory}, 47(7):2921--2933, 2001.

\bibitem[Ren05]{Renner2006}
Renato Renner.
\newblock Security of quantum key distribution.
\newblock {\em Ph.D. thesis}, 1, ETH Zurich 2005.

\bibitem[RFWA18]{Regula2017c}
Bartosz Regula, Kun Fang, Xin Wang, and Gerardo Adesso.
\newblock {One-shot coherence distillation}.
\newblock {\em Physical Review Letters}, 121(1):010401, July 2018.

\bibitem[SAJOS24]{Siddhu2024}
Vikesh Siddhu, Dina Abdelhadi, Tomas Jochym-O'Connor, and John Smolin.
\newblock Entanglement sharing across a damping-dephasing channel.
\newblock In {\em 2024 IEEE International Symposium on Information Theory (ISIT)}, pages 1432--1437, 2024.

\bibitem[SG21]{Vikesh2021}
Vikesh Siddhu and Robert~B. Griffiths.
\newblock Positivity and nonadditivity of quantum capacities using generalized erasure channels.
\newblock {\em IEEE Transactions on Information Theory}, 67(7):4533--4545, 2021.

\bibitem[Sho02]{Shor2002a}
Peter~W Shor.
\newblock {The quantum channel capacity and coherent information}.
\newblock In {\em lecture notes, MSRI Workshop on Quantum Computation}, 2002.

\bibitem[SJL22]{Sow2022}
Daouda Sow, Kaiyi Ji, and Yingbin Liang.
\newblock On the convergence theory for hessian-free bilevel algorithms, 2022.

\bibitem[Smi10]{Graeme2010}
Graeme Smith.
\newblock Quantum channel capacities.
\newblock In {\em 2010 IEEE Information Theory Workshop}, pages 1--5, 2010.

\bibitem[SN96]{Schumacher1996a}
Benjamin Schumacher and Michael~A Nielsen.
\newblock {Quantum data processing and error correction}.
\newblock {\em Physical Review A}, 54(4):2629--2635, oct 1996.

\bibitem[SS96]{Shor1996}
Peter~W. Shor and John~A. Smolin.
\newblock Quantum error-correcting codes need not completely reveal the error syndrome, 1996.

\bibitem[SS07]{Smith2007}
Graeme Smith and John~A. Smolin.
\newblock {Degenerate Quantum Codes for Pauli Channels}.
\newblock {\em Physical Review Letters}, 98(3):030501, jan 2007.

\bibitem[SSWR17]{Sutter2014approxepChannel}
David Sutter, Volkher~B Scholz, Andreas Winter, and Renato Renner.
\newblock {Approximate Degradable Quantum Channels}.
\newblock {\em IEEE Transactions on Information Theory}, 63(12):7832--7844, dec 2017.

\bibitem[SXON21]{shi2021numerical}
Hao-Jun~Michael Shi, Melody~Qiming Xuan, Figen Oztoprak, and Jorge Nocedal.
\newblock On the numerical performance of derivative-free optimization methods based on finite-difference approximations.
\newblock {\em arXiv preprint arXiv:2102.09762}, 2021.

\bibitem[TR19]{Takagi2019}
Ryuji Takagi and Bartosz Regula.
\newblock {General Resource Theories in Quantum Mechanics and Beyond: Operational Characterization via Discrimination Tasks}.
\newblock {\em Physical Review X}, 9(3):031053, sep 2019.

\bibitem[TWW17]{tomamichel2017strong}
Marco Tomamichel, Mark~M Wilde, and Andreas Winter.
\newblock Strong converse rates for quantum communication.
\newblock {\em IEEE Transactions on Information Theory}, 63(1):715--727, 2017.

\bibitem[Ume62]{Umegaki1962}
Hisaharu Umegaki.
\newblock {Conditional expectation in an operator algebra. IV. Entropy and information}.
\newblock {\em Kodai Mathematical Seminar Reports}, 14(2):59--85, 1962.

\bibitem[VB96]{Vandenberghe1996}
Lieven Vandenberghe and Stephen Boyd.
\newblock Semidefinite programming.
\newblock {\em SIAM review}, 38(1):49--95, 1996.

\bibitem[Wan21]{wang2021pursuing}
Xin Wang.
\newblock Pursuing the fundamental limits for quantum communication.
\newblock {\em IEEE Transactions on Information Theory}, 67(7):4524--4532, 2021.

\bibitem[WBHK20]{Wilde_2020}
Mark~M. Wilde, Mario Berta, Christoph Hirche, and Eneet Kaur.
\newblock Amortized channel divergence for asymptotic quantum channel discrimination.
\newblock {\em Letters in Mathematical Physics}, 110(8):2277–2336, June 2020.

\bibitem[WFD19]{Wang2019}
Xin Wang, Kun Fang, and Runyao Duan.
\newblock {Semidefinite Programming Converse Bounds for Quantum Communication}.
\newblock {\em IEEE Transactions on Information Theory}, 65(4):2583--2592, apr 2019.

\bibitem[Wil11]{wilde2011classical}
Mark~M Wilde.
\newblock From classical to quantum shannon theory.
\newblock {\em arXiv preprint arXiv:1106.1445}, 2011.

\bibitem[Wil13]{wilde2013quantum}
Mark~M Wilde.
\newblock {\em Quantum information theory}.
\newblock Cambridge university press, 2013.

\bibitem[Win16]{winterTightUniformContinuity2016}
Andreas Winter.
\newblock Tight {{Uniform Continuity Bounds}} for {{Quantum Entropies}}: {{Conditional Entropy}}, {{Relative Entropy Distance}} and {{Energy Constraints}}.
\newblock {\em Communications in Mathematical Physics}, 347(1):291--313, October 2016.

\bibitem[WJZ23]{Wang2023b}
Xin Wang, Mingrui Jing, and Chengkai Zhu.
\newblock {Computable and Faithful Lower Bound for Entanglement Cost}.
\newblock {\em arXiv preprint arXiv:2311.10649}, nov 2023.

\bibitem[WWS19]{WWS19}
Xin Wang, Mark~M Wilde, and Yuan Su.
\newblock {Quantifying the magic of quantum channels}.
\newblock {\em New Journal of Physics}, 21(10):103002, oct 2019.

\bibitem[WWS20]{Wang2018e}
Xin Wang, Mark~M Wilde, and Yuan Su.
\newblock {Efficiently Computable Bounds for Magic State Distillation}.
\newblock {\em Physical Review Letters}, 124(9):090505, mar 2020.

\bibitem[WZM25]{Wu2025}
Zhen Wu, Qi~Zhao, and Zhihao Ma.
\newblock Super-additivity of quantum capacity in simple channels, 2025.

\bibitem[Zhu25]{coderepo}
Chengkai Zhu.
\newblock Geometric optimization for quantum communication.
\newblock \url{https://github.com/Chengkai-Zhu/qcapacity-rieopt}, 2025.

\bibitem[ZLZW24]{Zhu2024}
Chengkai Zhu, Zhiping Liu, Chenghong Zhu, and Xin Wang.
\newblock {Limitations of Classically Simulable Measurements for Quantum State Discrimination}.
\newblock {\em Physical Review Letters}, 133(1):010202, jul 2024.

\bibitem[ZPGW25]{zhu2025R}
Chengkai Zhu, Renfeng Peng, Bin Gao, and Xin Wang.
\newblock Riemannian optimization for holevo capacity, 2025.

\bibitem[ZZAZ25]{Zhu_2025}
Xuanran Zhu, Chao Zhang, Zheng An, and Bei Zeng.
\newblock Unified framework for calculating convex roof resource measures.
\newblock {\em npj Quantum Information}, 11(1), April 2025.

\bibitem[ZZW{\etalchar{+}}21]{Zhao2021}
Xuanqiang Zhao, Benchi Zhao, Zihe Wang, Zhixin Song, and Xin Wang.
\newblock {Practical distributed quantum information processing with LOCCNet}.
\newblock {\em npj Quantum Information}, 7(1):159, dec 2021.

\bibitem[ZZW24]{Zhu_2024}
Chengkai Zhu, Chenghong Zhu, and Xin Wang.
\newblock Estimate distillable entanglement and quantum capacity by squeezing useless entanglement.
\newblock {\em IEEE Journal on Selected Areas in Communications}, 42(7):1850–1860, July 2024.

\end{thebibliography}

\end{document}